\documentclass[
letterpaper,
showpacs,floatfix, aps,pre,amsmath,amssymb,
twocolumn,
superscriptaddress,
,eqsecnum
]{revtex4}

\usepackage{graphicx}   

\usepackage{hyperref}   
\usepackage{amsmath}    
\usepackage{amsthm}
\usepackage{amssymb}

\graphicspath{
  {figs/}
  }

\newcommand{\be}{\begin{equation}}
\newcommand{\ee}{\end{equation}}
\newcommand{\bea}{\begin{eqnarray}}
\newcommand{\eea}{\end{eqnarray}}
\newcommand{\bs}{\begin{split}}
\newcommand{\bes}{\begin{equation}\begin{split}}
\newcommand{\ees}{\end{split} \end{equation}}
\newcommand{\es}{\end{split}}
\newcommand{\nl}{\nonumber\\}
\newcommand{\lpr}{\left(}
\newcommand{\rpr}{\right)}
\newcommand{\lbr}{\left[}
\newcommand{\rbr}{\right]}
\newcommand{\lcr}{\left\{}
\newcommand{\rcr}{\right\}}
\newcommand{\mylang}{\left\langle}
\newcommand{\myrang}{\right\rangle}
\newcommand{\areq}{&=&}

\newcommand{\req}[1]{Eq.~(\ref{#1})}
\newcommand{\reqs}[1]{Eqs.~(\ref{#1})}
\newcommand{\myref}[1]{(\ref{#1})}

\newcommand{\vc}[1]{{\mathbf{#1}}}
\newcommand{\vct}[1]{{\boldsymbol{#1}}}

\renewcommand{\Im}{\mathrm{Im}}
\renewcommand{\Re}{\mathrm{Re}}
\newcommand{\trans}{^{{\scriptscriptstyle \mathrm{T}}}}

\newcommand{\Tr}{\mathrm{Tr}\,}
\newcommand{\tr}{\mathrm{tr}\,}
\newcommand{\one}{\mathbf{1}}
\newcommand{\0}{\mathbf{0}}

\newcommand{\C}{\mathcal{C}}
\newcommand{\D}{\mathcal{D}}

\newcommand{\F}{\mathcal{F}}
\newcommand{\G}{\mathcal{G}}

\newcommand{\J}{\mathcal{J}}
\newcommand{\K}{\mathcal{K}}

\renewcommand{\S}{\mathcal{S}}
\newcommand{\T}{\mathcal{T}}
\newcommand{\U}{\mathcal{U}}
\newcommand{\V}{\mathcal{V}}

\newcommand{\vrr}{\vct{r}}

\newcommand{\ve}{\vc{e}}

\newcommand{\vi}{\vc{I}}

\newcommand{\vn}{\vct{n}}

\newcommand{\vx}{\vc{x}}
\newcommand{\vv}{\vct{v}}
\newcommand{\vcv}{\vc{v}}
\newcommand{\vu}{\vc{u}}

\newcommand{\vy}{\vc{y}}

\newcommand{\defin}{\equiv}

\newlength{\intwidth}

\newcommand{\zbar}{\bar{z}}

\newcommand{\Javg}[1]{\mylang #1 \myrang_{{\!\!J}}}
\newcommand{\Jstd}{\sigma}

\newcommand{\A}{A}

\newcommand{\sigp}{\sigma^{{ +}}}
\newcommand{\sigm}{\sigma^{{ -}}}

\newcommand{\Bb}{\boldsymbol{B}}
\newcommand{\tB}{B_{\scriptscriptstyle R}}
\newcommand{\tC}{C_{\scriptscriptstyle L}}
\newcommand{\rhob}{\boldsymbol{C}}
\newcommand{\Gb}{\boldsymbol{G}}
\newcommand{\origGF}{G(z;J)}

\newcommand{\tGb}{\tilde{\boldsymbol{G}}}
\newcommand{\GJ}[1]{\Gb(#1;J)}

\newcommand{\tH}{\tilde{H}}

\newcommand{\dbar}{\partial_{\zbar}}

\newcommand{\sL}{L}
\newcommand{\sR}{R}

\newcommand{\Fr}{_{\rm \scriptstyle F}}

\newcommand{\BB}{K}

\newcommand{\ignore}[1]{\index{ignore}}
\def\eg{{\it e.g.}}
\def\ie{{\it i.e.}}
\def\vectwo#1#2{\left(\begin{array}{c}  #1 \\ #2 \end{array}\right)}

\def\mat#1#2#3#4{\left(\begin{array}{cc} #1 & #2 \\ #3 & #4 \end{array}\right)}

\newcommand{\avgsig}[1]{\mylang #1 \myrang_{{\! \sigma}}}
\newcommand{\avgc}[1]{\mylang #1 \myrang_{{\! c}}}
\newcommand{\avgsigp}[1]{\mylang #1 \myrang_{{\! \sigma}}'}

\begin{document}

\title{Properties of networks with partially structured and partially random connectivity}
\author{Yashar Ahmadian}
    \email[Corresponding author.\\]{ya2005@columbia.edu}
\affiliation{Center for Theoretical Neuroscience, Department of Neuroscience,}
\affiliation{Swartz Program in Theoretical Neuroscience, and Kavli Institute for
  Brain Science, College of Physicians and Surgeons, Columbia
  University, NY, NY 10032}
\author{Francesco Fumarola}
\affiliation{Center for Theoretical Neuroscience, Department of Neuroscience,}
\author{Kenneth D.\ Miller}
\affiliation{Center for Theoretical Neuroscience, Department of Neuroscience,}
\affiliation{Swartz Program in Theoretical Neuroscience, and Kavli Institute for
  Brain Science, College of Physicians and Surgeons, Columbia
  University, NY, NY 10032}


\begin{abstract}

Networks studied in many disciplines, including neuroscience and mathematical biology, have connectivity that may be stochastic about some underlying mean connectivity represented by a nonnormal matrix. Furthermore the stochasticity may not be i.i.d. across elements of the connectivity matrix. 
More generally, the problem of understanding the behavior of stochastic matrices with nontrivial mean structure and correlations arises in many settings. We address this by characterizing large random $N\times N$ matrices of the form $A = M + LJR$, where $M$, $L$ and $R$ are arbitrary deterministic matrices and $J$ is a random matrix of zero-mean independent and identically distributed  elements. $M$ can be nonnormal, and $L$ and $R$ allow correlations that have separable dependence on row and column indices. We first provide a general formula for the eigenvalue density of $A$.
 For $A$ nonnormal, the eigenvalues do not suffice to
specify the dynamics induced by $A$, so we also provide general
formulae for the transient evolution of the magnitude of activity and frequency power spectrum in
an $N$-dimensional linear dynamical system with a coupling matrix given by
$A$. These quantities can also be thought of as characterizing 
the stability and the magnitude of the linear response of a nonlinear network to small perturbations about a fixed point.
We derive these formulae and work them out
analytically for some examples of $M$, $L$ and $R$ motivated by
neurobiological models. 
We also  argue that the persistence  as $N\rightarrow\infty$ of a finite number of randomly distributed outlying eigenvalues outside the support of the eigenvalue density of $A$, as previously observed, arises in regions of the complex plane $\Omega$ where there are nonzero singular values of $L^{-1} (z\one  - M) R^{-1}$ (for $z\in\Omega$) that vanish as $N\rightarrow\infty$. When such singular values do not exist and $L$ and $R$ are equal to the identity,  there is a correspondence in the normalized Frobenius norm (but not in the operator norm) between the support of the spectrum of $A$ for $J$ of norm $\sigma$ and the $\sigma$-pseudospectrum of $M$.

\end{abstract}
\date{\today}

\maketitle

\section{Introduction}
\label{sec-intro}

Knowledge of the statistics of eigenvalues and eigenvectors of random matrices has applications in the modeling of phenomena relevant to a wide range of disciplines
\cite{Mehta:2004,Guhr:1998,BaiSilverstein:2006}. 
In many applications, however, the matrices of interest are not entirely random, but feature substantial deterministic structure. Furthermore this structure, as well as the disorder on top of it, are in general described by nonnormal matrices. 

In neuroscience, for example, connections between neurons typically have restricted spatial range and
show specificity with respect to neuronal type, location and response properties.
Experience-based synaptic plasticity, which underlies learning and memory, naturally gives rise to synaptic connectivity
matrices that encode aspects of the statistical structure of the sensory environment, while containing significant randomness partly due to the inherent stochasticity of particular histories of sensory experience. 
Another simple example of  structured neural connectivity is due to what is known as Dale's principle \cite{Dale:1935,Eccles:1954aa,StrataHarvey:1999}: neurons come in two main types, excitatory and inhibitory. This empirical principle imposes a certain structure on the synaptic connectivity matrix, forcing all elements in each column of the matrix, describing the 
synaptic projections of a certain neuron, to have the same sign. 
Particularly when the typical weight magnitude is much larger
than typical differences between {the magnitudes of} excitatory and
inhibitory weights,  
such a matrix can be extremely nonnormal {by some measures}, much more so than a fully 
random matrix \cite{Murphy:2009a}. Similarly, biological knowledge
imparts a great deal of structure to models of biochemical  \cite{Jeong:2000aa,Barabasi:2004aa,Zhu:2007aa,Vidal:2011aa} or ecological networks \cite{May:1972aa, Camacho:2002aa, Valdovinos:2010aa, Vermaat:2009aa,Guimera:2010aa}, and matrices characterizing
such interactions are typically nonnormal. 
Yet our knowledge of connectivity or interactions is at best probabilistic. To describe realistic
biological behavior, we must generalize from the behavior of
a fixed, regular connectivity to the expected behavior of a typical sample from an appropriate
connectivity ensemble.

Furthermore, nonnormality can lead to important dynamical properties not
seen for normal matrices \cite{Trefethen-book}. 
In general, networks with a recurrent connectivity pattern described by a
nonnormal matrix can be described as having a hidden feedforward
connectivity structure between orthogonal activity patterns, each of
which can also excite or inhibit itself
\cite{Murphy:2009a,Ganguli:2008,Goldman:2009}.  In
neural networks such hidden feedforward connectivity arises
 from the natural separation of excitatory and inhibitory neurons, 
yielding so-called ``balanced amplification" of patterns of activity
without any dynamical 
slowing \cite{Murphy:2009a}.  Underlying this is the phenomenon of ``transient
amplification": a small perturbation  from a fixed
point of a stable system with nonnormal connectivity can lead to a
large transient response over finite time
\cite{Trefethen-book}. Transient amplification also 
yields unexpected results in ecological networks \cite{Neubert:1997aa,Chen:2001aa,Tang:2014aa}, and has  been conjectured to play a key role in many biochemical systems \cite{McCoy:2013aa}. 
Networks that yield long hidden feedforward chains can also generate
long time scales and provide a substrate for working memory
\cite{Ganguli:2008,Goldman:2009}. Systems with nonnormal connectivity can also exhibit pseudo-resonance frequencies in their power-spectrum at which the system responds strongly to external inputs, even though the external frequency is not close to any of the system's natural frequencies as determined by its eigenvalues \cite{Trefethen-book}.

While Hermitian random matrices and fully random non-Hermitian matrices with zero-mean, independent and identically distributed (iid) elements have been widely studied, there is a shortage of results on quantities of interest for nonnormal matrices that fall in between the two extremes of fully random or fully deterministic. 
A natural departure from a nonnormal deterministic structure,
described by a connectivity matrix $M$, is to additively perturb it
with a fully random matrix $J$ with zero-mean, iid elements. In many
important examples, however, the strength of disorder (deviations from
the mean structure) is not uniform and itself has some structure
(\eg\ for each connection it can depend on the types of the connected
nodes or neurons). Moreover, the deviations of the strength of
different connections or interactions from their average need not be independent. Hence it is important to move beyond a simple iid deviation from the mean structure. 
Here, we study ensembles of large $N\times N$ random matrices of the form $A = M + L J R$ where $M$, $L$  and $R$ are arbitrary {($M$) or
arbitrary invertible ($L$ and $R$)} deterministic matrices that are in general nonnormal, and $J$ is a completely random matrix with zero-mean iid elements of variance $1/N$. The matrix $M$ is thus the average of $A$, and describes average connectivity. 
Note that when $L$ and $R$ are diagonal, they specify variances that
depend separably on the row and column of $A$; while when they are
not diagonal, the elements of $A$ are not statistically independent.
As we show in Sec.~\ref{sec-res-merav}, this form arises naturally,
for example, in linearizations of dynamical systems involving simple
classes of nonlinearities.
This type of ensemble is also natural from the random matrix theory
viewpoint, as it describes a classical fully random ensemble -- an iid
random matrix $J$ -- modified by
the two basic algebraic operations of matrix multiplication and addition.

We study the eigenvalue distribution of such matrices, but also directly study the dynamics of a linear system of differential equations governed by such matrices. 
Specifically,   for matrices of the above type, using the Feynman diagram technique in the large $N$ limit (we follow the particular version of this method developed by  Refs.~\cite{FeinbergZee:1997,FeinbergZee97}), we have derived a general 
formula for the density of their eigenvalues in the complex plane,
which generalizes the well-known circular law for fully random matrices \cite{Ginibre:1965aa,Girko:84,Bai:1997,TaoVu:2008,GotzeTikhomirov:2010}. 
It also generalizes a result \cite{Khoruzhenko:1996aa} obtained for the case
where $L$ and $R$ are scalar multiples of  $\one$, the $N$-dimensional identity matrix 
 (the same result was obtained in 
\cite{BianeLehner:2001} using the methods and language of free probability theory; the eigenvalue density for the case $L \propto R \propto \one$ and a \emph{normal} $M$ was also calculated in Ref.~\cite{FeinbergZee97} in the limit $N\to\infty$, and that result was extended to finite $N$ in Ref.~\cite{Hikami:1998aa}).  
Apart from generalization to arbitrary invertible $L$ and $R$, we also provide a correct regularizing procedure for finding the support of the eigenvalue density in the limit $N\to\infty$, 
in certain highly nonnormal cases of $M$; the naive interpretation of the formulae fails in these cases, which were not previously discussed.
Furthermore, with the aim of studying dynamical signatures of nonnormal connectivity, we focused on the dynamics directly, 
deriving general formulae for the magnitude of the response of the system to a delta function pulse of input (which provides a measure of the time-course of potential transient amplification), as well as the frequency power spectrum of the system's  response to external time-dependent inputs. 

These general results are presented in the next section. There, we also present the explicit results of analytical or numerical calculations based on these general formulae  for some specific examples of $M$, $L$ and $R$. 
Sections~\ref{sec-deriv1} and \ref{sec-deriv2} contain the detailed derivations of our general formulae for the eigenvalue density and the {response magnitude} formulae, respectively. 
 Section \ref{sec-example} contains the detailed analytical calculations of these quantities for the specific examples presented in Sec.~\ref {sec-results}, based on the general 
formulae. We conclude the paper in Sec.~\ref{sec-concl}.

\section{Summary of results}
\label{sec-results}

We study ensembles of large $N\times N$ random matrices of the form 
\be\label{ensemble}
\A = M+ L J R,
\ee
 where $M$, $L$ and $R$ are arbitrary {($M$) or
arbitrary invertible ($L$ and $R$)} deterministic matrices \footnote{Since we will present results for the limit of the eigenvalue density, etc, as $N\to\infty$,   $M$, $L$, $R$ and $J$ must each be more precisely understood as an infinite sequence of matrices dependent on $N$.}, and $J$ is a random matrix of independent and identically distributed (iid)  elements with zero mean and  variance $1/N$. 
 Since $J$ and therefore $LJR$ have zero mean, $M$ is the ensemble average of $\A$. The random fluctuations of $A$ around its average are given by the matrix $LJR$, which for general $L$ and/or $R$ has dependent and non-identically distributed elements, due to the possible mixing and non-uniform scaling of the rows (columns) of the iid $J$ by $L$ ($R$). 
     
We are firstly interested in the statistics of the eigenvalues of \req{ensemble}. While the statistics of the eigenvalues and eigenvectors of $\A$ are of interest in their own right, we also
directly  consider certain properties of the linear dynamical system 
\be\label{ode}
\frac{d \vx(t)}{dt} = -\gamma\vx(t) +  \A \vx(t) + \vi(t),
\ee
for an $N$-dimensional state vector $\vx(t)$, when $A$ is a sample of the ensemble \req{ensemble}. Here, $\gamma$ is a scalar and $\vi(t)$ is an external, time-dependent input.
In studying this system, we generally assume that \req{ode} is asymptotically stable. This means that $M$, $L$, $R$ and $\gamma$ must be chosen such that for any typical realization of $J$,  no eigenvalue of $-\gamma\one + M + LJR$ has a positive real part;  this can normally be achieved, for example,  by choosing a large enough $\gamma>0$.

Using the diagrammatic technique in the non-crossing approximation,
which is valid for large $N$, we have derived general formulae for
several useful properties of such matrices involving their eigenvalues
and eigenvectors (see Sec.~\ref{sec-deriv1}--\ref{sec-deriv2} { for the details of the derivations and the definition of the non-crossing approximation}). We present these results in in this section. 
In our derivation of these results, we assume the random $J$ belongs to the complex Ginibre ensemble \cite{Ginibre:1965aa}, \ie\ the distribution of the elements of $J$ is complex Gaussian. 
However, we emphasize that universality theorems  ensure that, for given $M$, $L$ and $R$, the obtained result for the eigenvalue density in the limit $N \to \infty$ will not depend on the exact choice of the distribution of the elements of $J$, beyond its first two moments, and extend to \emph{any} iid $J$ (including, \eg, $J$ with real binary or log-normal elements) whose
elements have the same first two moments, \ie\ zero mean and variance $1/N$; the universality of the eigenvalue density for general $M$, $L$ and $R$ was established in Ref.~\cite{TaoKrishnapur:2010}, following earlier work on the universality of
the circular law established and successively strengthened in Refs.~\cite{Ginibre:1965aa,Girko:84,Bai:1997,TaoVu:2008,GotzeTikhomirov:2010}. 
 { Furthermore, empirically, from limited simulations, we have thus far found (but have not proved) such
universal behavior to also hold for the other quantities we compute
here (however, it is possible that universality for these
quantities might require
the existence of some higher moments beyond the
second, as has been found for universality of certain other properties of
random matrices; see \eg\ Ref.~\cite{TaoVu:2011}).} To demonstrate the universality of our results, we have used non-Gaussian and/or real $J$'s in most of the numerical examples below.

Hereinafter, we adopt the following notations. 
For any matrix $B$, we denote its operator norm (its maximum singular value) by $\| B\|$ and we define its (normalized) Frobenius norm via  
 \be\label{frobnorm}
 \|B\|_{\Fr}^2 \defin  {\frac{1}{N} \sum_{ij} |B_{ij}|^2}  =  {\frac{1}{N} \Tr\! (B B^\dagger)}
 \ee
  (equivalently, $\|B\|_{\Fr}$ is the root mean square of the singular values of  $B$). 
For general matrices, $A$ and $B$, 
\be
\tr\!(A) \!\defin\!  \frac{1}{N}\Tr\!(A),\,\, A^{-\dagger} \!\defin\!  (A^{\dagger})^{-1},\,\,
 \frac{1}{A} \!\defin\!  A^{-1},\,\,  \frac{A}{B}\!\defin\!  A B^{-1},
\nonumber
\ee
 and when adding a scalar to a matrix, it is implied that the scalar is multiplied  by the appropriate identity matrix. We denote the identity matrix in any dimension (deduced from the context) by $\one$. For a complex variable $z=x + iy$, the Dirac delta function is defined by $\delta^2(z) \defin  \delta(x) \delta(y)$, and we define  $\partial_{\zbar}\defin  \partial/\partial \zbar = (\partial/\partial x + i \partial/\partial y )/2$, and $\partial_z \defin \partial/\partial z = (\partial/\partial x - i \partial/\partial y )/2$. For simplicity, we use the notation $f(z)$ (instead of $f(z,\zbar)$) for general, nonholomorphic functions on the complex plane.
 We say a quantity is $O(f(N))$ (resp. $\Theta(f(N))$) when, for large enough $N$, the absolute value of that quantity is bounded above (resp. above \emph{and} {below}) by a fixed positive multiple of  $|f(N)|$. 
Finally, we say a quantity is $o(f(N))$ when its ratio to $|f(N)|$ vanishes as $N\to\infty$.

 The only conditions we impose on $M$, $L$ and $R$ are that $\| M \|_{\Fr}$, $\|L\|_{\Fr}$, $\| R\|_{\Fr}$, $\| L^{-1} M R^{-1}\|_{\Fr}$ and $\| (LR)^{-1}\|$ are bounded as $N\to \infty$. 
We use the bound on $\| (LR)^{-1}\|$ in Appendices \ref {app-nonxing} and \ref {app-rho}; the Frobenius norm conditions are assumptions in the universality theorem of Ref.~\cite{TaoKrishnapur:2010} which we use as discussed above.  
Finally, we assume that  for all $z\in \mathbb{C}$, the distribution of the eigenvalues of $M_z M_z^\dagger$, where $M_z$ is defined below in \req{Mzdef}, tends to a limit distribution as $N\to\infty$. This last condition simply makes precise the requirement that $M$, $L$ and $R$ are defined consistently as functions of $N$, such that a limit spectral density for $M+LJR$ is meaningful; in particular,  it  does not impose any further limits on the growth of the eigenvalues of $M_z M_z^\dagger$ with $N$, beyond the various norm bounds imposed above.

\subsection{Spectral density}
\label{subsec-specdensity}

\subsubsection{Summary of results}

The  density of  the eigenvalues   of $M+ L J R$ in the complex plane for a realization of $J$  (also known as the empirical spectral distribution) is   defined by
\be\label{rhoJdef}
\rho_J(z) = \frac{1}{N} \sum_\alpha \delta^2(z - \lambda_\alpha),
\ee
where $\lambda_\alpha$ are the eigenvalues of $M+ L J R$. 
It is known \cite{TaoKrishnapur:2010} that $\rho_J(z)$ is asymptotically self-averaging, in the sense that with probability one $\rho_J(z) - \rho(z)$ converges to zero (in the distributional sense) as $N\to \infty$, where $\rho(z) \defin  
\mylang \rho_J(z) \myrang_{{\!\!J}}$ is the ensemble average of $\rho_J(z)$. 
Thus for large enough $N$, any typical realization of $J$ yields an eigenvalue density $\rho_J(z)$ that is arbitrarily close to $\rho(z)$.

Our  general result is that  for large $N$,
with certain cautions and
excluding certain special
cases as described below (\reqs{Kdef1}--\myref{support-res-K} and
preceding discussion), 
$\rho(z)$ is nonzero in the region of the complex plane satisfying 
\be\label{support-res}
\tr\!\!\lbr {(M_z M_z^\dagger)^{-1}} \rbr \geq 1
\ee
where we defined
\be\label{Mzdef}
M_z \defin  \sL^{-1} (z  - M) \sR^{-1}.
\ee
Using the definition \req {frobnorm}, we can also express \req{support-res} as
\be\label{support-res-2}
 \left \| R \frac{1}{z -M} L \right \|_{\Fr} \geq 1. 
\ee
Inside this region, $\rho(z)$ is given by  
\be\label{rho-res}
 \rho(z) = \frac{1}{\pi}\frac{\partial \,}{\partial \zbar}\,\, \tr\!\!\lbr \frac{(\sR \sL)^{-1} M_z^\dagger}{M_z M_z^\dagger + g(z)^2} \rbr,
 \ee
 where $g(z)$ is a real, scalar function found by solving 
 \be\label{geqn}
    \tr\!\!\lbr \frac{1} {M_z M_z^\dagger + g^2} \rbr ={1},
 \ee
 for $g$ for each $z$. 
As a first example, for the well-known case of $M=0$, $\sL =  \one$, and $\sR = \Jstd \one$, we have $M_z = z/\Jstd$ and the circular law  follows immediately from \req{support-res}, which  yields $\Jstd^2/|z|^2 \geq 1$ or $|z| \leq \Jstd$ for the support, and from \reqs{rho-res}--(\ref{geqn}) which yield  the uniform $\rho(z) = 1/(\pi\Jstd^2)$ within that support.
As we noted in the introduction, formulae \myref{support-res}--\myref {geqn} generalize the results of Refs.~\cite{Khoruzhenko:1996aa} and \cite{BianeLehner:2001} for the special case $L \propto R \propto \one$ to arbitrary invertible $L$ and $R$ (the eigenvalue density for the case $L \propto R \propto \one$ and a \emph{normal} $M$ was also calculated in Ref.~\cite{FeinbergZee97}).

It is possible and illuminating to express \reqs{support-res-2}--\myref{geqn} exclusively in terms of 
the singular values of $M_z$, which we denote by $s_i(z)$ (we include possibly vanishing singular values among $s_i(z)$, so that we  always have $N$ of them). 
First, noting that the squared singular values of $M_z$ are the eigenvalues of the Hermitian $M_z M_z^\dagger$, we can evaluate the trace in \req{geqn} in the eigen-basis of the latter matrix, and rewrite this equation   as
\be\label{geqn-svd}
\frac{1}{N} \sum_{i=1}^N \frac{1}{s_i(z)^2 + g^2}  ={1}.
\ee
Similarly, \req{support-res}  can be equivalently rewritten as 
\be\label{support-res-3}
\frac{1}{N} \sum_{i=1}^N s_i(z)^{-2}  \geq {1}.
\ee
As we prove at the end of Sec.~\ref {sec-deriv1}, \req{rho-res} can also be written in a form that makes it explicit that the dependence of $\rho(z)$ on $M$, $L$ and $R$ is \emph{only} through the singular values of $M_z$ and their derivatives with respect to $z$ and $\zbar$. We have 
\be\label{rho-svd-res}
\rho(z) =  \frac{1}{\pi}\dbar\lbr \frac{1}{N} \sum_{i=1}^N \frac{\partial_z(s_i(z)^2)}{s_i(z)^2 + g(z)^2} \rbr.
\ee

For the special case of $M = 0$ and general $L$ and $R$,  
our formulas can be simplified considerably. The spectrum is isotropic around the origin in this case, \ie\ $\rho(z)$ depends only on $r\defin  |z|$, and its support is a disk centered at the origin with radius
\be\label{boundaryM0}
r_0 = \| RL \|_{\Fr} =  \lbr \frac{1}{N}\sum_{i=1}^N \sigma_i^2 \rbr^{1/2},
\ee
where $\sigma_i$ are the singular values of $RL$ (this follows from \req{support-res-3} by noting that for $M=0$, the singular values of $M_z = z (RL)^{-1}$  are  $s_i(z) = |z|/\sigma_i$).
Within this support the spectral density is given by 
\be\label{rhoM0}
\rho(r) = - \frac{1}{2\pi r} \partial_r\! \lpr  g(r)^2 \rpr,
\ee
where $g(r)^2>0$ is found by solving
\be\label{geqnM0}
1 =  \frac{1}{N} \sum_{i=1}^N \frac{1}{\sigma_i^{-2} r^2 + {g(r)^2} }.
\ee
Integrating \req{rhoM0},  we see that the proportion of eigenvalues lying a distance larger than $r$ from the origin is, in this case, given by 
\be\label{n>M0} 
n_>(r) = \lcr \begin{array}{l} g(r)^2 \qquad (r<r_0) 
\\
0 \quad\,\,\,\,\qquad (r\geq r_0). 
\end{array}\right.
\ee
In Sec.~\ref{sec-deriv1} we prove  that the eigenvalue density, given
by \reqs{rhoM0}--\myref{geqnM0}, is always a  decreasing function of
$r = |z|$, \ie\ for $r>0$ its derivative with respect to $r$ is
strictly negative,  as long as the limit distribution of the $\{\sigma_i\}$ as $N\to\infty$ has nonzero variance (otherwise $\rho(z)$ is given by the circular law with radius \req{boundaryM0}). The values of spectral density at $r=0$ and $r=r_0$ can be calculated explicitly for general $L$ and $R$: 
\bea\label{rhoM00r}
\rho(r=0) \areq  \frac{1}{\pi} \frac{1}{N} \sum_{i=1}^N \sigma_i^{-2}\\
\rho(r=r_0) \areq \frac{1}{\pi}  r_0^2 \lbr \frac{1}{N} \sum_{i=1}^N \sigma_i^4\rbr^{-1} \leq \rho(r=0).
\label{rhoM0r0}
\eea

As noted above, certain cautions apply in using
the above formulae for the eigenvalue density and its boundary
(\myref{support-res}--\myref{geqn}, or equivalently 
\reqs{geqn-svd}--\myref{rho-svd-res}, and for $M=0$,
\reqs{rhoM0}--\myref{geqnM0}). We have written these formulas for finite $N$ (assuming it is large). 
However, the non-crossing approximation used in deriving these formulas is only guaranteed to yield the correct result for the eigenvalue density in the limit, \ie\ $\lim_{N\to\infty}\rho(z)$ (see Appendix~\ref{app-nonxing}); finite-size corrections obtained from  \reqs{support-res}--\myref{geqn} are not in general correct, and $o(1)$ contributions to $g(z)^2$ or $\rho(z)$ obtained from \reqs{geqn} and \myref{rho-res} should be discarded.
 
Furthermore,  in general, the correct way of finding the support of $\lim_{N\to\infty}\rho(z)$ using \req{support-res} is by 
 setting the left side of the inequality
\myref{support-res} to $\lim g^2\to 0^+ \lim N\to\infty$ of the left
side of \req{geqn}, as discussed in Sec.~\ref {sec-deriv1}
and Appendix \ref{app-nonxing}. ÊHowever, in writing \req{support-res}
we have simply set $g=0$ in Eq.\ 2.9, and thus implicitly taken the
limit $g^2\to 0^+$ before the $N\to\infty$ limit. ÊTo correctly
express the support, we must first define the function
\bea\label{Kdef1}
\K(g,z) &\defin& \lim_{N\to \infty} \tr\!\!\lbr \frac{1}{M_z M_z^\dagger+g^2} \rbr
\nl
\areq \lim_{N\to \infty} \frac{1}{N} \sum_{i=1}^N \lbr \frac{1}{s_i(z)^2 + g^2}\rbr
\eea
for fixed, strictly positive $g$, which serves to regularize the denominators in \req{Kdef1} for $s_i(z)$ which are zero or vanishing in the limit $N\to\infty$. The generally correct way of expressing \req{support-res} or \req{support-res-3} is then
\be\label{support-res-K}
\K(0^+,z) \equiv \lim_{g\to 0^+} \K(g,z) \geq 1.
\ee
Let us denote the  the support of $\lim_{N\to\infty}\rho(z)$, given by \req{support-res-K}, by $\S_{0^+}$ and the region specified by the limit $N\to\infty$ of \req{support-res} or \req{support-res-3} by $\S_{0}$.
For many  examples of $M$, $L$ and $R$, the  limits ${N\to\infty}$ and $g\to 0^+$ commute everywhere and hence $\S_{0^+} = \S_{0}$.
However,  if there are $z$'s at which some of the smallest $s_i(z)$ are either zero or vanish in the limit $N\to \infty$, the two limits may fail to commute, and the naive use of \req{support-res}  can yield a region, $\S_0$, strictly larger than and  containing  $\S_{0^+}$, the correct support of $\lim_{N\to\infty}\rho(z)$.
For example, at $z$'s for which a $\Theta(1)$ number of $s_i(z)$ are zero or $o(1)$, these singular values do not make a contribution to $\K(g,z)$ for $g>0$ (their contribution to the sum in \req{Kdef1} is $O(N^{-1})$) and hence to $\K(0^+,z)$, but if they vanish sufficiently fast as $N\to\infty$ they can make a nonzero contribution to the left side of \req{support-res-3}; such $z$ may fall within $\S_{0}$, but not within  $\S_{0^+}$. 
 For finite $N$, the $s_i(z)$ can vanish exactly when $z$ coincides with an eigenvalue of $M$; thus the above situation can, \eg, arise close to eigenvalues of $M$ that are isolated and far from the rest of $M$'s spectrum, so that they fall outside the support of $\lim_{N\to\infty}\rho(z)$. 
In such cases, the spectrum of $M + LJR$ will nonetheless typically also contain isolated eigenvalues (which do not contribute to $\lim_{N\to\infty}\rho(z)$) with effectively deterministic location, \ie\ within $o(1)$ distance of corresponding isolated eigenvalues of $M$; examples of this phenomenon, for which $\S_0 - \S_{0^+}$ is not empty but has zero measure,   have been studied in Refs.~\cite{Tao:2013,ORourke:2014aa} (for symmetric matrices, outlier eigenvalues corresponding to eigenvalues of the mean matrix were first studied in Ref.~\cite{Edwards:1976aa}). 
For some choices of $M$, $L$ and $R$, however, a more interesting case
can arise such that for $z$ in a certain region of the complex plane
with nonzero measure, all $s_i(z)$ are nonzero at finite $N$ (hence
$M$ has no eigenvalue there), but a few $s_i(z)$ are $o(1)$  and vanish sufficiently fast as $N\to\infty$; 
in particular when $L\propto R\propto \one$, this can occur for certain highly nonnormal $M$ 
      \footnote{The designation ``highly nonnormal" can be motivated, when $L$ and $R$ are proportional to the
     identity matrix, as follows. Let us denote the (operator norm
     based) $\epsilon$-pseudospectrum of $M$, \ie\ the region of $z$'s
     over which $\|(z-M)^{-1}\| >\epsilon^{-1}$, by
     $\Sigma_\epsilon(M)$. For fixed $N$,  the true
     spectrum of $M$, which we denote by $\Sigma(M)$, is the set of
     points over which the smallest singular value of $(z-M)$ is
     exactly zero and hence $\| (z-M)^{-1}\| = \infty$. For
     finite $N$, $\lim_{\epsilon \to 0^+}\Sigma_\epsilon(M) =
     \Sigma(M)$ for any $M$. However, for nonnormal $M$ this approach
     could be much slower than in the normal case (see our discussion
     in Sec.~\ref{sec:pseudospec}, and the book \cite {Trefethen-book}
     for a complete discussion of pseudospectra and their relationship
     with nonnormality).   Now suppose that, as in the
     atypical cases under discussion, in a finite region of the
     complex plane the smallest singular value of $M_z$ is nonzero for
     finite $N$, but vanishes in the limit $N\to\infty$. This means
     that the operator norm of $(z-M)^{-1}\propto M_z^{-1}$ is finite
     over such a region but goes to infinity as $N\to\infty$. Hence,
     if we define $\Sigma^\infty_\epsilon(M) \defin
     \lim_{N\to\infty}\Sigma_\epsilon(M)$ and  $\Sigma^\infty(M) \defin
     \lim_{N\to\infty}\Sigma(M)$,  
     we see that in such cases $\lim_{\epsilon \to
       0^+}\Sigma^\infty_\epsilon(M) \neq \Sigma^\infty(M)$ (or equivalently,
     $\lim_{\epsilon \to 0^+}\lim_{N\to\infty}\Sigma_\epsilon(M) \neq
     \lim_{N\to\infty}\lim_{\epsilon \to
       0^+}\Sigma_\epsilon(M)$).
     More generally but less precisely, this indicates that at finite
     but large $N$, the $\epsilon$-pseudospectra of such matrices  can
     cover a significantly broader region than the spectrum even for
     very small $\epsilon$, indicating extreme nonnormality.}. 
In such cases the non-commutation of the two limits can lead to a  difference $\S_0 - \S_{0^+}$ with nonzero measure.
In cases we have examined this signifies that there exists a finite, non-vanishing region outside  the support of $\lim_{N\to\infty}\rho(z)$ (typically surrounding it) where, although $\rho(z)$ is $o(1)$, it nonetheless converges to zero sufficiently slowly that a $\Theta(1)$ number of ``outlier" eigenvalues  lie there (note that the vast majority of eigenvalues, \ie\ $(1-o(1))N$  of them, lie within the support of the limit density).
We will discuss examples of this phenomenon in Sec.~\ref {sec-examples} below; 
in one of the examples (discussed in Sec.~\ref {sec-res-rajanabbott}), the existence of such outlier eigenvalues was first noted in Ref.~\cite{Rajan:2006}, and their distribution was  quantitatively characterized in Ref.~\cite {Tao:2013}. 
 {However, the connection between such outlier eigenvalues and  nonzero but $o(1)$ singular values of $M_z$, which arise, \eg, for highly nonnormal $M$, were not noted before to the best of our knowledge.}
We have observed in simulations (and also supported by \cite{Tao:2013}) that the distribution of these outliers 
remains random
as $N\to\infty$,   is  in general less  universal than $\lim_{N\to\infty}\rho(z)$ (\eg\ it could  depend on the choice of real vs. complex ensembles for $J$), and its 
average 
behavior may not be correctly given  by the non-crossing approximation.

\subsubsection{Relationship to pseudospectra}
\label{sec:pseudospec}

Finally, we note a remarkable connection between our general result for the support of the spectrum \req{support-res} and the notion of pseudospectra, in the case in which the limits $g^2\to 0^+$ and $N\to\infty$ commute (so that \req{support-res} correctly describes the support). Pseudospectra are generalizations of eigenvalue spectra, which are particularly useful in the case of nonnormal matrices (see Ref.~\cite{Trefethen-book} for a review). The eigenvalue spectrum of matrix $M$ can be thought of as the set of points, $z$, in the complex plane where $(z-M)^{-1}$ is singular, \ie\ it has infinite norm. Given a fixed choice of matrix norm, $\|\cdot \|$, the pseudospectrum of $M$ at level $\Jstd$, or its ``$\Jstd$-pseudospectrum" in the given norm, is the set of points $z$ for which $ \| (z-M)^{-1}\| \geq \Jstd^{-1}$ (thus as $\Jstd\to 0$ we recover the spectrum). 
For the specific choice of the operator norm (\ie\ when $\| A\|$ is taken to be  the maximum singular value of 
$A$), the $\Jstd$-pseudospectrum can equivalently  be characterized as
the set of points, $z$, for which there exists a matrix perturbation
$\Delta M$, with $\| \Delta M\| \leq \Jstd$, such that $z$ is in the
eigenvalue spectrum of $M + \Delta M$
\cite{Trefethen-book}\footnote{This equivalence is true more generally
  for any matrix norm derived from a general vector norm; see
  Ref.~\cite{Trefethen-book} for a proof.}.  In words, in the operator
norm, the $\Jstd$-pseudospectrum of $M$ is the set to which its
spectrum can be perturbed by adding to it arbitrary perturbations of
size $\sigma$ or smaller.

In our setting we can think of $LJR$ as a perturbation of $M$. Let us
focus on the case where $L$ and $R$ are proportional to the identity,
\ie, we have $\Delta M = \Jstd J$, with a positive scalar
$\Jstd$. Our result \req{support-res-2}, in this case reads $\| \Jstd
(z-M)^{-1}\|_ {\Fr} \geq 1$ or $\| (z-M)^{-1}\|_ {\Fr} \geq \Jstd^{-1}$. In
other words, as $N\rightarrow\infty$, the spectrum of $M + \Jstd J$,
for an iid 	 random $J$
with zero mean and variance $1/N$, is the $\Jstd$-pseudospectrum of
$M$ in the normalized Frobenius norm defined by Eq.~\ref{frobnorm}.
Interestingly, the perturbation, $\Delta M = \Jstd J$, has
normalized Frobenius norm $\Jstd$ as $N\rightarrow\infty$: this norm is
$\Jstd \sqrt{\sum_{ij} J_{ij}^2/N }$, which,
by the law of large numbers, converges to $\Jstd$ for large $N$. 
That is, as $N\rightarrow\infty$, the spectrum
in response to the random perturbation $\Jstd J$, 
which has size $\Jstd$ (in normalized Frobenius norm), is the
$\Jstd$-pseudospectrum of $M$ in the normalized Frobenius norm. 

This result sounds similar to the equivalence of
the two definitions of pseudospectra for the operator norm which we noted above (one based on the
norm of $(z-M)^{-1}$, and one based on the spectra of bounded
perturbations), but it differs in two key respects. 
First, unlike in the case of the operator norm, the general equivalence of the two notions of pseudospectra noted above does {\em not}  hold for the normalized Frobenius norm.
Second, for the operator norm, it is {\em not} in general the case that the
$\Jstd$-pseudospectrum of $M$ is equivalent to the spectrum obtained from a single 
{\em random} perturbation of $M$ of size $\Jstd$, even in the limit
$N\rightarrow \infty$ (although the spectra arising from such random
perturbations are sometimes used as a ``poor man's version'' or
approximation of the pseudospectra \cite{Trefethen-book}). This can
be seen as follows. The operator norm of the random iid perturbation, $\Jstd J$, \ie\ its
maximum singular value, converges almost surely to $2 \Jstd $ as
$N\rightarrow\infty$ \cite{Yin:1988}. 
Condition \ref{support-res-2} for $z$ to be in the spectrum under this
random perturbation is $\left \| (z -M)^{-1}\right
\|_{\Fr} \geq \Jstd^{-1}$, or rms$(\{s_i(z)^{-1}\})\geq 1$
where the $s_i(z)$ are the singular values of
$\frac{z-M}{\Jstd}$ 
and rms$(\{x_i\})$ 
represents the root-mean-square of the set of values $\{x_i\}$. This
is not equivalent to the condition that $z$ be in the
$2\Jstd$-pseudospectrum of $M$
in the operator norm, \ie\ that
$\left \| (z -M)^{-1}\right \|  \geq (2\Jstd)^{-1}$ or 
$s_{\min}(z)^{-1}\geq \frac{1}{2}$, where $s_{\min}(z)$
is the minimum of the $s_i(z)$; in fact, noting that $s_{\min}(z)^{-1} \geq$ rms$(\{s_i(z)^{-1}\})$, it is easy to see that the spectrum under random iid perturbations with operator norm $\| \Jstd J\| = 2\Jstd$ is strictly a proper subset of the $2\Jstd$-pseudospectrum in the operator norm.  For example, for  $M=0$, the ``poor man's $2\Jstd$-pseudospectrum'' in the limit
$N\rightarrow\infty$ is a ball of radius
$\sigma$ about the origin (the circular law), while the true $2\Jstd$-pseudospectra of the zero matrix is the ball of radius $2\sigma$
about the origin.

In sum, in the operator norm, the $\Jstd$-pseudospectrum of $M$ for any $N$
is equivalent to the set of points $z$ for which  {\em some} perturbation $\Delta M$ with $\|\Delta M\|\leq \Jstd$ can be found such that $z$ is in the
spectrum of $M+\Delta M$ \cite{Trefethen-book}. 
In the normalized Frobenius norm  in the limit $N\rightarrow\infty$, however, 
the $\Jstd$-pseudospectrum of $M$  is equivalent to the spectrum 
of $M+\Delta M$ where $\Delta M$ is any {\em random} perturbation with
zero-mean iid elements with $\|\Delta M\|_{\Fr}=\Jstd$. This statement
for the normalized Frobenius norm holds when the two limits
$N\to\infty$ and $g\to 0^+$ commute; when the two limits do not
commute, the support of the spectral distribution of $M+\Delta M$ is a
subset of the $\Jstd$-pseudospectrum of $M$  in
the normalized Frobenius norm.

\subsection{Average norm squared and power spectrum}
\label{sec:genformdynamics}

As we mentioned in the introduction, an important phenomenon encountered in dynamics governed by nonnormal matrices, as described by \req{ode} with $\vi(t) =0$, 
is transient amplification in asymptotically stable systems. In any stable system, the size of the response to an initial perturbation eventually decays to zero, with an asymptotic rate set by the system's eigenvalues.
In stable nonnormal systems, however, after an initial perturbation, the size of the network activity, as measured, \eg, by its norm  squared $\| \vx(t)\|^2= \vx(t)\trans \vx(t)$, can nonetheless exhibit transient, yet possibly large and long-lasting growth, before it eventually decays to zero. 
By contrast, in stable normal systems, $\| \vx(t)\|^2$ can only decrease with time. 
The strength and even the time scale of transient amplification are set by properties of the matrix $A$ beyond its eigenvalues; they depend on the degree of nonnormality of the matrix, as measured, \eg, by the degree of non-orthogonality of its eigenvectors, or alternatively by  its hidden feedforward structure  (see \req{schur} for the latter's definition).

Nonnormal systems can also exhibit pseudo-resonances at frequencies that could be very different from their natural frequencies as determined by their eigenvalues; such pseudo-resonances will be manifested in the frequency power spectrum of the response of the system to time dependent inputs. 
$\|\vx(t)\|^2$ and the power spectrum of response are examples of quantities that depend not only on
the eigenvalues of $M + L J R$ but also on its eigenvectors. 

Here, we present a few closely related formulas for general $M$, $L$ and $R$. These include a formula for $\Javg{\|\vx(t)\|^2}$, \ie\ the ensemble average of the norm squared of the state vector, $\vx(t)$, as it evolves under \req{ode} with $\vi(t) = 0$, as well as a formula for the ensemble average of the power spectrum of the response of the network to time-varying inputs.
The results of this section are valid, and in the case of the power
spectrum meaningful, when the system \req{ode} is asymptotically
stable. As we mentioned after \req{ode},  this means that $M$, $L$,
$R$ and $\gamma$ must be chosen such that for any typical realization
of $J$,  all eigenvalues of $-\gamma\one + M + LJR$ have negative real
part. 
In particular, the entire support of the eigenvalue density of
$M+LJR$, as determined by \req{support-res}, must fall to the left of
the vertical line of $z$'s with real part $\gamma$; this is a
necessary condition, but may not be sufficient either at finite $N$ or
in cases where an $O(1)$ number of eigenvalues remain outside this
region of support even as $N\rightarrow\infty$.

First, we consider the time evolution of the squared norm,   
$\| \vx(t)\|^2$, of the response of the system to an impulse input, $\vi(t) = \vx_0\delta(t)$, at $t=0$,  before which we assume the system was in its stable fixed point $\vx = 0$ (for $t>0$ this is equivalent to the squared norm  of the activity as it evolves according to \req{ode} with $\vi(t) = 0$, starting from the initial condition $\vx(0) = \vx_0$). 
We provide a formula for the ensemble average of  the more general quadratic function, $\vx(t)\trans B \vx(t)$, where $B$ is any $N\times N$ {symmetric} matrix; the norm squared corresponds to $B = \one$. 
The result for general $B$, $M$, $L$ and $R$ is given as a double inverse Fourier transform
\bea\label{res-avgnormsq}
&&\Javg{\vx(t) \trans B\vx(t) }
=
\\
&& \qquad\quad
\!\int\!\!\!  \int\! \frac{d\omega_1}{2\pi}\frac{d\omega_2}{2\pi} e^{it(\omega_1-\omega_2)}
\Tr\!\!\lbr B\,  \C^{\vx}(\omega_1,\omega_2;\vx_0\vx_0\trans) \rbr,   
\nonumber
\eea
in terms of the $N\times N$ Fourier-domain ``covariance matrix," $\C^{\vx}(\omega_1,\omega_2;\vx_0\vx_0\trans) \defin \Javg{\tilde\vx(\omega_1)\tilde\vx(\omega_2)^\dagger} $ (where $\tilde\vx(\omega)$ is the Fourier transform of $\vx(t)$). The expression for the latter is given by 
\be\label{res-FBcdc}
\C^{\vx}(\omega_1,\omega_2;C^{\vi}) = \C^{\vx}_0(\omega_1,\omega_2;C^{\vi}) + \Delta \C^{\vx}(\omega_1,\omega_2;C^{\vi})
\ee 
where
\be\label {res-FBdiscon1}
\C^{\vx}_0(\omega_1,\omega_2;C^{\vi}) \equiv \frac{1}{\gamma + i\omega_1 -M} C^{\vi} \frac{1}{\gamma -i\omega_2 -M^\dagger},
\ee
yields the result obtained by ignoring the randomness in the connectivity (\ie\ by setting $A=M$),  and 
\bea\label {res-FBcon1}
&& \Delta \C^{\vx}(\omega_1,\omega_2;C^{\vi}) \equiv 
 \frac{1}{\gamma +i\omega_1 -M} L L^\dagger \frac{1}{\gamma -i\omega_2 -M^\dagger} \quad
  \nl
&&
 \qquad \quad\times \frac {\tr\!\!\lpr R^\dagger R \frac{1}{\gamma +i\omega_1 -M} C^{\vi} \frac{1}{\gamma -i\omega_2 -M^\dagger}    \rpr    } {1- \tr\!\!\lpr R^\dagger R \frac{1}{\gamma +i\omega_1-M} L L^\dagger \frac{1}{\gamma -i\omega_2 - M^\dagger}     \rpr},
\eea
is the contribution of the random part of connectivity $LJR$. 
For later use, we have provided these expressions for a general third argument in $\C^{\vx}(\cdot,\cdot;\cdot)$; for use in \req {res-avgnormsq} $C^{\vi}$ must be substituted with $\vx_0\vx_0\trans$.
In the special case of $\Javg{\|\vx(t)\|^2}$ corresponding to $B = \one$, 
and iid disorder ($L= \one$, $R= \Jstd\one$), the contributions from \reqs{res-FBdiscon1}--\myref{res-FBcon1} can be more compactly combined into
\bea\label{res-F1}
\Javg{\|\vx(t)\|^2} && =  \!\int\!\!\!  \int\! \frac{d\omega_1}{2\pi}\frac{d\omega_2}{2\pi} e^{it(\omega_1-\omega_2)} 
\\
&& 
\qquad\qquad
\frac{\vx_0\trans \frac{1}{\gamma-i\omega_2 - M^\dagger}  \frac{1}{\gamma  + i\omega_1-M}\vx_0}{1-\Jstd^2\, \tr\!\!\lpr  \frac{1} {\gamma  + i\omega_1-M} \frac{1} {\gamma-i\omega_2 - M^\dagger}    \rpr}
\nonumber
\eea
(we used $\Tr\!( \frac{1}{z_1-M}\vx_0 \vx_0\trans \frac{1}{z_2 - M^\dagger}) = \vx_0\trans \frac{1}{z_2 - M^\dagger}\frac{1}{z_1-M}\vx_0 $  to write the numerator in \req {res-F1}).

Next, we look at the power spectrum of the response of the system to a noisy input, $\vi(t)$, that is temporally white, with zero mean and covariance
\be\label{CovI}
\overline{{\rm I}_i(t_1) {\rm I}_j(t_2)} = \delta(t_1 - t_2) C^{\vi}_{ij}.
\ee
Here the bar indicates averaging over the input noise (or by ergodicity, over a long enough time). Our general result for the ensemble average of the matrix power spectrum of the response, which by definition is the Fourier transform of the steady-state response covariance,
\be\label{Covxomega}
C^{\vx}_{ij}(\omega) \defin  \int\! d\tau\, e^{-i\omega \tau}\, \overline{\vx_i(t + \tau) \vx_j(t)} ,
\ee
is given by
\be\label{Cnoise}
\Javg{C^{\vx}(\omega)} = C^{\vx}_0(\omega) + \Delta C^{\vx}(\omega).
\ee
Here we defined
\be\label{C0noise}
C^{\vx}_0(\omega) \equiv \C^{\vx}_0(\omega,\omega;C^{\vi})
\ee
and 
\bea\label{delCnoise}
 \Delta C^{\vx}(\omega) \equiv  \Delta \C^{\vx}(\omega,\omega; C^{\vi})
\eea
are the power spectrum matrices obtained by ignoring the randomness in connectivity (\ie\ by setting $A=M$),  and  the contribution of quenched randomness $LJR$, respectively.

A closely related quantity is the total power of the steady-state response of the system to a sinusoidal input $\vi(t) = \vi_0 \sqrt{2} \cos\omega t$ (the $\sqrt{2}$ serves to normalize the average power of $\sqrt{2} \cos\omega t$ to unity, so that the total power in the input is $\| \vi_0\|^2$). For such an input, the steady-state activity, which we denote by $\vx_\omega(t)$, is also sinusoidal (with a possible phase shift). 
By total power of the steady-state response we mean  the time average of the squared norm of the activity,  ${\overline{\| \vx_\omega(t)\|^2}}$, where now the bar indicates temporal averaging  (we call this \emph{total} power, because the squared norm sums the power in all components of $\vx_\omega(t)$). 
As in \reqs {res-avgnormsq}--\myref{res-FBcon1}, we present a formula for the ensemble average of the more general quantity ${\overline{\vx_\omega\trans B\, \vx_\omega}}$.
  We have 
\be\label{res-powerspec}
\Javg{\overline{\vx_\omega\trans B\, \vx_\omega}}  =  \Tr\!\!\lpr B \Javg{C^{\vx}(\omega)} \rpr,
\ee
where $ \Javg{C^{\vx}(\omega)}$ is given by \reqs {Cnoise}--\myref {delCnoise} with $C^{\vi}$ replaced by $\vi_0\vi_0\trans$.
For the special case of $B=\one$, corresponding to the {total} power of the response at frequency $\omega$, using \req {res-FBdiscon1}--\myref {res-FBcon1} with $\omega_1 = \omega_2 = \omega$, this formula 
can be simplified into 
\bea\label{res-powerspecB1}
&& \Javg{\overline{\| \vx_\omega\|^2}} = 
 \\
&&    
\qquad\qquad\left \|  \frac{1}{z -M} \vi_0 \right \|^2 +   \frac{\left\|  \frac{1}{z -M} L \right\|_{\Fr}^2  \,\, \left \| R \frac{1}{z -M} \vi_0 \right \|^2}{1 - \left\| R \frac{1}{z -M} L \right\|_{\Fr}^2},
\nonumber
\eea
where $z=\gamma + i\omega$, $\|\cdot\|$ denotes the vector norm, and  $\| \cdot \|_{\Fr}$ denotes the Frobenius norm defined in \req{frobnorm}. Finally, for the case that the random part of the matrix is iid, \ie\ $L= \Jstd\one$ and $R = \one$, we can further simplify \req {res-powerspecB1} into
\bea\label{res-powerspecBLR1}
&& \Javg{\overline{\| \vx_\omega\|^2}} =      \frac{\left \|  (\gamma + i\omega - M)^{-1} \vi_0 \right \|^2 }{1 - \Jstd^2 \left\| (\gamma + i\omega - M)^{-1}\right\|_{\Fr}^2}.
\eea
The stability of the $\vx=0$ fixed point guarantees the positivity of the expressions \reqs{res-powerspecB1}--\myref {res-powerspecBLR1} for the power spectrum. This is true because, as we noted above, stability requires that the support of the eigenvalue density of $A$ is entirely to the left of the vertical line  $\Re(z) =\gamma$. By our result \req{support-res-2} for that support, this can only be true if the denominators of the last terms in \req {res-powerspecB1} --\myref {res-powerspecBLR1} are positive, which guarantees the positivity of  the full expressions.

Note that the first term in Eq.~\myref{res-powerspecB1} and the
numerator in Eq.~\myref {res-powerspecBLR1} represent the power
spectrum in the absence of randomness, \ie\ if $A$, in \req{ode} is
replaced with $M$.
Thus, formulae \myref{res-powerspecB1}--\myref {res-powerspecBLR1} show
that the correct average power spectrum is always strictly larger than
the naive power spectrum obtained by assuming that random effects will
``average out''.
Furthermore, due to the denominators of the last terms in \reqs{res-powerspecB1}--\myref {res-powerspecBLR1}, the power spectrum will be larger for frequencies where the support of the eigenvalue density, \req{support-res-2}, is closer to the vertical line  with $\Re(z) =\gamma$. 
Similar, but less precise statements can also be made about the strength of transient amplification using formulae \myref{res-avgnormsq}--\myref{res-F1} for the squared norm of the impulse response.  One measure of the strength of transient amplification up to time $T$ is $\int_0^T \| \vx(t)\|^2 dt$. Integrating formulae \req{res-avgnormsq} (with $B=\one$) or \req{res-F1}  over $t$, one obtains formulae for $\int_0^T \| \vx(t)\|^2 dt$ that are the same as \reqs{res-avgnormsq}--\myref{res-F1}, except for the factor $e^{it (\omega_1 - \omega_2)}$ in the integrands of \reqs{res-avgnormsq} and \myref{res-F1} being replaced by $\frac{i [1 - e^{iT (\omega_1 - \omega_2+ i\epsilon)}]}{\omega_1 - \omega_2 + i\epsilon}$ 
(with $\epsilon \to 0^+$).
 Due to the denominator in this factor (for $T$ sufficiently large the numerator is constant),  the main contribution to the integrals over $\omega_1$ and $\omega_2$ should typically arise for $\omega_1 \approx \omega_2$. On the other hand, note that for $\omega_1 = \omega_2$ the denominators in \reqs{res-FBcon1}--\myref{res-F1} reduce to the those in \reqs{res-powerspecB1}--\myref {res-powerspecBLR1}, with the connection to the support of the spectral density noted above.  Thus this dominant  contribution to  $\int_0^T \| \vx(t)\|^2 dt$ must be larger, the closer the support of the eigenvalue density, \req{support-res-2},  is to the vertical line  with $\Re(z) =\gamma$. This also suggests that, as in the case of the power spectrum, the strength of transient amplification would typically be underestimated if randomness of connectivity is ignored and only its ensemble average $M$ is taken into account in solving \req{ode}.
 
  Numerical simulations indicate that the quantities ${{\| \vx(t)\|^2}}$ and ${\overline{\| \vx_\omega\|^2}}$ are {self-averaging} in the large $N$ limit; that is, for large $N$, $\|\vx(t)\|^2$ or ${\overline{\| \vx_\omega\|^2}}$ for any typical random realization of $J$ will be very close to their ensemble averages, given by \req{res-F1} and \req {res-powerspecB1} respectively, with the random deviations from these averages approaching zero as $N$ goes to infinity (see Figs.~\ref {fig-bidiag-powerspec},  \ref{fig-bidiag-avgnormsq} and \ref {fig-rajantransamp}, below). This conclusion is also corroborated by rough estimations {(not shown)} based on Feynman diagrams (the diagrammatic method is introduced in Secs. \ref{sec-deriv1} and \ref{sec-deriv2})  of the variance of fluctuations of these quantities for different realizations of $J$. 
   
{Finally, we note that the general formulae presented  in this section are valid only for cases where
   the initial condition, $\vx_0$, or the input structure, $\vi_0$ or $C^{\vi}$, are chosen independently of the particular realization of the random matrix
   $J$ (\eg, cases where $\vx_0$ is itself random but independent of
   $J$, or when $\vx_0$ is chosen based on properties of
   $M$, $L$ or $R$). 
     In particular,    our results do not apply to cases in which the initial condition or the input is tailored or
   optimized for the particular realization of the quenched
   randomness, $J$, in which case the true result could be
   significantly different from those given by the formulae of this section. }

\subsection{Some specific examples of $M$, $R$ and $L$}
\label{sec-examples}

In this section we present the results of explicit calculations of the eigenvalue density 
\req{rho-res}, the average squared norm of response to impulse \reqs{res-avgnormsq} and \myref{res-F1}, and the total power in response to sinusoidal input \req {res-powerspecBLR1}, for specific examples of $M$,  $L$ and $R$ (the details of the calculations for the results presented here can be found in Sec.~\ref{sec-example}). 
For many of the examples presented here,
$L$ and $R$
are both proportional to the identity matrix; thus in these examples
the full matrix is of the form $M+\Jstd J$ where $\Jstd>0$ determines
the strength of disorder in the matrix. In
Secs.~\ref{sec-res-rajanabbott} and \ref{sec-res-merav}, we also
present examples with nontrivial $L$ and/or $R$.

Any matrix, $M$, can be turned into an upper-triangular form by a unitary transformation, \ie\
\be\label{schur}
M = U T U^\dagger,
\ee
where $U$ is unitary and $T$ is upper-triangular (\ie\ $T_{ij} = 0$ if $i>j$) with its main diagonal consisting of the eigenvalues of $M$. The difference between nonnormal and normal matrices is that for the latter, $T$ can be taken to be strictly diagonal. Equation \myref{schur} is referred to as a Schur  decomposition of $M$ \cite{HornJohnson:1990}, and we refer to the orthogonal modes of activity represented by the columns of $U$ as Schur modes. 
The Schur decomposition provides an intuitive way of characterizing the dynamical system \req{ode}. Rewriting \req{ode}, with $J$ and $\vi(t)$ set to zero, in the Schur basis by defining $\vy = U^\dagger \vx$ (\ie\, $y_i$ is the activity in the $i$-th Schur mode), we obtain $\frac{d\vy}{dt} = -\gamma \vy +   T \vy$. 
 We see that activity in the $j$-th Schur mode provides an input to the equation for the $i$-th mode only 	when $i\leq j$ (as $T_{ij} = 0$ for $i>j$). Thus the coupling between modes is feedforward, going only from higher modes to lower ones, without any feedback. 
We refer to $T_{ij}$'s for $j>i$ as feedforward weights. {As these vanish for normal matrices, we can say a matrix is more nonnormal the stronger its feedforward weights are.}

Due to the invariance of the trace, the norm, and the adjoint
operation under unitary transforms,  our general formulae for the
spectral density \req{rho-res} and the average squared norm in time
and frequency space,  \reqs{res-F1} and \myref{res-powerspecBLR1},
take the same form in any basis, so in particular we can work in
the Schur basis of $M$. Hence $M$ can be
replaced by $T$, provided $L$ and $R$ are also expressed in
$M$'s Schur basis and $\vx_0$ or $\vi_0$ are replaced by $U^\dagger \vx_0$
or $U^\dagger \vi_0$, respectively  
 \footnote{The unitary invariance of these formulae is in turn a consequence of the invariance of both the corresponding quantities ($\rho(z)$ and $\| \vx(t)\|^2$), as well as the statistical ensemble for $J$, \req{jmeas}, and hence that of $LJR$ when $L\propto R\propto \one$, under unitary transforms like \req{schur}.}.
Thus we use the feedforward structure of the Schur decomposition to characterize the different examples we consider below. Our examples are chosen to demonstrate interesting features of nonnormal matrices in the simplest possible settings.

\subsubsection{Single feedforward chain of length $N$}
\label{sec-res-longffwd}

In the first example, each and every Schur mode is only connected to its  lower adjacent mode, forming a long feedforward chain of length $N$. For simplicity, we take all  feedforward weights in this chain to have the same value $w$,  so that 
\be\label{Mbidiag0-res}
M = T =  \begin{pmatrix}
     \lambda_1 & w & 0 & \cdots   \\
     0 & \lambda_2 & w & \cdots   \\
     \vdots & \vdots & \vdots & \ddots   \\     
\end{pmatrix}
\ee
or more succinctly $M_{nm}= w\, \delta_{n+1,m} + \lambda_n \delta_{nm}$.

\begin{figure}[!t]\hspace{0cm}
\includegraphics[height=3.06in,angle=0]{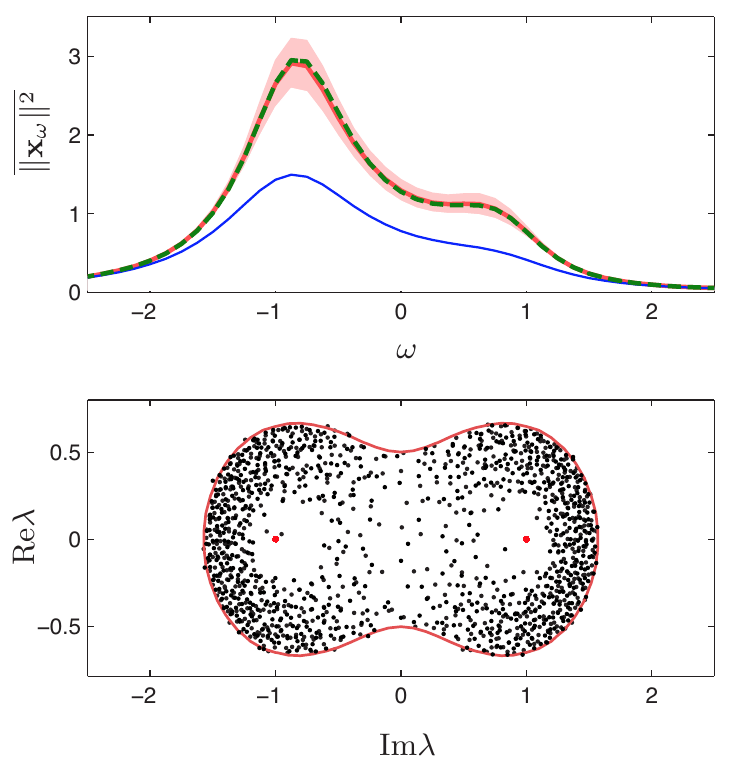}
\vspace{-.3cm}
\caption{(Color online) Top panel: the total power spectrum of steady state response ${\overline{\|\vx_\omega\|^2}}$ as a function of input frequency $\omega$, \req{res-powerspecBLR1}, for the system \req{ode} with $A = M + \sigma J$, and $M$ given by \req{Mbidiag0-res} with $w=1$ and $\lambda_n=\pm i$ (with $+i$ and $-i$ alternating), respectively. Here, $N= 700$, $\Jstd = 0.5$, and $\gamma = 0.8$. The input was fed into 
the last component of $\vx$ (the beginning of the feedforward chain characterized by \req{Mbidiag0-res}), which for the matrix $M$ has natural frequency -1. That is, the input was $\vi_0 \sqrt{2} \cos\omega t$ where $\vi_0$ was 1 for the last component and 0 for all other components.
 The green (thick dashed) curve is the ensemble average of the total power spectrum,  $\Javg{\overline{\|\vx_\omega\|^2}}$, calculated numerically using the general formula \req{res-powerspecBLR1}, which is compared with an empirical average over 100 realizations of real Gaussian $J$ (solid red line, mostly covered by the dashed green line). 
 The pink (light gray) area shows the standard deviation  among these 100 realizations around this average. 
The blue (thin) line shows the result when disorder, $\sigma J$, is ignored, \ie\ $A$ is replaced by its ensemble average $M$.  
Bottom panel: the eigenvalue spectrum of $M+\Jstd J$ (black dots). Red big dots at $\pm i$ show the eigenvalues of $M$.  The red curve is the outer boundary of the eigenvalue spectrum of $A$ as computed numerically using  \req{support-res}. The real and imaginary axes of the complex plane are interchanged, so that the frequency axis in the top panel can be matched with the imaginary  part of the eigenvalues, \ie\ the natural frequencies of \req{ode}.  }
\label{fig-bidiag-powerspec}
\end {figure} 	

Figure~\ref {fig-bidiag-powerspec} shows the power spectrum of response (top panel) and the eigenvalue distribution (bottom panel)  of $A = M + \Jstd J$  for an example $M$ of the form \req{Mbidiag0-res}  with alternating imaginary eigenvalues,  $\lambda_n=(-1)^{n+1} i$. 
The black dots in the bottom panel of Fig.~\ref {fig-bidiag-powerspec} show the eigenvalues of $A$ for one realization of $J$, scattered around the highly degenerate spectrum of $M$ at $\pm i$ (red dots). 
The top panel 
shows the ensemble average of the total power spectrum of response,  $\Javg{\overline{\|\vx_\omega\|^2}}$,  of the system \req{ode} to sinusoidal stimuli as given by our general formula \req{res-powerspecBLR1} (green curve), showing that
it perfectly matches the empirical average (red curve) over a set of 100 realizations of $J$ (the latter was obtained by generating 100 realizations of $J$, calculating  ${\overline{\|\vx_\omega\|^2}}$ for each realization,   which is given by the numerator of \req {res-powerspecBLR1} with $M$ replaced by $M + \Jstd J$, and then averaging the results over the 100 realizations).
 The pink (light gray) shading shows the standard deviation of the power spectrum over these 100 realizations. This will shrink to zero as $N$ goes to infinity, so that for large $N$ the power spectrum of any single realization of $A = M + \Jstd J$ will lie very close to the ensemble average. 
The system \myref{ode} in  the zero disorder case, $\Jstd =0$, has two highly degenerate resonant
frequencies (imaginary parts of the eigenvalues of $M$), $\omega_0^\pm =  \pm 1$, leading to possible peaks in the power spectrum  at these frequencies. The smaller the decay of these modes (in this case given by $\gamma$) is, \ie\ the closer the eigenvalues {of the combined matrix $-\gamma + M$} are to the imaginary axis, the sharper and stronger are the resonances. 
Comparing the zero disorder power spectrum (blue curve) with that for $A = M + \Jstd J$, we see that the disorder has led to strong but unequal amplification of the two resonances relative to the case without disorder. 
This is partly due to the disorder scattering some of the eigenvalues {of $-\gamma + A$} much closer to the imaginary axis, creating larger resonances.

\begin{figure}[!t]\hspace{-.5cm}
\includegraphics[height=2.1in,angle=0]{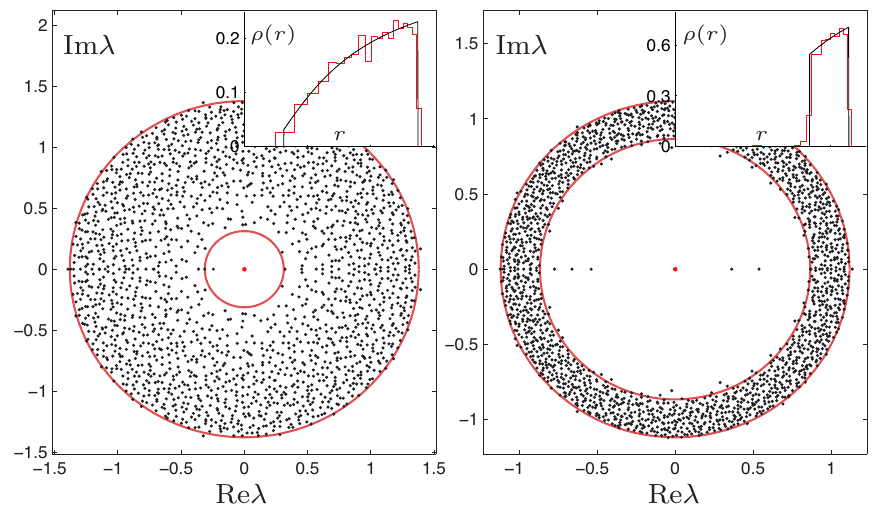}
%
\caption{(Color online) The eigenvalue spectra of $A = M+\Jstd J$ for  $N=2000$ and $M$ given by \req{Mbidiag0-res} with $\lambda_n = 0$, $w = 1$ for single realizations of real Gaussian $J$. $\Jstd= 0.95$ and 0.5 in the left and rights panels, respectively.  The red circles mark the circular boundaries of the spectral support given by \req {suppbidiag-res}. The insets show a comparison of the analytic formula \req{rho-bidiag-res} for the spectral density (black smooth trace) and histograms corresponding to the particular realization shown in the main plot (red jagged trace). 
}
\label{fig-bidiag-spectrum}
\end{figure}

For  $M$ of the form \myref{Mbidiag0-res} with all eigenvalues zero we have analytically calculated the eigenvalue density, \req{rho-res},
 the magnitude of response to impulse \req {res-F1}, and the power-spectrum \req {res-powerspecBLR1}. 
In this case, using \req{support-res} naively yields $|z| \leq \sqrt{|w|^2 + \Jstd^2}$ for the support of the eigenvalue density.
However, using the correct procedure, \reqs{Kdef1}--\myref{support-res-K}, we find that this formula is only
correct for $\sigma \geq |w|$, while for $\sigma < |w|$, 
the true support of the eigenvalue density in the limit $N \to \infty$ is  the annulus 
\be\label{suppbidiag-res}
\sqrt{|w|^2 - \Jstd^2} \leq |z| \leq \sqrt{|w|^2 + \Jstd^2},
\ee
(this result was obtained in Ref.~\cite{Khoruzhenko:1996aa}). 
Within this support the eigenvalue density in either case is 
\be\label{rho-bidiag-res}
\rho(z) =  \frac{1}{\pi\Jstd^2}\lbr 1 - \frac{|w|^2}{\sqrt{4 |w|^2|z|^2 + \Jstd^4  } } \rbr.
\ee
Figure~\ref {fig-bidiag-spectrum} demonstrates the close agreement of \reqs{suppbidiag-res}--\myref {rho-bidiag-res}   with the empirical spectrum of $M+ \Jstd J$ for a single realization of $J$, for $N = 2000$ and two different values of $\Jstd$. 
The discrepancy between the results obtained by the naive use of \req {support-res} and \req {suppbidiag-res} is due to the fact that  for $|z|<|w|$,  $M_z = (z - M)/\Jstd$ has an exponentially small, $O(e^{-c N})$, singular value (see next paragraph) which makes the result of \reqs{Kdef1}--\myref{support-res-K} dependent on the order of the two limits $N\to\infty$ and $g\to 0^+$.  As we discussed after \req {rho-svd-res},  such a discrepancy can signify the existence of an $O(1)$ number of outlier eigenvalues outside the support of $\lim_{N\to\infty}\rho(z)$. 
Simulations show that this is the case for $|z| <  \sqrt{|w|^2 - \Jstd^2}$ (see Fig.~ \ref {fig-bidiag-spectrum}).

The most striking aspect of these results is revealed in the limit $\Jstd \to 0$. For $\Jstd = 0$, the spectrum is that of $M$, which is concentrated at the origin. 
Remarkably, however, as seen from  \reqs{suppbidiag-res}--\myref {rho-bidiag-res}, for very small but nonzero $\Jstd$ the bulk of the eigenvalues are concentrated in the narrow ring with modulus $|z| \approx |w|$. Thus in the limit $N\to\infty$ the spectrum has a discontinuous jump at $\Jstd =0$. 
This is a consequence of the  extreme nonnormality  of $M$, which manifests itself in  the extreme sensitivity of its spectrum to small perturbations, which is well-known (see Ref.~\cite{Trefethen-book}, Ch.~7). The notion of  pseudospectra quantifies this sensitivity: the (operator norm)  $\epsilon$-pseudospectrum of $M$ is the region of complex plane to which its spectrum can be perturbed by adding to $M$ a matrix of operator norm no larger than $\epsilon$. As we mentioned in Sec.~\ref {subsec-specdensity},   this is precisely  the set of complex values $z$ for which $\| (z-M)^{-1}\| > \epsilon ^{-1}$  \cite{Trefethen-book}, and therefore by the definition of the operator norm $\|\cdot\|$, the region in which $\| (z-M)^{-1}\|^{-1} = s_{\min}(z-M) <\epsilon $, where $s_{\min}(z-M)$ is the least singular value of $z-M$. { As noted above,} for $|z|<|w|$, $s_{\min}(z-M)$ is exponentially small: $s_{\min}(z-M) \leq |w| |\frac{z}{w} |^N$ (for a proof see after \req {suppbidiag} in Sec.~\ref{subsec-bidiagcalcs}). Thus the $\epsilon$-pseudospectrum of $M$ contains the set of points $z$ satisfying $|w||\frac{z}{w} |^N < \epsilon$, \ie\ the centered disk with radius $|w| (\frac{\epsilon}{|w|})^{1/N}$ which approaches $|w|$ as ${N\to\infty}$. In other words, for large enough $N$, any point $|z|<|w|$ is in the $\epsilon$-pseudospectrum for any \emph{fixed} $\epsilon$, no matter how small. It has been stated \cite{Trefethen-book} that dense random perturbations, of the form $\Jstd J$ considered here, tend to trace out the entire $\epsilon$-psuedospectrum (where $\epsilon = \Jstd \| J\| \approx 2 \Jstd$). Our result shows that, for  $\epsilon, \Jstd \ll |w|$, the spectrum of such perturbations traces out the  $\epsilon$-psuedospectrum in quite an uneven fashion; the vast majority ($\Theta(N)$) of the perturbed eigenvalues only trace out the boundary of the pseudospectrum, $|z| \approx |w|$, while only a few ($O(1)$) eigenvalues lie in its interior. Thus, dense random perturbations can fail as a way of visualizing (operator norm based) pseudospectra.

\begin{figure}[!t]\hspace{.8cm}
\includegraphics [width=3.4in,angle=0]{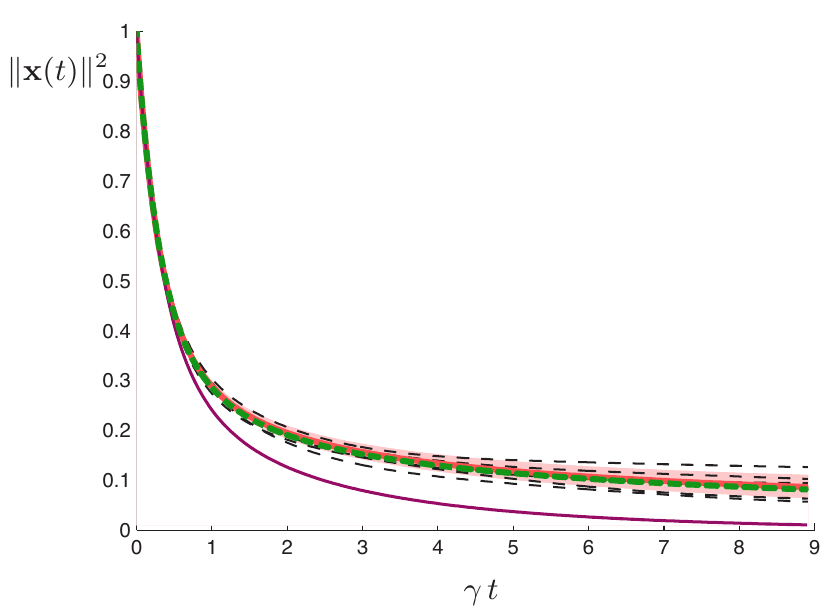}
\vspace{-.3cm}
\caption{(Color online) The norm squared of the response to impulse, ${\|\vx(t)\|^2}$, of the system \req{ode}, for $A = M + \Jstd J$, with binary $J$, and $M$ given by \req{Mbidiag0-res} (with $\lambda_n=0$) describing a $N$-long feedforward chain with uniform weights $w$. Here, $w = 1$, $\Jstd = 0.5$, $\gamma = 1.005\sqrt{\Jstd^2 + w^2} \simeq 1.124$, and $N=700$. The 
 green (thick dashed) curve shows our result, \req {avgnormsq-bidiag-res},  for the average squared impulse response,  $\Javg{\|\vx(t)\|^2}$, which lies on top of the red (thick solid) curve showing the empirical average of ${\|\vx(t)\|^2}$ over 100 realizations of binary $J$. The five thin dashed black curves show the result for five particular realizations of $J$, and the pink (light gray) area shows the standard deviation among the 100 realizations. The standard deviation shrinks to zero as $N\to\infty$, and ${\|\vx(t)\|^2}$ for any realization lies close to its average for large $N$. For comparison the purple (thin, lowest) curve shows ${\|\vx(t)\|^2}$ obtained by ignoring the effect of quenched disorder, \ie\ by setting $A = M$.
}
\label{fig-bidiag-avgnormsq}
\end{figure}
    

We now turn to the dynamics. 
We have explicitly calculated  the average evolution  of the magnitude of $\vx(t)$,  \req {res-F1},
and the total power spectrum of steady-state response, \req {res-powerspecBLR1}, for the case where the initial condition is (or the input is fed into) the last Schur mode, \ie\ the beginning of the feedforward chain: $\vx_0=(0,\cdots,0,1)\trans$ ( or  $\vi_0 \propto (0,\cdots,0,1)\trans$). For the evolution of the average norm squared, with the initial condition $\vx_0=(0,\cdots,0,1)\trans$, we obtain
\be\label{avgnormsq-bidiag-res}
\Javg{\| \vx(t)\|^2}  = e^{-2\gamma t} I_0(2 t \sqrt{|w|^2 + \Jstd^2} ), \qquad (t\geq 0)
\ee
where $I_\nu(x)$ is the $\nu$-th modified Bessel function. Figure~\ref {fig-bidiag-avgnormsq} plots the function \req {avgnormsq-bidiag-res} and compares it with the result obtained by ignoring the disorder (corresponding to $\Jstd = 0$). 
The main difference between the two curves is the slower asymptotic decay of the $\Jstd\neq 0$ result (green) compared with the zero-disorder case (purple). This is the result of the disorder spreading some of the eigenvalues {of $-\gamma + A$} closer to the imaginary axis, creating modes with smaller decay. Importantly, in neither case do we see transient amplification. By contrast, in the $\Jstd = 0$ and for small enough decay, \ie\ for $\gamma <|w|$, the system \req{ode} exhibits very strong transient amplification. In this case, starting from the initial condition $\vx_0=(0,\cdots,0,1)\trans$, the solution for the $(N-n)$-th Schur component is $x_{N-n}(t) = \frac{(wt)^n}{n!}e^{-\gamma t}$ (for $0\leq n \leq N-1$), which is maximized at $t = n/\gamma$ with a value $\max |x_{N-n}| \sim (\frac{|w|}{\gamma})^n$ for $n\gg 1$. Thus up to time $t\sim N/\gamma$ the norm of the activity grows exponentially; $\| \vx(t)\|^2 \gtrsim (\frac{|w|}{\gamma})^{2\gamma t}$ for $t\lesssim N/\gamma$. For larger times the activity reaches the end of the $N$-long feedforward chain and starts decaying to zero; asymptotically $\| \vx(t)\|^2 \sim e^{-2\gamma t}$ for $t\gg N/\gamma$. However, as we have seen, the spectrum of $M$ is extremely sensitive to perturbations; even for very small but nonzero $\Jstd$, the  spectrum of $-\gamma \one + A$ has eigenvalues with real part as large as $|w|-\gamma$. Therefore,  in the limit $N\to \infty$, the system \req{ode} is unstable for $|w|>\gamma$, as soon as $\Jstd \neq 0$.  Conversely, in the presence of disorder (even infinitesimally small disorder in the $N\to \infty$ limit), as long as the system is stable (which from \req{suppbidiag} requires  $\gamma > \sqrt{|w|^2 + \Jstd^2}$) it exhibits no transient amplification for the initial condition along the last Schur mode. Let us note, however, that as we mentioned after \req{res-powerspecBLR1}, \req{res-F1} and hence \req {avgnormsq-bidiag-res} do not yield the correct answer when the direction of the impulse is optimized for the specific realization of the quenched disorder $J$; such  disorder-tuned initial conditions can yield significant transient amplification even for the stable $\Jstd\neq 0$ system. 

Incidentally, we can also read the result for $M=0$ from \req {avgnormsq-bidiag-res}, by setting $w = 0$, obtaining $\Javg{\| \vx(t)\|^2}  = e^{-2\gamma t} I_0(2 \Jstd t )$. Since in this case, all directions are equivalent, this is the answer for the (normalized) initial condition along any  direction, again as long as the direction is chosen independently of the specific realization of $J$. 
  
Finally, the total power of response to a sinusoidal input with amplitude  $\vi_0 =  (0,\cdots,0,{\rm I}_0)\trans$ is given by 
\be \label{res-bidiag-powerspec}
\Javg{\overline{\| \vx_\omega\|^2}} =  \frac{\| \vi_0\|^2}{\omega^2 + \gamma^2 - |w|^2 - \Jstd^2} .
\ee
The main effect of the disorder is to reduce the width of the resonance (the peak of $\Javg{\overline{\| \vx_\omega\|^2}}$ at $\omega = 0$) and increase its height. This is partly a consequence of the scattering of the eigenvalues {of $-\gamma + A$} closer to the imaginary line by the disorder, creating modes with smaller decay.

\subsubsection{Examples motivated by Dale's law: 1 or $N/2$ feedforward chains of length $2$}
\label {sec-res-rajanabbott}

In this section we consider examples motivated by 
Dale's law \cite{Dale:1935,Eccles:1954aa,StrataHarvey:1999} in neurobiology.  Dale's law is the
observation (which holds generally but with some exceptions
\cite{JonasSandkuhler:1998,Root:2014aa}) that individual neurons release the same
neurotransmitter at all of their synapses. In the context of many
theoretical papers including this one, it refers more specifically to
the fact that an individual neuron either makes only excitatory
synapses or only inhibitory synapses; that is, each column of the synaptic connectivity matrix has a fixed sign, positive for excitatory neurons and negative for inhibitory ones.
We will first consider two examples of connectivity matrices respecting Dale's law which take the form \req{ensemble} with $L = \Jstd^{-1} R = \one$, and a scalar $\sigma$. At the end of this subsection we consider an example with nontrivial $L$ and $R$.

In the first example, we consider a matrix $M$,  which as we will
show, has a Schur form $T$ that is composed of $N/2$ disjoint
feedforward chains, each connecting only two modes (we assume $N$ is even). 
For simplicity we will focus on the case where all eigenvalues are zero.
Thus in the Schur basis we have 
\be\label{Tex2}
T 
= 
\begin{pmatrix}
     0 & w_1 & 0   & 0      & \cdots   \\
     0 & 0      & 0   & 0      & \cdots   \\
     0 & 0      & 0   & w_2 & \cdots   \\
     0 & 0      & 0   & 0      & \cdots   \\
     \vdots & \vdots & \vdots & \vdots & \ddots   \\     
\end{pmatrix}  
=
 W \otimes \begin{pmatrix}
     0 & 1   \\
     0 & 0 
\end{pmatrix}
\ee
where we defined $W$ to be the $N/2\times N/2$ diagonal matrix of Schur weights $W = \mathrm{diag}(w_1,w_2,\ldots,w_{N/2})$. 
$T$ in \req{Tex2} arises as the Schur form of a mean matrix of the form 
\be\label{Mex2}
M 
=
\frac{1}{2} \begin{pmatrix}
     \BB & \,\, -\BB  \\
     \BB & \,\,-\BB 
\end{pmatrix}
=
\frac{1}{2} \begin{pmatrix}
     1 & -1   \\
     1 & -1 
\end{pmatrix} \otimes \BB
\ee
where  $\BB$ is a \emph{normal} (but otherwise arbitrary) ${N}/{2}\times
{N}/{2}$  matrix (note that $M$ is nonetheless nonnormal). The feedforward weights
in \req{Tex2} are then the eigenvalue of $\BB$. When $K$ has only positive entries, matrices of the form
\req{Mex2} satisfy Dale's principle, and were studied in Ref.~\cite{Murphy:2009a}, in the context of
networks of excitatory and inhibitory neurons. We imagine a grid of
$N/2$ spatial positions, with an excitatory and an inhibitory neuron at each
position. $\frac{1}{2}\BB$, a
matrix with positive entries, describes the mean connectivity {strength} between
spatial positions, which is taken to be identical regardless of
whether the projecting, or receiving, neuron is excitatory or inhibitory. The sign of the weight, on the other hand, depends on the excitatory or the inhibitory nature of the projecting or presynaptic neuron;
 the first (last) $N/2$
columns of $M$ represent the projections of the excitatory
(inhibitory) neurons and are positive (negative). Since $\BB$ is normal it can be diagonalized by a
unitary transform: $\BB = EWE^\dagger$, where $W$ is as above, and $E =
(\ve_1,\ve_2,\ldots)$ is the matrix of the orthonormal eigenvectors
$\ve_b$ of $\BB$,  $b=1,\ldots,N/2$ (with eigenvalues $w_b$). Then
transforming to the basis $\left \{\vectwo{\ve_1}{\0},\vectwo{\0}{\ve_1},
\vectwo{\ve_2}{\0},\vectwo{\0}{\ve_2},\ldots,\vectwo{\ve_{N/2}}{\0},
  \vectwo{\0}{\ve_{N/2}}\right \}$ (where $\0$ represents the
$N/2$-dimensional vector of 0's)
transforms the matrix to being
  block-diagonal with the $2\times 2$ matrices
  $\frac{1}{2}\mat{w_b}{-w_b}{w_b}{-w_b}$,
  $b=1,\ldots,N/2$, along the diagonal.  The $b^{th}$ block becomes
$\mat{0}{w_b}{0}{0}$ in its Schur basis 
$\left \{\frac{1}{\sqrt{2}} \vectwo{\ve_b}{\ve_b},\frac{1}{\sqrt{2}} \vectwo{\ve_b}{-\ve_b}\right \}$, so
the full matrix takes the form \req{Tex2}.
Thus, the $b$-th difference mode $\frac{1}{\sqrt{2}} \vectwo{\ve_b}{-
  \ve_b}$ feeds
forward to the $b$-th  sum mode $\frac{1}{\sqrt{2}} \vectwo{\ve_b}{\ve_b}$
with weight $w_b$.
This feedforward structure  leads to a specific form of
nonnormal transient amplification, which the authors of
Ref.~\cite{Murphy:2009a}  dubbed ``balanced amplification"; small
differences in the activity of excitatory and inhibitory modes
feedforward to and cause possibly large transients in modes in which
the excitatory and inhibitory activities are balanced.

\begin{figure}[!t]
\hspace{-.3cm}
\includegraphics[height=2.1in,angle=0]{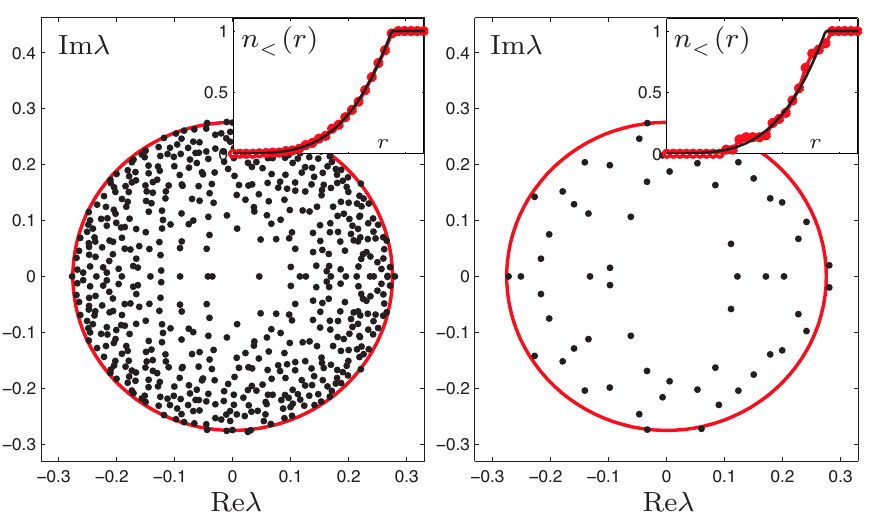}
%
\vspace{-.3cm}
\caption{(Color online) The eigenvalue spectra of $A = M+\Jstd J$ for a binary $J$ with $\Jstd= 0.1$ and $M$ given by \req{Mex2} with $\BB =  \one$ (corresponding to $w_b = 1$  for all the diagonal $2\times 2$ blocks in \req{Tex2}). The main panels show the eigenvalues for single realizations of $J$, with $N=600$ (left) and $N=60$ (right).   The red circles mark the  boundaries of the spectral support, \req {spectralrad-rajan-res}. 
Since $A$ is real in this case, its eigenvalues are either exactly real, or come in complex conjugate pairs; the spectrum is symmetric under reflections about the real axis. 
However, such signatures of the reality of the matrix  appear only as subleading corrections to the spectral density $\rho(z)$; they are finite size effects which vanish as $N\to\infty$. 
The insets show a comparison of the analytic formula \req {eigdens-rajan-res} (black curve) and the empirical result, based on the eigenvalues of the realizations in the main panels, for the proportion, $n_{_<}(r)$, of eigenvalues lying within a radius $r$ of the origin (red dots). The random fluctuations  and the average bias of the empirical $n_{_<}(r)$  are both already  small for $N=60$, and negligible for $N=600$. 
}
\label{fig-diagEI-spectrum}
\end{figure} 

\begin{figure}[!t]\hspace{-0.1cm}
\includegraphics[height=3in,angle=0]{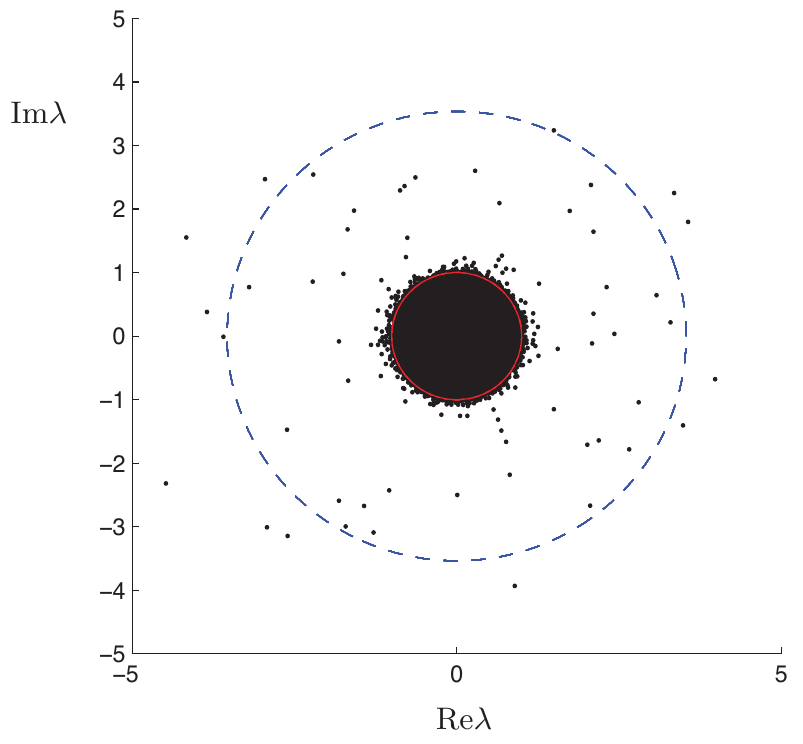}
\vspace{-.2cm}
\caption {(Color online) The eigenvalue spectra of $A = M+\Jstd J$ for  the $M$ given by \req{rajanM} in the balanced case, $\vcv\trans \vu = 0$. Here, $N=800$,  $\Jstd= 1$ and $\mu = 12$ (see equation \req{muRajan}).  The black dots are the superimposed eigenvalues of $A$ for 20 different realizations of complex Gaussian $J$.
The small red circle enclosing the vast majority of the eigenvalues has radius $\Jstd=1$, corresponding to the standard circular law \req {res-rajanrho}. A $\Theta(N)$ number of eigenvalues lie within this circle. A $\Theta(\sqrt{N})$ number lie just outside of this circle in a thin boundary layer which shrinks to zero as $N\to \infty$. Finally, a $\Theta(1)$ number of eigenvalues lie at macroscopic distances outside the unit circle. {The dashed blue circle shows radius $r_0$ given by \req {spectralrad-rajan-res}; outliers can even lie outside this boundary.  
}
}
\label{fig-rajanspectrum}
\end{figure}

Another interesting example of Dale's law is that in which $M$ simply
captures the 
differences between the mean inhibitory and mean excitatory synaptic
strengths and between the numbers of excitatory and inhibitory neurons,
with no other structure assumed (uniform mean connectivity), as
studied in Ref.~\cite{Rajan:2006}. 
Thus, all excitatory projections have the same mean $\mu_E/\sqrt{N}$,
and all inhibitory ones have the mean $-\mu_I/\sqrt{N}$.  
 If we assume a fraction $f$ of all neurons are excitatory, then we can write $M$ as 
 \be\label{rajanM}
M =  \vu \vcv\trans
\ee
where $\vu  = N^{-1/2} (1,\ldots,1)\trans$ is a unit vector, and the vector $\vcv$ has components $v_i = \mu_E$ or $v_i = - \mu_I$ for $i \leq  fN$ and $i > fN$, respectively (for $f=1/2$ and $\mu_E = \mu_I$, \req{rajanM} is a special case of \req{Mex2}). The single-rank matrix $M$ has only one non-zero eigenvalue given by $\vcv\cdot\vu = \frac{1}{\sqrt{N}} \sum_i v_i$, with eigenvector $\vu$. 
The case in which the excitatory and inhibitory weights are \emph{balanced on average}, in the sense that $\sum_i v_i = 0$, is of particular interest; mathematically it is  in a sense the least symmetric and most nonnormal case as $\vcv\cdot\vu  = 0$. In this case all eigenvalues of $M$ are equal to zero. Furthermore, since in this case $\vu$ and $\vcv$ are orthogonal, we can readily read off the Schur decomposition of $M$ from \req{rajanM}. The normalized Schur modes are given by $\vu$, $\vcv/\| \vcv\|$ and $N-2$ other unit vectors spanning the subspace orthogonal to both $\vu$ and $\vcv$. All feedforward Schur weights are zero, except for one very large weight, equal to $\|\vcv\| \propto \sqrt{N}$, which feeds from $\vcv/\| \vcv\|$ to $\vu$. Thus the Schur representation  of $M$ has the form \req{Tex2} with $w_1 = \|\vcv\| \defin  \mu\sqrt{N}$ and $w_{b\neq 1} = 0$, where we defined 
\be\label{muRajan}
\mu^2 \defin \tr(M^\dagger M) =
  \|\vcv\|^2/{N} = f \mu_E^2 + (1-f)\mu_I^2.
\ee
{Note that this is again a case of balanced amplification: differences between excitatory and inhibitory activity, represented by $\vcv$, feed forward to balanced excitatory and inhibitory activity, represented by $\vu$, with a very large weight.}
In the following we present results only for this balanced case of \req{rajanM}, which as just noted is a special case of  \reqs{Tex2}.

\begin{figure}[!t]\hspace{0.5cm}
\includegraphics[height=2.45in,angle=0]{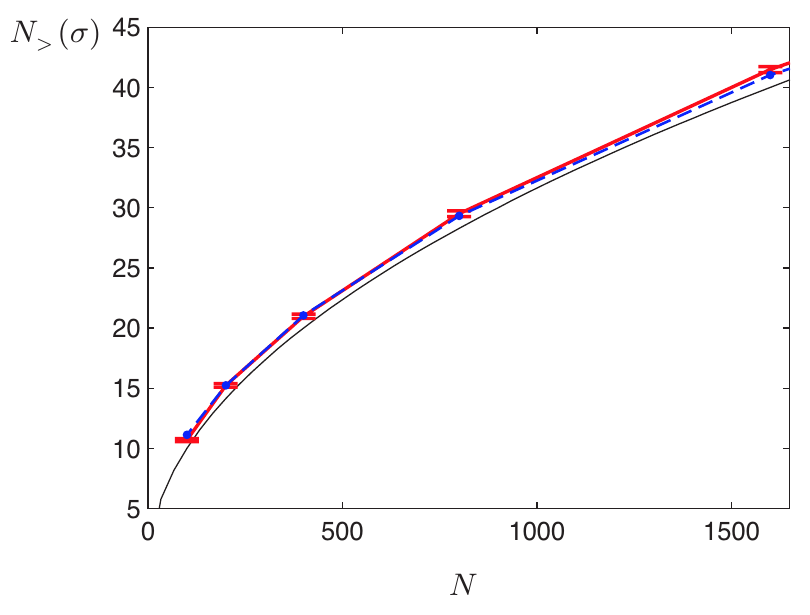}
\vspace{-.3cm}
\caption{(Color online) The number of eigenvalues of $M + \Jstd J$, for the $M$ given by \req{rajanM}, lying outside the circle of radius $\Jstd$ vs. $N$ (red line). Here, $\Jstd = 1$, $\mu = 12$ and $\vcv\trans \vu = 0$.  The numbers (red points connected by solid red lines) are obtained by numerically calculating the eigenvalues and counting the outliers for 200 realizations of $J$,  and taking the average of the counts over all realizations, for $N = 100,200,400,800,1600$ (error bars show standard error of mean). The black dashed line plots $\sqrt{N}$ for comparison with our theoretical result \req{rajanoutliers}; the (dashed) blue line which includes subleading corrections to $\sqrt{N}$, is obtained by numerically solving  \req{geqnrajan} and substituting  the result in \req{n>}  (these formulae are in turn obtained from  \reqs {rho-res}--\myref {geqn} in Sec.~\ref {sec-rajanabbott}).
}
\label{fig-rajanN_>(1)}
\end{figure}

We start by presenting the results for the eigenvalue density.
For general diagonal $W$ in \req{Tex2} (or equivalently, for general
normal $\BB$ in \req{Mex2}), the eigenvalue density, $\rho(z)$, of $A = M + \Jstd J$ is
isotropic around the origin $z=0$, and depends only on $r= |z|$. The
spectral support is a disk centered at the origin. 
In cases in which all the weights $w_b$ are $O(1)$, the radius of this disk can be found directly from \req{support-res}, which yields
\be\label{spectralrad-rajan-res}
r_0 = {\Jstd}\lbr \frac{1}{2}  + \sqrt{\frac{1}{4} + \frac{\langle |w_b|^2\rangle_b}{2\Jstd^2}}\, \rbr^{1/2}.
\ee
Here, $\langle |w_b|^2\rangle_b$ is the average of the squared feedforward weights over all blocks of \req{Tex2}; equivalently, $\langle |w_b|^2\rangle_b = 2\,\tr\!(M^\dagger M) \equiv 2\mu^2$. 
As long as some $w_b$ are nonzero,   $r_0$ is larger than the radius of the circular law, $\Jstd$, with the difference an increasing function of $\langle |w_b|^2\rangle_b$; thus the spreading of the spectrum of $M$ (originally concentrated at  the origin) after the random perturbation by $\Jstd J$, is larger the more nonnormal $M$  is. 
In cases in which the feedforward weights of some of the $2\times 2$ blocks of \req{Tex2} 
grow without bound as $N\to\infty$, there is a  corresponding singular value of $M_z\propto z-M$ for every such block which is nonzero for $z\neq 0$ but vanishes in the limit, scaling like $\sim\frac{|z|^2}{|w_b|}$ where $w_b$ is the unbounded weight of that block (see  \req{svrajan} and its preceding paragraph).
 (Note that as stated after \req {frobnorm} we assume $\|M\|_{\Fr} = \mu = \sqrt{\langle |w_b|^2\rangle_b/2}$ is $O(1)$, so that at most $o(N)$ number of weights can be unbounded, and each can at most scale like $O(N^{1/2})$.) 
 In line with the general discussion  after \req {rho-svd-res}, in such cases the naive use of \req{support-res} may yield an area larger than the true support of $\lim_{N\to\infty}\rho(z)$; the correct support must be found by using \reqs{Kdef1}--\myref{support-res-K}, which in this case can yield a support radius strictly smaller than  \req {spectralrad-rajan-res}. 
  We have calculated the explicit results for $\lim_{N\to\infty}\rho(z)$ for two specific examples of $M$ with the Schur form \req{Tex2}. The first example belongs to the first case (bounded $w_b$'s) where $\lim_{N\to\infty}\rho(z)$ is $\Theta(1)$ within the entire disk $r\leq r_0$, while the second belongs to the second case (unbounded $w_b$'s) where the limit density is only nonzero in a proper subset of that disk.

In the first example, we take all the Schur weights in \req{Tex2} to have the same value, which we denote by $ w$. 
 In this case,  
the eigenvalue density is given by $ \rho(r)  =\frac{1}{2 \pi r}  \frac{\partial n_{_<}(r)}{\partial r}$, 
where $n_{_<}(r)$ is the proportion of eigenvalues within a distance $r$ from the origin and is given by
 \be\label{eigdens-rajan-res}
   n_{_<}(r) = \frac{r^2}{\Jstd^2}\lbr 1  - \frac{|w|^2}{\Jstd^2 + \sqrt { {\Jstd^4} + |w|^4 + 4  |w|^2 r^2}} \rbr.
   \ee
$n_{_<}(r)$ reaches unity exactly at $r  = r_0$ given by \req {spectralrad-rajan-res}, and $\rho(r)$ is $\Theta(1)$ for any smaller $r$. 
Figure~\ref{fig-diagEI-spectrum} shows the close agreement of \req {eigdens-rajan-res} with empirical results based on  single binary  realizations of $J$, for $N$ as low as 60.

\begin{figure}[!t]\hspace{0.5cm}
\includegraphics[height=2.54in,angle=0]{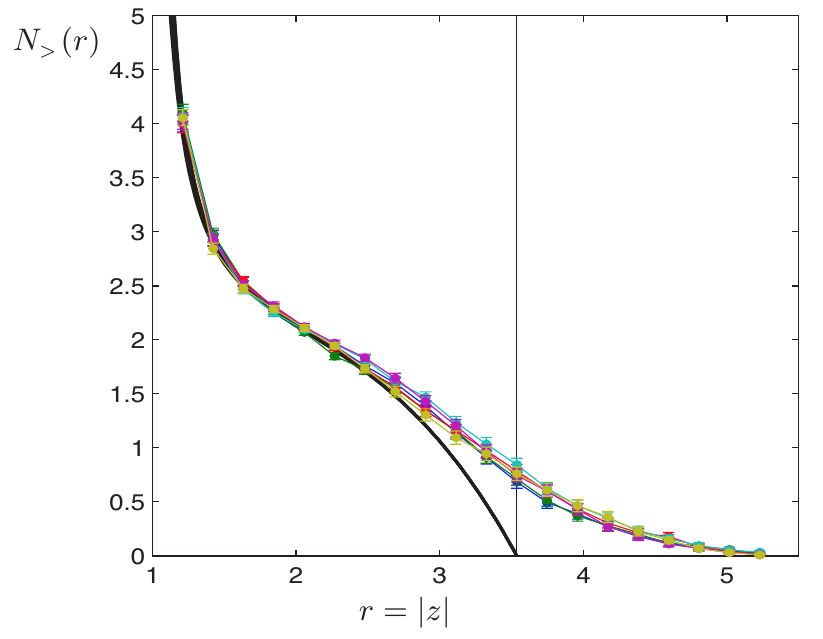}
\vspace{-.2cm}
\caption{(Color online) The number, $N_{_{>}}(r)$, of outlier eigenvalues of $ A = M + \Jstd J$, for the $M$ given by \req{rajanM}, lying farther from the origin than $r$, as a function of $r$. Here, $\Jstd = 1$, $\mu = 12$ and $\vcv\trans \vu = 0$. The vertical line marks  $|z| = r_0\simeq 3.54$ where $r_0$ is given by \req {spectralrad-rajan-res}. The colored (shades of gray) connected points are  $N_{_{>}}(r)$ for realizations of $A$, based on 200 samples of $J$, each color for a different $N$, for $N = 100,200,400,800,1600$ and 3200 (error bars show standard error of sample mean). Note the lack of scaling of $N_{_>}(r)$ with $N$.
}
\label{fig-rajanoutliers}
\end{figure}

The second example is that of the balanced \req{rajanM} with $\vu\cdot\vcv = 0$.  As we saw, all $w_b$ are zero in this case except for one very large, unbounded weight $w_1 = \mu\sqrt{N}$. As discussed above, in this case $M_z\propto z-M$  has an $o(1)$ smallest singular value, approximately given by $\frac{|z|^2}{\mu\sqrt{N}}$.
    Using \reqs{Kdef1}--\myref{support-res-K}, we find that the support of $\lim_{N\to \infty} \rho(z)$ is the disc with radius $\Jstd$ (within the annulus $\Jstd <  |z| \leq r_0$ the eigenvalue density is $o(1)$), and solving  \reqs {rho-res}--\myref {geqn} for $|z|\leq \Jstd$, we find that the spectral density  is in fact identical with the circular law (the eigenvalue density for the $M=0$ case), \ie\
\be\label{res-rajanrho}
\rho(r) = \lcr \begin{array}{l}  \frac{1}{\pi\Jstd^2} + o(1),
\qquad\qquad (r<\Jstd)
\\
\\
o(1) \qquad\qquad\qquad \quad (r>\Jstd).
\end{array}
\right.
\ee
It was shown in Refs.~\cite{Tao:2013,ChafaiChafai:2010} that more generally, for any $M$ of   rank $o(N)$ and bounded $\|M\|_{\mathrm{F}}$, the eigenvalue density of $A = M + \Jstd J$ is given by the circular law in the limit $N\to\infty$. For single rank $M$ (as in the present case) and a diagonal $R$, it was shown in Ref.~\cite{Wei:2012} that the eigenvalue density of $M + JR$ agrees with that of $JR$ as $N\to\infty$.
In the present example, it  was observed in Ref.~\cite {Rajan:2006} that even though the majority of the eigenvalues are distributed according to the circular law, there also exist a number of ``outlier" eigenvalues spread outside the circle $|z| = \Jstd$, which unlike in the $M=0$ case, may lie at a  significant distance away from it (see Fig.~ \ref{fig-rajanspectrum}). 
As we mentioned in Sec.~\ref {subsec-specdensity}, the non-crossing approximation cannot be trusted to correctly yield the $o(1)$ contributions to $\rho(z)$ by these outliers for $|z|>\Jstd$. However, we found that if we ignore this warning and use  \reqs {rho-res}--\myref {geqn},  keeping track of finite-size, $o(1)$ contributions, we obtain results that agree surprisingly well (though not completely) with simulations.  
First, for the total number of outlier eigenvalues lying outside the circle $|z| = \Jstd$ we obtain
\be\label{rajanoutliers}
N_{_>}(\Jstd) \defin  N n_{_>}(\Jstd)  = \sqrt{N} + O(1)
\ee
(here we defined $n_{_>}(r) = 1 - n_{_<}(r)$ to be the proportion of eigenvalues lying outside the radius $r$);
see Fig.~\ref{fig-rajanN_>(1)} for a comparison of \req{rajanoutliers} with simulations. The vast majority of the outlier eigenvalues counted in \req {rajanoutliers} lie in a narrow boundary layer immediately outside the circle $|z|=\Jstd$, the width of which shrinks with growing $N$. In addition to these, however,  there are a $\Theta(1)$ number of eigenvalues lying at macroscopic, $\Theta(1)$ distances outside the circle $|z|=\Jstd$. Using  \reqs {rho-res}--\myref {geqn} we have calculated $N_{_>}(r)$, the  number of outlier eigenvalues lying outside radius $r$ for $r> \Jstd$. 
{Figure~\ref {fig-rajanoutliers}} shows a plot of $N_{_>}(r)$ and compares it with the results of simulations for different $N$. For roughly the inner half of the annulus $\Jstd <  |z| < r_0$,  $N_{_>}(r)$ agrees well with simulations, but as $r$ increases it deviates significantly from the empirical averages. 
In particular, $N_{_>}(r)$  calculated from \reqs {rho-res}--\myref {geqn} vanishes at $r_0$ given by \req {spectralrad-rajan-res}, while the empirical average of the number of outliers is nonzero well beyond $r_0$. 
Finally, we note that the distribution of these eigenvalues is not self-averaging, and depends on the real vs. complex nature of the random matrix $J$ \cite{Tao:2013}. In the real case, their distribution has been recently characterized  as that of the inverse roots of a certain random power series with iid standard real Gaussian coefficients  \cite{Tao:2013}.

\begin{figure}[!t]\hspace{0.6cm}
\includegraphics[width=3.35in,angle=0]{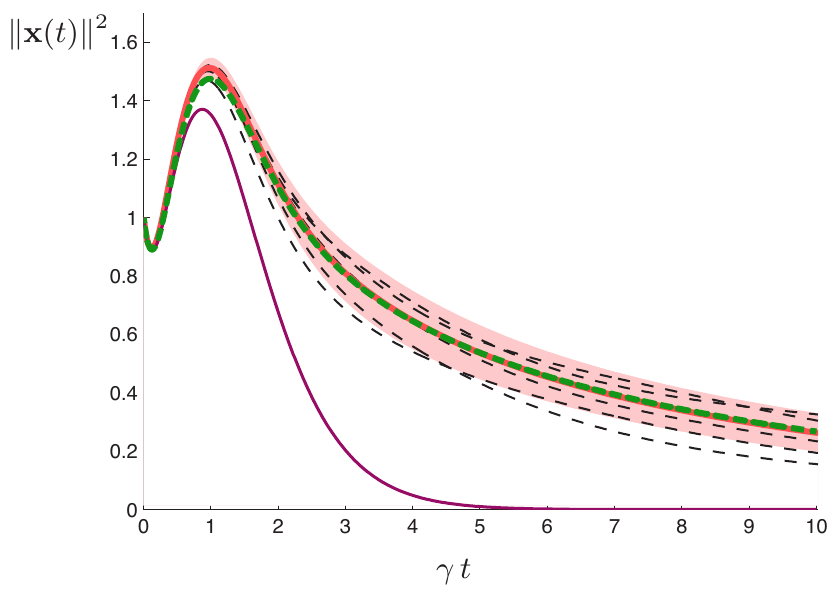}
\vspace{-.3cm}
\caption{(Color online) The squared norm of response to impulse, ${\|\vx(t)\|^2}$, of the system \req{ode}, for $A = M + \Jstd J$, with log-normal $J$, and $M$ given by \req{Tex2} describing $N/2$ doublet feedforward chains weights $w_b$. Here, $w_a = \sqrt{\langle |w_b|^2\rangle_b} = 3$, $\Jstd = 0.4$,  $\gamma = 1$, and $N=1400$. The green (thick dashed) curve shows our result,  \reqs {res-rajantransamp}--\myref{Cbdef},  for the average norm squared which, except for a small window around its peak, lies on top of the red (thick solid) curve showing the empirical average of ${\|\vx(t)\|^2}$ over 100 realizations of binary $J$. The five thin dashed black curves show the result for five particular realizations of $J$, and the pink (light gray) area shows the standard deviation among the 100 realizations. The standard deviation shrinks to zero as $N\to\infty$ and ${\|\vx(t)\|^2}$ for any realization lies close to its average for large $N$. For comparison the purple (thin, lowest) curve shows ${\|\vx(t)\|^2}$ obtained by ignoring the effect of quenched disorder, \ie\ by setting $A = M$.
}
\label{fig-rajantransamp}
\end{figure}

As for the dynamics, we have analytically calculated the magnitude of impulse response,
 \req {res-F1}, as well as the power-spectrum of steady-state response \req {res-powerspecBLR1}, for $A = M + \Jstd J$ with $M$ given by \reqs{Tex2}--\myref{Mex2} with general $w_b$, when the  (impulse or sinusoidal) input feeds into the 
second Schur mode in one of the $N/2$ chains/blocks of \req{Tex2}; we denote the index for this block by $a$. For the average magnitude of impulse response we find
\be\label{res-rajantransamp}
\Javg{\| \vx(t)\|^2}=  
\lbr 
 \frac{1 + C_a}{2} I_0(2 r_0 t) +  \frac{1 - C_a}{2}  J_0(2 r_1 t)
\rbr e^{-2 \gamma t}
\ee
where $J_0(x)$ ($I_0(x)$) is the (modified) Bessel function, $r_0$ is given by \req {spectralrad-rajan-res}, we defined $r_1^2 = r_0^2 - \Jstd^2$, and
\be\label{Cbdef}
C_a \defin  \frac {1+ {2 |w_a|^2}/{\Jstd ^2}}{\sqrt {1+ {2 \langle |w_b|^2\rangle_b }/{\Jstd ^2}}},
\ee
with $ \langle |w_b|^2\rangle_b = 2 \,\tr\!(M^\dagger M)$ denoting the average squared feedforward weight among all the blocks of \req{Tex2}. 
In Fig.~\ref {fig-rajantransamp} we plot \req {res-rajantransamp} and compare it with the result obtained by ignoring the disorder (\ie\ by setting $\Jstd = 0$); in the latter case, the block $a$ is decoupled from the rest of the network, and solving the $2\times 2$ linear system governed by the matrix $\mat{-\gamma}{w_a}{0}{-\gamma}$, we obtain $\| \vx(t)\|^2 = (1 + w_a^2 t^2)e^{-2\gamma t}$.   From the figure, we see that the $\Jstd\neq 0$ result (green) has a slower asymptotic decay compared with the zero-disorder case (purple); this is due to the disorder having spread some eigenvalues closer to the imaginary axis, creating modes with smaller decay,
 along with the fact that the coupling between the $2\times 2$ blocks induced by the disorder insures that these more slowly decaying modes will be activated.
 Indeed, for large $t$, $\|\vx(t)\|$ decays like $e^{-\gamma t}$ when $\Jstd = 0$, while in the $\Jstd> 0$Ê case, based on \req {spectralrad-rajan-res}  it must decay like $e^{-(\gamma-r_0)t}$, \ie\ by a rate set  by the largest real part of the spectrum shifted by $-\gamma$ (this is indeed what we obtain  from  \req {res-rajantransamp} using the asymptotics of Bessel functions).  In addition, both curves exhibit transient amplification where the magnitude of activity initially grows to a maximum, before it decays asymptotically to zero. The $\Jstd\neq 0$ curve shows larger and longer transient amplification, which is most likely attributable both to the  eigenvalues being closer to the $\Re(z) =\gamma$ line and to augmented nonnormal effects (\eg\ larger effective feedforward weights, or longer chains).
We also mention that, as in our previous examples, if the input direction is optimized for the particular realization of $J$, significantly larger transient amplification may be achieved.   

Finally, the total power spectrum of response  to a sinusoidal input, \req {res-powerspecB1}, is   given by the explicit formula 
\be\label{res-powerspecrajan}
\Javg{\overline{\| \vx_\omega\|^2}} = \frac{\omega^2 + \gamma^2 + |w_a|^2 } {(\omega^2 + \gamma^2)^2 -  \Jstd^2(\omega^2 + \gamma^2 + \mu^2)} \| \vi_0\|^2.
\ee
where $\mu^2 \defin \tr\!(M^\dagger M) = {\langle |w_b|^2\rangle_b/2}$ and, as noted above, the direction of $\vi_0$ is that of the second Schur mode in block $a$.

The example \req{rajanM} motivated by Dale's law with neurons  of either excitatory or inhibitory types, 
can be generalized to a network of neurons belonging to one of $C$
different types (these could be subtypes of excitatory or inhibitory
neurons), in which not only the mean but also the variance of
connection strengths depends on the pre- and post-synaptic types. When this dependence is factorizable, in a way we will now describe, the connectivity matrix of such a network will be of the form \req{ensemble} with non-trivial $L$ and $R$. Let $c(i) \in \lcr 1, \ldots, C\rcr$  denote the type  of neuron $i$, and let $f_c$ denote the fraction of neurons of type $c$ (so  $\sum_{c=1}^C f_c = 1$); we assume $C$ and $f_c$ are all $\Theta(1)$. 
Assume further that each synaptic weight is a product of a pre- and a post-synaptic factor, and that in each synapse these factors are chosen independently from the same distribution, except for a deterministic sign and overall scale that depend only on the type of the pre and post-synaptic neurons, respectively. Thus if $A_{ij}$ denotes the weight of the synaptic projection from neuron $j$ to neuron $i$, we have 
\be\label{typeensemble}
A_{ij} = \frac{1}{\sqrt{N}} (l_{c(i)} x_{ij})  (r_{c(j)} y_{ij})
\ee
where $x_{ij}$'s and $y_{ij}$ are positive random variables chosen iid from the distributions $P_x(x)$ and $P_y(y)$, respectively. Here,  $l_c$ and $r_c$ determine the sign and the scale (apart from the overall $\frac{1}{\sqrt{N}}$)  of the pre and post-synaptic factors of the neurons in cluster $c$, respectively. Note that when all $l_c$ are positive, $A_{ij}$ satisfies Dale's law. 
By absorbing appropriate constants into $l_c$'s and $r_c$'s we can assume that $\mathrm{Var}[xy] = \langle x^2 \rangle \langle y^2\rangle  - \langle x \rangle^2 \langle y\rangle^2  = 1$. Then it is easy to see that $A$ can be cast in the form \req{ensemble} with 
\bea\label{typesMLR}
L_{ij} \areq l_{c(i)} \delta_{ij}
\\
R_{ij} \areq r_{c(i)} \delta_{ij}
\label{typesR}
\\
J_{ij} \areq \frac{1}{\sqrt{N}} (x_{ij} y_{ij} - \xi)
\label{typesJ}
\\
M \areq s\, L\,  \vu \vu\trans\, R
\label{typesM}
\eea
where $\vu$ is the unit vector $\frac{1}{\sqrt{N}} (1,\ldots,1)\trans$,  
\be
s\equiv \xi \sqrt{N}, 
\ee
and $\xi \equiv \langle x\rangle \langle y\rangle$ is dimensionless and $\Theta(1)$
(note that $J$, given by Eq.~\myref{typesJ}, indeed has iid elements with zero mean and variance $N^{-1}$). Being single-rank, $M$ has $N-1$ zero eigenvalues; its only (potentially) non-null eigenvector is $L\vu$, with a generically large eigenvalue 
\be\label{lamdaM-typed}
\lambda_M = s \vu\trans R L \vu = s \frac{1}{N}\sum_{i=1}^N r_{c(i)}l_{c(i)} = \xi \sqrt{N} \langle \sigma_c \rangle_c
\ee
where we defined 
\bea
\sigma_c &\equiv & l_c r_c,
\\
\langle X_c\rangle_c &\equiv& \sum_{c=1}^C f_c X_c.
\label{typesavgdef}
\eea

As for the example \req{rajanM}, we will focus on the balanced case in which $\lambda_M  \propto  \langle \sigma_c \rangle_c = 0$.
From Eq.~\myref{typesM}, $M=\tilde\vu\tilde\vcv\trans$ with 
$\tilde\vu=L\vu$ and $\tilde\vcv=sR\vu$. The balanced condition is
equivalent to $\tilde\vu\cdot\tilde\vcv=0$ (see
Eq.~\myref{lamdaM-typed}). Thus, similar to 
Eq.~\myref{rajanM}, the Schur representation of $M$ has the form 
\myref{Tex2} with $w_1=\|\tilde\vu\|\|\tilde\vcv\|$ and $w_b=0$ for $b>1$.

\begin{figure}[!t]
\hspace{-.4cm}
\includegraphics[height=2.25in,angle=0] 
{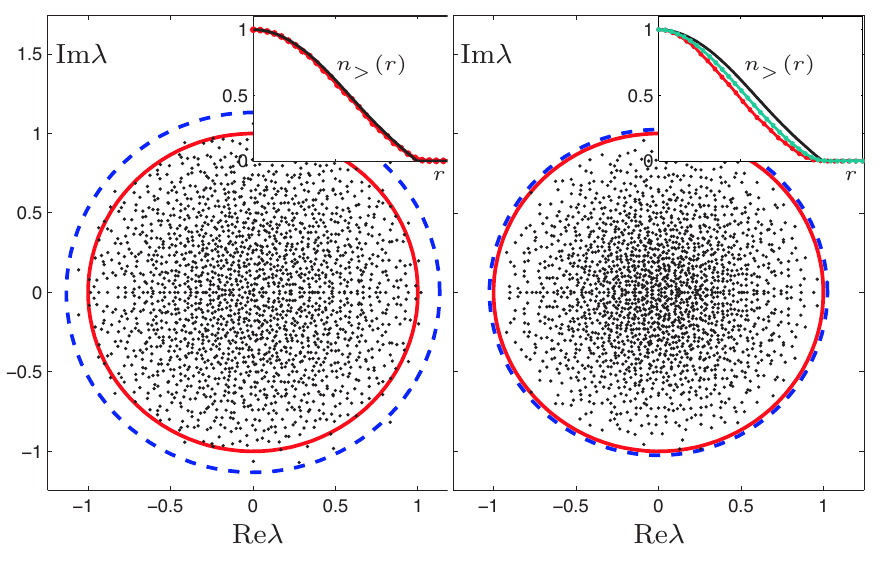}
%
\vspace{-.35cm}
\caption{(Color online) The eigenvalue spectra of $A = M+L J R$ with $M$, $L$ and $R$ given by \reqs{typesMLR}--\myref{typesM} with neurons belonging to one of three different types ($C=3$). The main panels show the eigenvalues for two  particular realizations of $J$. In both panels, $N=2000$, $f_1 = 0.6$, $f_2 = f_3 = 0.2$, $l_c =1$, $\sigma_1 = r_1 = 0.76$, $\sigma_2 = r_2 = -0.57$, $\sigma_3 = r_3 = -1.71$ (so $\avgc{\sigma_c} = 0$ and $r_0^2 = \avgc{\sigma_c^2} = 1$), and $J_{ij}$ had real entries with log-normal distribution; in the left (right) panel, the normally distributed $\log_{10} J_{ij}$ had standard deviation 0.5 (0.75). The solid red circles mark the boundaries of the spectral support as given by \req{boundary-types}, and the dashed blue circles show the radii given by \req{wrongradius-types}.
The insets compare $n_>(r)$ based on the numerically calculated eigenvalues shown in the main panels (connected red dots), with that found by solving \req{geqn-types} (black curve). In the right panel's inset we have also plotted (green connected dots lying slightly above the red connected circles)  the empirically calculated $n_>(r)$ for a single realization with the same ensemble parameters, but with $N=8000$; the convergence to the universal limit at $N\to\infty$ is significantly slower in the right panel in which the distribution of $J_{ij}$ had a considerably heavier tail.
}
\label{fig-CtypesNet-spectrum}
\end{figure}

 In Sec.~\ref{sec-typenet} we prove that, as for \req{rajanM},   for the ensemble \reqs{typesMLR}--\myref{typesM} 
   the limit of the eigenvalue distribution, $\lim_{N\to \infty}\rho(z)$,  is also not affected by the nonzero mean matrix \req{typesM}; hence we can obtain $\lim_{N\to \infty}\rho(z)$ for that example by safely setting $M$ to zero, and using formulae \reqs{boundaryM0}--\myref{n>M0} with $L$ and $R$ given by \reqs{typesMLR}--\myref{typesR}. Thus $\lim_{N\to \infty}\rho(z)$
 is isotropic and its support is the disk with radius 
 \be\label{boundary-types}
 r_0 = \| RL \|_{F} = \sqrt{\langle \sigma_c^2\rangle_c}.
 \ee
As in the previous example, when the balance condition
$\avgc{\sigma_c} = 0$ holds, use of the naive formula
\req{support-res} with $M=\tilde\vu\tilde\vcv\trans$ would have yielded
\be\label{wrongradius-types}
\tilde r_0 = r_0 \lbr  \frac{1}{2}  + \sqrt{ \frac{1}{4} + {\xi^2}} \rbr^{1/2},
\ee
which is larger than the correct result  \req{boundary-types}. As discussed above, this result is not correct, but it  indicates the existence of $\Theta(1)$ number of outlier eigenvalues lying outside the boundary of $\lim_{N\to\infty}\rho(z)$ given by \req{boundary-types}.
For $r<r_0$, the $N\to\infty$ limit of the proportion, $n_>(r)$, of eigenvalues lying farther than distance $r$ of the origin is given by $g^2(r)$ which is found by solving \req{geqnM0}, or equivalently 
 \be\label{geqn-types}
  \left\langle  \frac{1}{ g^2   + \sigma_c^{-2} r^2}\right\rangle_c = 1.
  \ee
The results \reqs{rhoM00r}--\myref{rhoM0r0} also hold, wherein the normalized sums over $i$ can be replaced with appropriate averages $\langle \cdot\rangle_c$. 
In the case of two neuronal types a closed solution can be obtained for $n_>(r)$ and $\rho(r)$. Identifying the two types with excitatory and inhibitory neurons, and assuming that $l_c = 1$, $\sigma_E\equiv \sigma_1 >0$ and $\sigma_I\equiv \sigma_2 <0$ (we will use $E$ and $I$ as indices instead of $c=1$ and $2$, respectively) the ensemble \reqs{typesMLR}--\myref{typesM} describes a synaptic connectivity matrix in which all excitatory (inhibitory) connections are iid with mean $\xi \sigma_E N^{-\frac{1}{2}}$ ($-\xi |\sigma_I | N^{-\frac{1}{2}}$) and variance $\sigma_E^2 N^{-1}$ ($\sigma_I^2 N^{-1}$). 
In this case,  \req{geqn-types} yields
 a quadratic equation. 
Differentiating 
the solution of that equation with respect to $r^2$ we obtain the explicit result
\be
\rho(r) = \frac{\sigma_E^{-2} + \sigma_I^{-2}}{2\pi} \lbr 1 - \frac{\frac{({\sigma_E^{-2} + \sigma_I^{-2}})r^2 - 1}{2} + \frac{r_0^2 - 2r^2}{\sigma_E^{2} + \sigma_I^{2}} }{\sqrt{\frac{\lpr(\sigma_E^{-2} + \sigma_I^{-2}) r^2 - 1\rpr^2}{4} + \frac{r^2(r_0^2 - r^2)}{ (\sigma_E\sigma_I)^{2}}}}\rbr
\ee
This result was first obtained (in a less simplified form) in Ref.~\cite{Rajan:2006}.
Figure~\ref{fig-CtypesNet-spectrum} shows two examples of spectra for single realizations of matrices of the form  \req{typeensemble}, with three neural types ($C = 3$), where $x_{ij}$ and $y_{ij}$, and hence $J_{ij}$, have log-normal distributions. The insets compare $n_>(r)$ based on the numerically calculated eigenvalues, with those found by solving \req{geqn-types}.  In the right panel, the normally distributed $\log J_{ij}$ have a higher standard deviation, and hence the distribution of $J_{ij}$ has a heavier tail.  The right panel's inset demonstrates that the convergence to the universal, $N\to\infty$ limit  can be considerably slow when the distribution of $J_{ij}$ is heavy-tailed.

\subsubsection{Linearizations of nonlinear neural and ecological networks}
\label{sec-res-merav}

In neuroscience applications, \req{ode} can arise as a linearization of nonlinear firing rate equations for a recurrent neural network of $N$ neurons, around some stationary background. 
The nonlinear dynamical equations for the evolution of the network activity typically take the form
\footnote{It is also common to write the firing rate equations in the different form, $T \frac{d \vrr(t)}{dt} =  - \vrr(t) + f( W \vrr(t)  + \vi^{\scriptscriptstyle r}(t))$. At least in the case where all neurons have equal time constants, \ie\ $T\propto \one$, the two formulations are equivalent and are related by the change of variable $\vv = W \vrr +  \vi^{\scriptscriptstyle r}$ \cite{MillerFumarola:2012}.
  }  
\be\label{nonlinerate}
T \frac{d \vv(t)}{dt} =  - \vv(t) + W f( \vv(t))  + \vi^{\scriptstyle v}(t).
\ee
Here $\vv(t)$ is the vector of state variables of all neurons at time
$t$; its $i$-th component, $v_i(t)$, is commonly thought of as the voltage of  the $i$-th neuron, or the total synaptic input it receives. 
 $f(\cdot)$ is the neuronal nonlinear input-output function, which
is imposed element-by-element on its vector argument, with $f(\vc
v)_i\equiv f(v_i)$ giving the output, \ie\ the firing rate, of neuron $i$;
$\vi^ {\scriptstyle v}(t)$ is the external input vector; $T=
\mathrm{diag}(\tau_1,\tau_2,\cdots, \tau_N)$ is a $N\times N$ diagonal
matrix whose diagonal elements are the positive
time-constants of the neurons (hence $T$ is invertible); and $W$ is
the $N\times N$ synaptic connectivity matrix.

Suppose that for a constant external input, $\vi^{\scriptstyle v}_*$, \req{nonlinerate} has a fixed point $\vv_*$. Then, given a small perturbation in the input,  $\vi^ {\scriptstyle v}(t) = \vi^ {\scriptstyle v}_* + \delta \vi^ {\scriptstyle v}(t)$, we can write $\vv(t) = \vv_* + \vx(t)$, and linearize the dynamics around the fixed point by expanding \req{nonlinerate} to first order in $\vx(t)$ and $ \delta \vi^ {\scriptstyle v}(t)$. This yields  the set of  linear differential equations 
\be\label{rateeqn}
T \frac{d \vx(t)}{dt} =  - \vx(t) + W \vct{\Phi}\, \vx(t)  + \delta \vi^ {\scriptstyle v}(t),
\ee
for the (small) deviations, where we defined the diagonal Jacobian
\be\label{jacobian_v}
\vct{\Phi} = {\rm diag}( f'(\vv_*)).
\ee
Now suppose that the original connectivity matrix can be written as $W
= \langle W \rangle + \delta W$, with a quenched disorder part that is an
iid random matrix: $\delta W = \Jstd J$.  
Then multiplying \req{rateeqn} by $T^{-1}$, we can convert \req{rateeqn} into the form \req{ode} with  $\gamma = 0$ and $A = M + LJR$ with
\bea\label{rateeqn_MLR_v}
M \areq T^{-1}(-\one +  \langle W \rangle \vct{\Phi})
\\
L \areq T^{-1}
\\
R \areq \Jstd   \vct{\Phi} 
\label{rateeqn_R}
\eea
and input 
\be\label{input_v}
\vi(t) = T^{-1}\, \delta \vi^ {\scriptstyle v}(t). 
\ee

This observation is not limited to neuroscience applications, and  can  also apply to many other frameworks,  \eg\  those used in mathematical biology. Generalized Lotka-Volterra (GLV) equations \cite{Hofbauer-LotkaVolterra-book} used in modeling the dynamics of food webs provide an example. Let $\vn(t) = (n_1(t),\ldots,n_N(t))\trans$ denote the vector of population sizes of $N$ species. The GLV equations take the form $\frac{d n_i}{dt}= n_i (r_i  + \sum_{j} W_{ij} n_j)$  or 
\be\label{GLV}
\frac{d \vn}{dt} = \mathrm{diag}(\vct{r} + W \vn)\vn
\ee
where $r_i>0$ are the species' intrinsic growth rates and $W$ is the interaction matrix. Linearizing \req{GLV} around a fixed point, $\vn_*$, yields again a linear system of the form \req{ode} with $\gamma = \vi(t) = 0$.  Starting with the same simple model $W = \langle W\rangle + \Jstd J$, we find that $A$ can be written in the form \req{ensemble} with 
\bea\label{GLV_L}
&& R = \Jstd\one, \qquad\qquad L = \mathrm{diag}(\vn_*),
\\
&& M =  \mathrm{diag}(\vct{r} + W \vn_*) +  L \langle W \rangle.
\label{GLV_M}
\eea
Note that if no species is extinct in the fixed point, \ie\ if all $n_{i*} >0$, then $M =  L \langle W \rangle.$

Assuming the linear systems thus obtained, \ie\ the fixed points $\vv_*$ or $\vn_*$, are stable, we can therefore think of our results for $\|\vx(t)\|^2$ and $\|\vx_\omega\|^2$ as characterizing the temporal evolution and the spectral properties of the linear response of the \emph{nonlinear} system \req{nonlinerate} (\req{GLV}) in its fixed point $\vv_*$ ($\vn_*$) to perturbations.

\begin{figure}[!t]\hspace{0.4cm}
\includegraphics [height=2.4in,angle=0]{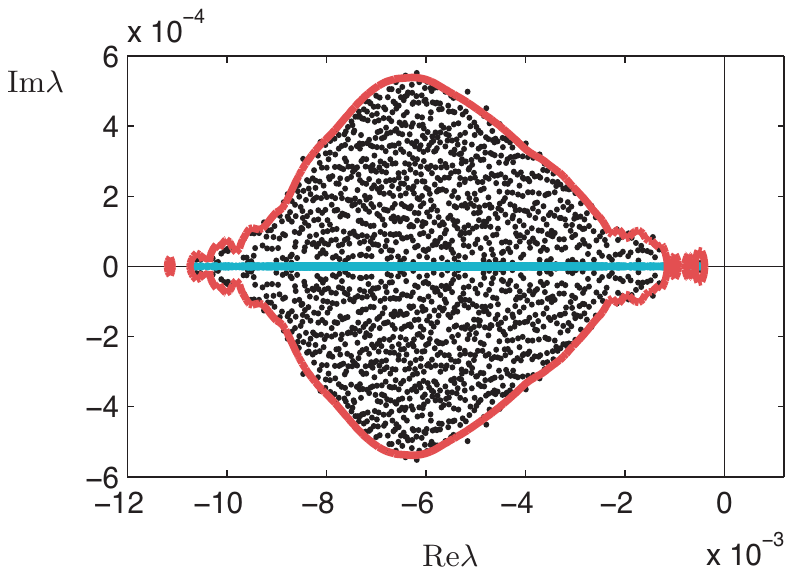}
\vspace{-.2cm}
\caption{(Color online) The eigenvalues (black dots) of $A = M + JR$, with $M$ and $R$ given by \reqs{gapjuncM}--\myref {gapjuncR} with $g = 0.01$, $a=1.02$ and $N=2000$. This matrix governs the dynamics of small perturbations away from a non-trivial random fixed point in a clustered network of neurons (see \req{gapjunc}), studied in Ref.~\cite{SternAbbott:2012}.
The cyan dots on the real line are the eigenvalues of $M$, and the red curve is the boundary of support of the eigenvalue distribution, as calculated numerically from \req{support-res}.  }
\label{fig-Gapjuncnet}
\end{figure}

The necessary and sufficient condition for the stability of a fixed point (without any change in the external input) is that all eigenvalues of the corresponding $A$ have negative real parts. Our formula for the boundary of the eigenvalue distribution, \req{support-res}, can be applied in these cases to map out the region in parameter space (parameters here mean the time constants or intrinsic growth rates in $T$ or $\vct{r}$, or the connectivity parameters determining the random ensemble for $W$, \ie\ $\Jstd$ and the parameters of $\langle W \rangle$) in which a particular fixed point is stable. 
Recently our general formula \req{support-res} was used in this way by colleagues \cite{SternAbbott:2012}  to determine the phase diagram of a clustered network of neurons, in which intra-cluster connectivity is large, but inter-cluster connectivity is random and weak. 
Because of the strong
  intra-cluster connectivity, each cluster behaves as a unit with a
  single self-coupling $a$. Letting the random
  inter-cluster couplings between $N$ clusters have zero mean and variance $g^2/N$, their
  analysis starts from the equation
  \be\label{gapjunc}
\frac{d \vct{v}(t)}{dt} =  - \vct{v}(t) + a  \tanh(\vct{v}(t))  + g J  \tanh(\vct{v}(t))
\ee
where $J$ is an iid random matrix  as above. Here, $\vct{v}$ is a vector whose $i$-th component is the  mean
voltage of \emph{cluster} $i$, while the nonlinear function
$\tanh(\vct{v}(t))$ (with the hyperbolic tangent acting
component-wise) represents the vector of mean firing rates of the
clusters. The analysis of Ref.~\cite{SternAbbott:2012} shows that there is
a region of the phase plane $(a,g)$ where the self-connectivity, $a$,
is excitatory and sufficiently strong, in which the system eventually
relaxes to non-zero random attractor fixed points $\vct{v}_*$; for
smaller values of $a$, the dynamics is chaotic (chaos in the $a=0$
case was established in Ref.~\cite{SompolinskySommers:1988}). The form of
these fixed points (the distribution of the elements of $\vct{v}_*$ as
$N\rightarrow\infty$ for a given $(a,g)$) 
can be obtained using mean-field theory, and the
linearization about $\vct{v}_*$ leads to an equation in the form of \req{ode}, with $A =
M + J R$, where $M$ and $R$ are the diagonal matrices 
\bea\label
{gapjuncM} M \areq {\rm diag}(-1 + a \tanh'(\vct{v}_*))
\\
R \areq {\rm diag}(g \tanh'(\vct{v}_*)).
\label {gapjuncR}
\eea 
Given this form, it can
be shown that the fixed point $\vct{v}_*$ is stable if $z=0$ is outside and to the right of
the spectrum of the Jacobian matrix of the linearization, $A$. 
The mean field solution for $\vct{v}_*$ determines the statistics of
the elements of $R^2 M^{-2}$ for a given $(a,g)$. From these it can be
determined if $z=0$ is outside the spectrum using our formula for the
boundary of spectrum \req{support-res}, which yields the requirement
$\tr\!\!\left(\frac{R^2}{M^2}\right)<1$. In this way, the region of stability
of the fixed points in the $(a,g)$ plane can be mapped (see Ref.~\cite{SternAbbott:2012} for the results, and a complete discussion of the analysis outlined here).
Figure \ref {fig-Gapjuncnet} shows a numerical example of the
eigenvalue distribution for $A$ for a given $(a,g)$ and the
superimposed boundary calculated using \req{support-res}.  

In closing we note a potential caveat in the applicability of our formulae to the linearization analysis of systems like \req{rateeqn} and \req{GLV}. We have derived the general formulae of Secs.~\ref{subsec-specdensity}--\ref{sec:genformdynamics} assuming that $M$, $L$ and $R$ are independent of $J$. However, $M$ and $R$ as given by \reqs{rateeqn_MLR_v} and \myref{rateeqn_R} (or $M$ and $L$ in \reqs{GLV_L}--\myref{GLV_M}) depend on $J$ via their dependence on $\vv_*$ ($\vn_*$). However, in our experience this dependence is often too weak and indirect to render our formulae inapplicable; an example is provided by the excellent agreement of the empirical spectrum and the red boundary  given by our formula in Fig.~\ref {fig-Gapjuncnet}, which also held for other parameter choices of the model of Ref.~\cite{SternAbbott:2012}.

\section{Derivation of the formula for the spectral density}
\label{sec-deriv1}

In this section we will derive the formulae \reqs{support-res}--\myref{rho-res} for the average spectral density, $\rho(z)$, of random matrices of the form $\A = M+ LJR$ where $M$, $L$ and $R$ are deterministic matrices, and $J$ is random with iid elements of zero  mean and variance $1/N$. 
We will use the Hermitianized diagrammatic method developed in Refs.~\cite{FeinbergZee:1997,FeinbergZee97} (and reviewed in Ref.~\cite{FeinbergFeinberg:2006}), which we will recapitulate here for completeness.
As mentioned in Sec.~\ref{sec-results}, the spectral density is self-averaging for large $N$. Furthermore, as established in  Ref.~\cite{TaoKrishnapur:2010}, it is also universal in the large $N$ limit, in the sense that it is independent of the details of the distribution of the elements of $J$ as long its mean and variance are as stated. The same universality theorem also ensures that the real or complex nature of $J$ does not by itself affect $\rho(z)$ to leading order. 
Therefore, for simplicity we consider the case where $J$ is 
a {zero-mean complex Gaussian} random matrix with $\mylang J_{ab} J_{cd}\myrang = 0$, and 
\be\label{jcov}
\mylang J_{ab} {J}^*_{cd}\myrang = \frac{1}{N} \delta_{ac}\delta_{bd}.
\ee
Thus $\langle |J_{ab}|^2\rangle = \frac{1}{N}$, and all other first and second moments of $J$ (including $\langle J_{ab}^2\rangle$)  vanish. 
The measure on $J$ can be written as
\be\label{jmeas}
d\mu(J) \propto e^{- N \Tr\!\lpr J J^\dagger\rpr} \prod_{ab}d\Im J_{ab}d\Re J_{ab}.
\ee
In this form, and by the invariance of the trace, it is clear that the measure is symmetric with respect to the group $U(N)\otimes U(N)$, acting on $J$ by $J\mapsto U J V^{\dagger}$ where $U$ and $V$ are arbitrary $N\times N$ unitary matrices.

For  a particular realization of $J$, we define the ``Green's function" $\origGF$ by
\be\label{Gj}
\origGF \equiv  \frac{1}{M_z- J}, 
\ee
where $M_z=L^{-1}(z-M)R^{-1}$ (Eq.~\myref{Mzdef}). 
In the case $L , R \propto \one$, $\origGF$ will be proportional to the resolvent of $A$, $\frac{1}{z-A}$. More generally we have 
\be\label{GFresolvant}
\frac{1}{z-A}  =  R^{-1} \origGF L^{-1}. 
\ee
Following Ref.~\cite{FeinbergZee97}, we will use the identity 
\be
\delta^2(z) 
= 
 \frac{1}{\pi}\dbar \partial_z \ln |z|^2 
=
 \frac{1}{\pi}\dbar \lpr \frac{1}{z} \rpr
\label{poisson}
\ee
where the first identity follows by noting that $4 \dbar \partial_z = \nabla^2$, where $\nabla^2$ is the 2-D Laplacian, and recalling from electrostatics that the solution of Poisson's equation for a point charge at origin, \ie\ $\nabla^2 \phi(z) = 4\pi \delta^2(z)$,   in 2-D  is given by the potential field $\phi(z) = \ln |z|^2$; the second identity follows from $\partial_z \ln|z|^2 =  \partial_z (\ln z + \ln \zbar) = \frac{1}{z} + 0$. Using \req{poisson} we can write the empirical spectral density, defined in  \req{rhoJdef}, as
\be
\rho_J(z)  = \frac{1}{\pi} \dbar  \,\frac{1}{N}  \sum_\alpha \frac{1}{z - \lambda_\alpha} 
= \frac{1}{\pi}\dbar  \,\,\tr \frac{1}{z-A}. 
 \ee
Performing the ensemble average we obtain
\be\label{rho}
\rho(z) \equiv \Javg{ \rho_J(z)} = \frac{1}{\pi}\dbar  \,\,\tr\![(RL)^{-1} \Javg{\origGF}],
\ee
where we used \req{GFresolvant}, and the linearity and cyclicity of the trace. 
Thus, to calculate $\rho(z)$, our task boils down to calculating $ \Javg{\origGF}$.

The diagrammatic technique provides a method for calculating averages of products of $G(z;J)$'s. However, this method in its standard form relies on $\A$ being  a  Hermitian matrix.   It starts by 
an expansion of $G(z;J)$ in powers of $J$, which is only valid when $z$ is far enough from the spectrum of $A$, \ie\ away from the points we are most interested in. For Hermitian matrices, this is no problem as the spectrum is confined to the real line, and therefore $\origGF$ and $\Javg{\origGF}$ will be analytic outside the real line.
Thus one can use the expansion for $z$ far away outside the real line, perform the averaging over $J$, and sum up the most dominant contributions to obtain a result analytic in $z$. 
This result can then be analytically continued to $z$ arbitrarily close to the spectrum on the real line, yielding information about the spectrum.
All this would seemingly fail in the case of a  nonnormal (and in particular non-Hermitian) $\A$, with eigenvalues that in general cover a two dimensional region in the complex plane.  However, using a trick  introduced by Ref.~\cite{FeinbergZee97}, we can turn this problem to an auxiliary problem of averaging the Green's functions for a Hermitian matrix. By doubling the degrees of freedom, one defines a $z$-dependent, $2N\times 2N$ Hermitian ``Hamiltonian"
\be
H(z) \defin  
\begin{pmatrix}
 0     &   M_z-J \\
M_z^\dagger - J^\dagger     &  0
\end{pmatrix},
\ee
and the corresponding $2N\times 2N$ resolvent matrix or Green's function depending on a new complex variable $\eta$:
\bea\label{GbJ}
&& \GJ{\eta,z} \defin  \lpr\eta-H(z)\rpr^{-1}
\\
&& \qquad = 
- \begin{pmatrix}
\frac{ \eta}{ (M_z-J)(M_z-J)^\dagger - \eta^2}   &    \frac{M_z-J}{ (M_z-J)^\dagger(M_z-J) - \eta^2} \\
\frac{(M_z-J)^\dagger}{ (M_z-J)(M_z-J)^\dagger - \eta^2}     &  \frac{ \eta} { (M_z-J)^\dagger(M_z-J) - \eta^2}
\end{pmatrix}. \nonumber
\eea
For ${\eta\to i 0}$, we see that
\be\label{G0z}
\GJ{0,z}  =  - \begin{pmatrix}
 0     &    (M_z-J) ^{-\dagger}\\
(M_z-J) ^{-1}     &  0
\end{pmatrix}.
\ee
and thus from \req{Gj}, for any realization of 
$J$
\be\label{Geta-z}
\origGF = - \lim_{\eta\to i 0} \Gb^{21}(\eta,z;J)
\ee
{Here, we have used the notation 
\be \Gb(\eta,z;J) = \begin{pmatrix}
   \Gb^{11}(\eta,z;J)   & \Gb^{12}(\eta,z;J)    \\
   \Gb^{21}(\eta,z;J)   &  \Gb^{22}(\eta,z;J)
\end{pmatrix}, 
\ee
where $\Gb^{\alpha\beta}$ (with $\alpha ,\beta \in \{1,2\}$) are $N\times N$ matrices, forming the four blocks of $\Gb$. }
We have written the limit in \req{Geta-z} as $\eta \to i0$ to emphasize that until the end of our calculations $\eta$ is to retain a nonzero imaginary part, which serves to regularize the denominators in \req{GbJ}; c.f. the discussion after \req {Gscba}.
We will be carrying out a perturbation expansion in powers of $J$, so we decompose the Hamiltonian according to 
 \bea\label{H0J}
&& H(z) = H_0(z) - \J, 
\\
\label{jsig}
&& \J   \defin   \begin{pmatrix}
   0   & J   \\
   J^\dagger   &  0
\end{pmatrix},
\quad
H_0(z) \defin   \begin{pmatrix}
   0   & M_z   \\
   M_z^{\dagger}   &  0
\end{pmatrix}\! .   \qquad
\label{H0def}
\eea
We will sometimes use a tensor product  notation to denote matrices in this doubled up space, \eg\ writing $\J= \sigp \otimes J + \sigm \otimes J^\dagger$,
  where we defined the $2\times 2$ matrices
\be
 \sigp = \begin{pmatrix}
   0   & 1    \\
   0   &  0
\end{pmatrix}\qquad\qquad  \sigm  = \begin{pmatrix}
   0   & 0    \\
   1   &  0
\end{pmatrix}.
\ee
By a slight abuse of notation we also denote $2N\times 2N$ matrices $\sigma^{\pm}\otimes \one_{_{N\times N}}$ by $\sigma^{\pm}$, and we will denote the identity matrix in any space by $\one$. 
From \reqs{Geta-z} we obtain 
$
\tr [ (RL)^{-1} \origGF] = - \tr\! \lbr \left(\sigp \otimes
(RL)^{-1}\right) \GJ{i0^+,z} \rbr,  
$
and  from \req{rho}
\bea\label{rho1}
\rho(z) 
\areq - \lim_{\eta\to i 0} \frac{1}{\pi}\,\partial_{\zbar}\,\tr\!\!\lpr \left(\sigp
\otimes (RL)^{-1}\right) \G(\eta,z) \rpr\qquad
\\
\areq  - \lim_{\eta\to i 0}\frac{1}{\pi}\,\partial_{\zbar}\,\tr\!\!\lpr (RL)^{-1} \G^{21}(\eta,z) \rpr,
\label{rho11}
\eea
where we defined
\be\label{gbar}
\G(\eta,z) \defin  \Javg{\Gb(\eta,z;J)}.
\ee
Having expressed $\rho(z)$ in terms of the ensemble average of the Green's function for a Hermitian matrix, we now develop the diagrammatic method for calculating ensemble averages of products of $\Gb(\eta,z;J)$ (including $\G(\eta,z)$).
Note that, being the Green's function of a Hermitian matrix, $\Gb(\eta,z;J)$ and hence $\G(\eta,z) =  \Javg{\Gb(\eta,z;J)}$ are analytic functions of $\eta$ for $\eta$ outside the real line, and therefore  analytic continuation can be used to take the limit $\eta\to i 0$ after obtaining the average over $J$ for $\eta$ sufficiently away from the real line.  

We will denote the elements of a generic $2N\times 2N$ matrix $A$  by $A^{\alpha\beta}_{ab}$, where the Greek indices range in $\lcr 1,2\rcr$ and the Latin indices range in $\lcr 1,\ldots,N\rcr$. Using this notation, the definition \req{jsig}, and \req{jcov}, we can write the covariance for the components of $\J$ as
\be\label{jpair}
\Javg{ \J^{\alpha\beta}_{ab} \J^{\gamma\delta}_{cd} } = \frac{1}{N}\,\delta_{ad}\delta_{bc} \lpr \sigp_{\alpha\beta}\sigm_{\gamma\delta} +  \sigm_{\alpha\beta}\sigp_{\gamma\delta}  \rpr
\ee
(the terms proportional to $\sigp\sigp$ and $\sigm\sigm$ involve $\langle J_{ab}J_{cd}\rangle$, or its complex conjugate, {which vanish  for the complex Gaussian ensemble}). 
 It will be more handy to rewrite the parenthesis on the right side of \req{jpair}
as $ \pi^1_{\alpha\delta}\pi^2_{\gamma\beta} +  \pi^2_{\alpha\delta}\pi^1_{\gamma\beta}$,  where 
\be\label{pidef}
\pi^1  \defin  \begin{pmatrix}
     1 &  0  \\
     0 &  0 
\end{pmatrix} 
\qquad \qquad
\pi^2  \defin  \begin{pmatrix}
    0  &  0  \\
     0  &  1
\end{pmatrix},
\ee
yielding
\be\label{jpair2}
\Javg{ \J^{\alpha\beta}_{ab} \J^{\gamma\delta}_{cd} } = \frac{1}{N} \sum_{r=1}^2 (\pi^r_
{\alpha\delta}\delta_{ad}) \,\, (\pi^{3-r}_{\gamma\beta}\delta_{cb}).
\ee
Also, since $J_{ab}$ have zero mean, we have $\Javg{\J} = 0$.

\begin{figure}[!t]\hspace{-0cm}
\includegraphics[height=2.23in,angle=0]{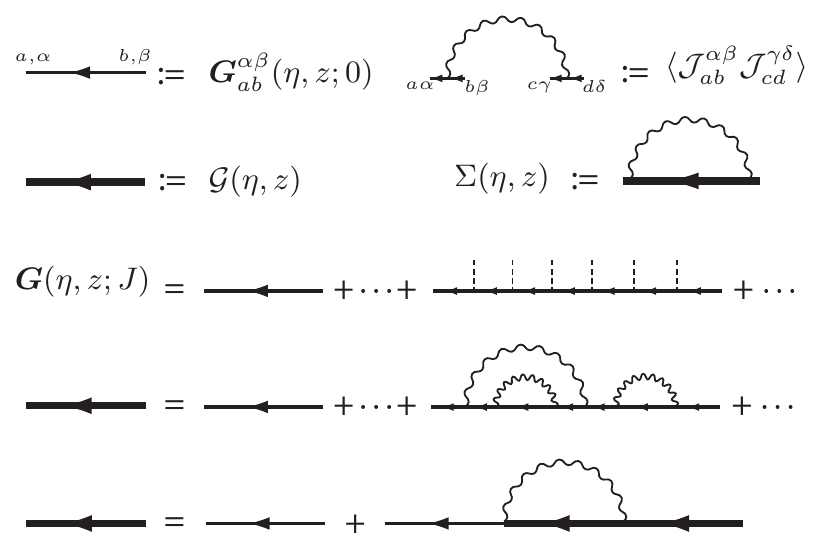}
\vspace{-.1cm}
\caption{The first two lines define different elements of Feynman diagrams: the Green's function for $J=0$ (zero disorder), $\Gb_{ab}^{\alpha\beta}(\eta,z;0)$, the covariance of two $\J$ elements, the ensemble averaged Green's function, $\G(\eta,z) \defin  \Javg{\GJ{\eta,z}}$, and the self-energy{ $\Sigma(\eta,z)$ }, \req {selfendef} (the matrix indices for $\G(\eta,z)$ and $\Sigma(\eta,z)$ are arranged as for $\Gb_{ab}^{\alpha\beta}(\eta,z;0)$). The third line is the diagrammatic representation of the expansion \req{PT} of { $\GJ{\eta,z}$ } before averaging over $J$, where the $\J$'s are represented by dashed lines. Averaging over \req{jmeas} is performed by pairing all $\J$'s and connecting them with the wavy lines representing $\langle \J \J \rangle$.  
In the large $N$ limit, the contribution of crossing pairings is suppressed by negative powers of $N$; the sum of all non-crossing diagrams, shown on the fourth line, yields the leading contribution to $\G(\eta,z)$ for large $N$.  
The last line shows the diagrammatic representation of \req{scba0}, which if iterated generates all the non-crossing diagrams. 
Alternatively, $\G(\eta,z)$ can be found by solving this self-consistent equation directly. 
}
\label{fig-dyson}
\end{figure}

The starting point of the diagrammatic method is the perturbation expansion of $ \GJ{\eta,z}= (\eta-H_0(z) -\J)^{-1}$ in powers of $\J$
\bes\label{PT} 
 \GJ{\eta,z}  
= \Gb(\eta,z;0) \sum_{n=0}^\infty \lbr \J \Gb(\eta,z;0) \rbr^n
\end{split}
\ee
where $\Gb(\eta,z;0)$ is given by \req{GbJ} with the $J$'s set to zero. This equation is represented diagrammatically in the third line of Fig.~\ref{fig-dyson}; the thin arrows defined in the first line of the figure represents $\Gb(\eta,z;0)$, and the dashed lines represent a power of $\J$ before ensemble averaging. 
To obtain the average resolvent, $\G(\eta,z)$, we then average  \req{PT}, term by term, with respect to the ensemble \req{jmeas}. Since the measure is Gaussian with zero mean, according to Wick's formula, the average of \emph{each} term  of \req{PT} involving $n$ factors of $\J$ is given by a sum over the contributions of all possible complete pairings of the $\J$'s in that term (in particular, since $\Javg{\J} = 0$, terms in \req{PT} with odd powers of $\J$ vanish after averaging). 
Each pairing can be represented as a Feynman diagram, as shown in Fig.~\ref{fig-dyson}, the first two lines of which define the diagram elements. For example, the last diagram in the fourth line of Fig.~\ref{fig-dyson} shows one possible pairing of the term in \req{PT} corresponding to $n=6$. 
The contribution of each pairing diagram is given by a product of factors, one per each pair, given by \req{jpair2} (represented by wavy lines)  with the right indices for that pair, as well as the factors of $\Gb(\eta,z;0)$ (represented by thin arrows), with all the intervening Greek and Latin matrix  indices summed over their proper ranges.
We show in Appendix \ref{app-nonxing} that for $\Im\, \eta\neq 0$, and so long as $\| (RL)^{-1}\|$ remains bounded as $N\to\infty$,  only  non-crossing pairings need to be retained  in the large $N$ limit, as crossing pairings are suppressed by inverse powers of $N$ and do not contribute in the limit (a pairing diagram is non-crossing if  it can be drawn on a plane, with the wavy lines drawn only on the half-plane above the straight arrow line, without any wavy lines crossing). 
As the last two lines of Fig.~\ref{fig-dyson} demonstrate, all non-crossing diagrams can be generated by iterating the equation 
\be\label{scba0}
\G(\eta,z) = \Gb(\eta,z;0) + \Gb(\eta,z;0) \Sigma(\eta,z) \G(\eta,z),
\ee
starting from $\G_{(0)}(\eta,z) = \Gb(\eta,z;0)$.
This equation is represented diagrammatically in the last line of Fig.~\ref{fig-dyson},  
with the ``self-energy" matrix, $\Sigma(\eta,z)$, defined by the diagram in the second line of that figure, \ie\ 
\be\label{selfendef}
\Sigma(\eta,z)\defin \Javg{\J \G(\eta,z) \J}.
\ee
 Using \req{jpair2} we obtain
\be
\Sigma^{\alpha\delta}_{ad} (\eta,z) = \delta_{ad} \sum_{r=1}^2 \pi^r_
{\alpha\delta} \,\,\frac{1}{N}\, \Tr (\pi^{3-r} \G(\eta,z)),
\ee
which using \req{pidef} we can write as 
\be\label{selfen}
\Sigma(\eta,z) = \begin{pmatrix}
	- i g_2(\eta,z)\one   & 0    \\
   0  & - i g_1(\eta,z)\one
\end{pmatrix},
\ee
where we defined the scalar functions
\be\label{selfcons}
g_\alpha(\eta,z) \defin  i \, \tr \G^{\alpha \alpha}(\eta,z).
\ee
Using \req{selfen}  we can solve \reqs{scba0}--\myref{selfen}  for $\G(\eta,z)$ at once, in terms of $g_\alpha(\eta,z)$, and then use \req{selfcons} to obtain a self-consistency equation, which can be solved for $g_\alpha(\eta,z)$. To this end, we multiply \req{scba0} by $\Gb^{-1}(\eta,z;0)$ on the left, and by $\G^{-1}(\eta,z)$ on the right,  to obtain
\bea\label{scba}
\G(\eta,z) \areq \lbr \Gb^{-1}(\eta,z;0)  - \Sigma(\eta,z)\rbr^{-1} \nl
\areq \lbr  \eta - H_0(z)  - \Sigma(\eta,z)\rbr^{-1}.
\eea
Using this expression with \reqs{jsig} and \myref{selfen}, it can be easily checked  that 
\bea\label{G2b2}
\G(\eta,z)  &=& - \begin{pmatrix}
 (\eta + ig_1)  K_1^{-1}   &    (z-M)K_2^{-1} \\
(\zbar - M^\dagger) K_1^{-1}     &   (\eta +  ig_2) K_2^{-1}
\end{pmatrix}, 
\eea
where $K_1 \equiv { M_z M_z^\dagger +  (g_1 - i\eta) (g_2 - i\eta)}$ and $K_2 \equiv {  M_z^\dagger M_z +  (g_1 - i\eta) (g_2 - i\eta)}$, and we dropped the arguments of $g_\alpha(\eta,z)$ for succinctness. 
Imposing \req{selfcons} we obtain the self-consistency equations
\bea\label{g1g2}
g_1 \areq  (g_1 - i\eta)\,  \tr\!\!\lpr K_1^{-1} \rpr, 
\\
g_2 \areq  (g_2 - i\eta)\,  \tr\!\!\lpr K_2^{-1} \rpr.
\eea
Before solving these equations for $g_1$ and $g_2$, we first show that $\tr\!\!\lpr K_1^{-1} \rpr = \tr\!\!\lpr K_2^{-1} \rpr$. One way to see this is to use the singular value decomposition  (SVD) of $M_z$ in the form
\be\label{svd}
M_z = U_z S_z V_z^\dagger,
\ee
where $S_z$ is a nonnegative diagonal matrix with the singular values of $M$, $s_i(z)$ ($i=1, \cdots, N$), on the diagonal, and $U_z$ and $V_z$ are unitary matrices (as in Sec.~\ref {subsec-specdensity} we include possibly vanishing singular values among $s_i(z)$, so that $S_z$, $U_z$ and $V_z$ are always $N\times N$ matrices). Using  the invariance of trace under similarity transforms, we obtain $\tr\!\!\lpr K_1^{-1} \rpr = \tr\!\!\lpr K_2^{-1} \rpr =  \tr\!\!\lpr { S_z^2 +  (g_1 - i\eta) (g_2 - i\eta)} \rpr^{-1} $.  Given this equality,
it is not hard to see that \reqs{g1g2} cannot be simultaneously satisfied unless $g_1(\eta,z) = g_2(\eta,z) \defin  g(\eta,z)$, with $g(\eta,z)$ satisfying 
\be
g =(g - i\eta)  \, \tr\!\!\lbr \frac{1}{S_z^2  + (g - i\eta)^2} \rbr,
\ee
or as written in the original basis
\be\label{geq}
g =  (g - i\eta)  \tr\!\!\lbr \frac{1}{M_z M_z^\dagger+(g -i\eta)^2} \rbr.
\ee
Noting from \reqs{selfen}, that the self-energy is thus proportional to the $2N\times 2N$ identity matrix, from \reqs{scba} and \myref{GbJ} (for $J=0$)  we obtain
\bea\label{Gscba}
\G(\eta,z) \areq \Gb(\eta+ig(\eta,z), z;0)
\\
\areq  - \begin{pmatrix}
\frac{ i\gamma}{ M_z M_z^\dagger + \gamma^2}   &    \frac{M_z}{ M_z^\dagger M_z + \gamma^2} \\
\frac{M_z^\dagger}{ M_z M_z^\dagger + \gamma^2}     &  \frac{ i\gamma}{ M_z^\dagger M_z + \gamma^2}
\end{pmatrix}
\nonumber
\eea
where $\gamma \defin g(\eta,z) - i\eta$.

According to \req{Geta-z}, for our  case of interest we must solve \req{geq} in the limit $\eta \to i0$. 
Note, however, that as shown in  Appendix \ref{app-nonxing}, the non-crossing approximation is in general guaranteed to work only for $\Im\, \eta\neq 0$; hence the limit $\eta \to i0$  must be taken after the limit $N\to\infty$ (as already pointed out in Sec.~\ref {sec-results}, taking the limits in this order is important in cases where some of the singular values in $S_z$ vanish in the limit $N\to\infty$).   For our purposes, it suffices to let $\eta = i\epsilon$ for some real positive $\epsilon$, and take the limit $\epsilon\to 0^+$ at the end. In this case one must seek a positive solution for $g(i\epsilon,z)$  in \req{geq};  this is because by definition,   
$
g(\eta,z) = i\tr\G_{11}(\eta,z) = \Javg{\tr i\Gb_{11}(\eta,z;J)}
$
and from \req{GbJ} we obtain 
$
g(i\epsilon,z)  =  \Javg{\tr\frac{\epsilon}{ (M_z-J)(M_z-J)^\dagger + \epsilon ^2} },
$
which for  $\epsilon >0$,  is the ensemble average of the trace of a positive definite matrix and hence positive. 
Taking the limit $N\to \infty$ while keeping $\epsilon$ (and hence $\epsilon + g$) positive and nonzero, we define
\be\label{Kdef0}
\K(\gamma,z)\defin \lim_{N\to \infty} \tr\!\!\lbr \frac{1}{M_z M_z^\dagger+\gamma^2} \rbr = \lim_{N\to \infty} \tr\!\!\lbr \frac{1}{S_z^2 + \gamma^2}\rbr
\ee
for $\gamma = g + \epsilon >0$. We can then rewrite \req{geq} as 
\be\label{gKeq0}
\gamma (1 - \K(\gamma,z)) = \epsilon,
\ee
with $\gamma = g + \epsilon$.
Since $\epsilon$ and $\gamma = g + \epsilon$ are positive, it follows that $1 - \K(\gamma,z)$ must also be positive. In the limit $\epsilon\to 0^+$ there are two possible situations: 1) $g,\gamma \to 0^+$, in which case we must have 
\be \label{reg1}
\lim_{\gamma\to 0^+} \K(\gamma,z) < 1,
\ee 
or $\lim_{\gamma\to 0^+} \K(\gamma,z) = 1$, or 2) the solution for $g$  stays finite and positive in the limit, while $\K(\gamma,z)\to 1^-$ as $\gamma \to g^+$. 
Thus in the second case  $g(z) \defin  \lim_{\epsilon \to 0^+} g(\epsilon ,z)$ must satisfy
$
\K(g(z),z) = 1,
$
\ie\ 
 \be\label{2ndeq}
1 =  \lim_{N\to \infty} \tr\!\!\lbr \frac{1}{S_z^2+g(z)^2} \rbr.
\ee
Note further that since $\K(\gamma,z)$ is a decreasing function of $\gamma$, in the second case  we have 
$ \K(0^+,z) \geq \K(g(z),z) =1$,
 \ie\ 
\be\label{reg2}
\lim_{\gamma\to 0^+} \K(\gamma,z) \geq 1.
\ee
Thus the two possible solutions are realized in complimentary regions (with a shared boundary) of the complex plane for $z$, respectively  given by \reqs{reg1} and \myref{reg2}.

Let us substitute the $g(z) = 0$ solution for the case \myref{reg1} in \req{Gscba}, and naively set $\eta = i\epsilon$ (and thus $\gamma$) to zero, to obtain
\be\label{sol1G}
\G(\eta=i0^+,z) = \Gb(\eta=i0^+, z;0).
\ee
From \reqs{G0z}--\myref{Geta-z}, this solution yields $\Javg{G(z;J)} = - \G^{21}(\eta=i0^+,z) 
= M_z^{-1} = R (z-M)^{-1} L$ which
is analytic outside the spectrum of $M$. 
Hence from \req{rho}, it yields $\rho(z) = 0$, at least outside the spectrum of $M$;  
a more careful analysis presented in Appendix \ref{app-rho}, in which we correctly take the limit $N\to\infty$ in \req{rho11} before taking $\epsilon \to 0^+$, confirms that in the region \req{reg1}, $\lim_{N\to\infty} \rho(z)$ always vanishes. 
We conclude that the support of $\lim_{N\to\infty} \rho(z)$ is  where  \req{reg2} holds (which is \req{support-res-K} of Sec.~\ref{sec-results}); here $g(z)$ is to be found by solving \req{2ndeq}, or equivalently \req{geqn} or \req{geqn-svd}.  
 In this region, we obtain $\rho(z)$ by substituting \req{Gscba}, with the solution of \req{2ndeq}, into \reqs{rho11}. This yields  \req{rho-res}, which we rewrite here as 
\bea\label{rho3}
 \rho(z) \areq  \frac{1}{\pi}\dbar\, E(z)\\
  E(z) &\defin & \tr\!\!\lbr \frac{(RL)^{-1} M_z^\dagger}{M_z M_z^\dagger+g(z)^2} \rbr,
\label{Edef0}
 \eea
with $g(z)$ given by \req{2ndeq}, or equivalently \req{geqn}.

We will now obtain an alternative expression for $\rho(z)$, equivalent to \reqs{rho3}--\myref{Edef0}, which explicitly shows that it depends only on the singular values of $M_z$. Noting that, from \req{Mzdef}, $\partial_z \lpr M_z M_z^\dagger \rpr  = (RL)^{-1} M_z^\dagger$, we can write \req{Edef0} as
\be\label{E0}
E(z) = \tr\!\!\lbr \frac{\partial_z \lpr M_z M_z^\dagger \rpr}{M_z M_z^\dagger+g(z)^2} \rbr.
\ee
On the other hand, we have
\bea
\partial_z \tr \ln\!\lbr {M_z M_z^\dagger+g(z)^2} \rbr \areq 
\tr\!\!\lbr \frac{\partial_z \lpr M_z M_z^\dagger + g^2(z) \rpr}{M_z M_z^\dagger+g(z)^2} \rbr, 
\nl
\areq E(z)  + \partial_z\!  \lpr g^2(z) \rpr,
\eea
where to write the last term we used \req{geqn}. Thus we obtain 
\bea\label{Ephi}
E(z) \areq \partial_z \varphi(z),
\\
\varphi(z) & \defin  & - g^2(z) +  \tr \ln\!\lbr {M_z M_z^\dagger+g(z)^2} \rbr.
\label{phidef00}
\eea
or using the SVD, \req{svd}, 
\bea
\varphi(z) \areq - g^2(z) +  \tr \ln\!\lbr {S_z^2+g(z)^2} \rbr,
\\
\areq  - g(z)^2  + \frac{1}{N} \sum_{i=1}^N \ln\! \lbr s_i(z)^2 + g(z)^2\rbr.
\label{phi-res}
\eea
Finally, substituting \req{phi-res} in \req{Ephi}, and using \req{geqn-svd}, we obtain \req{rho-svd-res}.

For the special case of $M = 0$, we have $M_z = z (RL)^{-1}$. If we let $\sigma_i$ to be the singular values of $RL$,  then the singular values of $M_z$ will be given by $s_i(z)= |z| \sigma_i^{-1}$.  
Substituting this in \req{geqn-svd} and multiplying both sides by $r^2 = |z|^2$, 
we obtain \req{geqnM0}.
We see immediately that $g(z)$, $\varphi(z)$ and $\rho(z)$ depend only on the radius $r=|z|$. 
Similarly we can rewrite \req{phi-res} as
\be\label{phiM0}
\varphi(r) =  - g(r)^2  +  \frac{1}{N}\sum_{i=1}^N\, \ln\! \lbr {r^2}{ \sigma_i^{-2}} + g(r)^2\rbr.
\ee
To find the spectral radius (boundary of the spectrum) $r_0$ we have to solve \req {geqnM0}  for $r$, setting $g(r) = 0$. This yields 
$r_0^{2} =  \frac{1}{N}\sum_{i=1}^N \sigma_i^2= \| RL \|_{\Fr}^2$, 
yielding \req {boundaryM0}.
 Let us define the proportion of eigenvalues lying outside a radius $r$ from the origin by $n_>(r)$.  To obtain \reqs{rhoM0} and \myref{n>M0}, first note that 
\be\label{rhoM01}
\rho(r) = \frac{1}{\pi}\dbar\partial_z\, \varphi(z) = \frac{1}{4\pi} \nabla^2\varphi(z) = \frac{1}{4\pi r} \frac{\partial}{\partial r} \lpr r \partial_r \varphi(r)\rpr,
\ee
where we used the expression of Laplacian, $\nabla^2 = \partial_x^2 + \partial_y^2$, in 2-D polar coordinates in the last equality.
Using this with the definition $n_>(r) = 2\pi \int_r^\infty \rho(r) r dr$, we obtain $n_>(r) = \lbr \frac{r}{2} \partial_r \varphi(r)\rbr^\infty_r$. 
For the limit at $r\to \infty$, note that for $r> r_0$,  $g(r) = 0$ and we have 
$\varphi(r) = \frac{1}{N}\sum_{i=1}^N \ln({r^2}{ \sigma_i^{-2}})  = 2 \ln r - \frac{2}{N}\ln \det (RL) $, 
and hence $\frac{r}{2} \partial_r \varphi(r) \to 1$ as $r\to\infty$. Thus we obtain 
\be\label{n>deriv}
n_>(r) = 1 - \frac{r}{2} \partial_r \varphi(r).
\ee
Differentiating \req {phiM0} and using \req{geqnM0} we obtain
\be
\partial_r \varphi(r) = 2r \frac{1}{N}\sum_{i=1}^N \frac{1}{r^2  + \sigma_i^{2} g(r)^2},
\ee
and
\bea
n_>(r) \areq 1 - r^2 \frac{1}{N}\sum_{i=1}^N \frac{1}{r^2  + \sigma_i^{2} g(r)^2},
\\
\areq g(r)^2 \frac{1}{N}\sum_{i=1}^N \frac{\sigma_i^{2}}{r^2  + \sigma_i^{2} g(r)^2}.
\eea
Using \req{geqnM0} once again we obtain \req{n>M0}. Finally, using the latter together with \reqs{rhoM01}--\myref {n>deriv} yields \req{rhoM0}.

We will prove further general properties for the eigenvalue density for $M=0$. Let us first define  
\be\label{InkdefM0}
I_{n,k}(g,r) \equiv \avgsig{\frac{\sigma^{-k} }{(g^2 + \sigma^{-2} r^2)^n} } 
\ee
and 
\be
\avgsig{f(\sigma)}  \equiv \lim_{N\to\infty} \frac{1}{N} \sum_{i=1}^N f(\sigma_i).
\ee
(We assume $\sigma_i$ have a  limit density,  $\rho_\sigma(\sigma)$, such that $
\avgsig{f(\sigma)} = \int_{0}^\infty f(\sigma)\rho_\sigma(\sigma) d\sigma$ is well-defined for $f(\sigma)$ with sufficiently fast decay at infinity. Note that since we assumed that $\| (RL)^{-1}\| = (\min_i \sigma_i)^{-1} = O(1) $, this density has no measure at $\sigma=0$ and hence the averages in \req{InkdefM0} are non-singular for $n,k\geq 0$. Also $\avgsig{f(\sigma)} $ is finite as long as $f(\sigma) = O(\sigma^2)$ as $\sigma\to\infty$, as we are assuming that the $\| RL\|_{F} = O(1)$ and $\lim_{N\to\infty} \| RL \|_F^2 = \avgsig{\sigma^2}$.)
First, we obtain  general expressions for $\rho(r=0)$ and $\rho(r = r_0)$, with  $r_0$  given by \req{boundaryM0}. From \req{rhoM0}, $\pi\rho(r) = -\frac{\partial n_>(r)}{\partial (r^2)}$, which using \req{geqnM0}, re-expressed as $I_{1,0}(g,r)=1$, we can write as 
\be\label{rhoinIsM0}
\pi\rho(r) = \frac{\frac{\partial I_{1,0} }{\partial (r^2)}}{\frac{\partial I_{1,0} }{\partial (g^2)}} = \frac{I_{2,2}(g,r)}{I_{2,0}(g,r)}.
\ee
Using the facts that at $r=0$, $g=1$,  and at $r=r_0$, $g=0$, we obtain
\bea
\rho(r=0) \areq \frac{1}{\pi} \frac{I_{2,2}(1,0)}{I_{2,0}(1,0)} = \frac{1}{\pi} \avgsig{ \sigma^{-2} }\\
\rho(r=r_0) \areq \frac{1}{\pi} \frac{I_{2,2}(0,r_0)}{I_{2,0}(0,r_0)} =\frac{1}{\pi}\frac{  \avgsig{ \sigma ^{2} }}{ \avgsig{ \sigma^{4} }}.
\eea
Using the fact that $\sigma^4$ and $\sigma^{-2}$ are anti-correlated and that $\sigma^{2} = \sigma^4 \sigma^{-2} $, we see that $  \avgsig{ \sigma ^{2} } \leq \avgsig{ \sigma ^{4} }  \avgsig{ \sigma ^{-2} }$ or 
\be
\rho(r=r_0) \leq \rho(r=0), 
\ee
with equality if and only if $\rho(\sigma)$ is deterministic, \ie,   a delta-function. This can happen if all but an $o(1)$ fraction of  the $\sigma_i$'s have the same limit as $N\to\infty$; in that case the eigenvalue distribution is given by the circular law. 
More generally, we can prove that $\rho(r)$ is a decreasing function of $r$ for any choice of $L$ and $R$ (with $M=0$). Using $\frac{d \rho(r)}{dr} = 2 r \frac{ d\rho(r)}{d(r^2)}$, and  \req{rhoinIsM0} we obtain
\be
\frac{d \rho(r)}{dr} = 2 r 
 \frac { \frac{ d I_{2,2}}{d(r^2)}  I_{2,0}  -  I_{2,2}  \frac{ d I_{2,0}}{d(r^2)}  }{ I_{2,0}^2},
\ee
and using
$\frac{d}{d(r^2)}  = \frac{\partial}{\partial(r^2)}  + \frac{\partial(g^2) }{\partial(r^2)} \frac{\partial}{\partial(g^2)} = \frac{\partial}{\partial(r^2)}  -\pi\rho(r) \frac{\partial}{\partial(g^2)} $ 
and $ \frac{\partial I_{n,k}}{\partial(g^2)} = - n I_{n+1,k}$ and $ \frac{\partial I_{n,k}}{\partial(r^2)} = - n I_{n+1,k+2}$ (we will drop the explicit $(g,r)$ dependence of $I_{n,k}$'s when convenient)
we find
\be
\frac{d \rho(r)}{dr} = -4 r 
\frac {   I_{2,0}^2 I_{3,4} - 2  I_{2,2}  I_{2,0} I_{3,2} + I_{2,2}^2 I_{3,0}}{I_{2,0}^3}
\ee
Defining
\be
\avgsigp{f(\sigma)} \equiv \frac{\avgsig{\frac{f(\sigma)}{(g^2 + \sigma^{-2} r^2)^2} }}{ \avgsig{\frac{1}{(g^2 + \sigma^{-2} r^2)^2}}},
\ee
($\avgsigp{f(\sigma)} $ is a bonafide expectation operator) we can write 
\bea
&& \frac{d \rho(r)}{dr} = -4 r 
\lbr 
\avgsigp{ \frac{\sigma^{-4}}{g^2 + \sigma^{-2}r^2}   } 
\right. 
\\
&& \left.
- 2 \avgsigp{ \frac{\sigma^{-2}}{g^2 + \sigma^{-2}r^2}   } \avgsigp{\sigma^{-2}}
+ \avgsigp{ \frac{1}{g^2 + \sigma^{-2}r^2}   } \avgsigp{\sigma^{-2}}^2
\rbr
\nonumber
\eea
or 
\bea\label{rhoderivM0}
&& \frac{d \rho(r)}{dr} = -4 r 
\lbr
 \mathrm{Cov}'[\frac{\sigma^{-2}}{g^2 + \sigma^{-2} r^2}, \sigma^{-2}]
\right. 
\nl
&& \left.
\qquad\qquad- \avgsigp{\sigma^{-2}} \mathrm{Cov}'[\frac{1}{g^2 + \sigma^{-2} r^2}, \sigma^{-2}]
 \rbr
 \eea
where $\mathrm{Cov}'[f,g] \equiv \avgsigp{fg} - \avgsigp{f}\avgsigp{g}$ is the covariance under $\avgsigp{\cdot}$. Now since 
$\frac{\sigma^{-2}}{g^2 + \sigma^{-2} r^2}$
 and $\sigma^{-2}$ are both strictly decreasing functions of $\sigma$ (since $g>0$ for $r<r_0$), while  $\frac{1}{g^2 + \sigma^{-2}r^2}$ is a strictly increasing function of $\sigma$ (for $r>0$), the first covariance on the right hand side of \req{rhoderivM0} is positive, while the second one is negative, and therefore 
\be
\frac{d \rho(r)}{dr} \leq 0.
\ee
This slope is zero at $r=0$ and strictly negative for $r>0$ as long as $\mathrm{Var}[\sigma]>0$ (again when  $\mathrm{Var}[\sigma]=0$ we obtain the circular law). 
At $r=r_0$ we obtain
\bea
\rho'(r_0) \areq -\frac{4}{r_0} \avgsigp{\sigma^{-2}} \lpr \avgsigp{\sigma^{-2}} \avgsigp{\sigma^{2}} -1\rpr
\nl
\areq -\frac{4}{r_0} \frac{\avgsig{\sigma^{2}} } {\avgsig{\sigma^{4}}^3} 
\lpr
  \avgsig{\sigma^{2}} \avgsig{\sigma^{6}} - \avgsig{\sigma^{4}}^2
\rpr
\eea
The curvature of $\rho(r)$ at zero can also be evaluated by taking the limit $r\to 0$ of the bracket in \req{rhoderivM0}, noting that $g\to 1$ as $r\to 0$. We obtain
\be
\rho''(r=0) = -4 \mathrm{Var}'[\sigma^{-2}]  = -4 \frac{\mathrm{Var}[\sigma^{2}]}{\avgsig{\sigma^4}^2} 
 \leq 0.
\ee

\section{Derivation of the  formula for the average norm squared}
\label{sec-deriv2}

In this section, we focus on the dynamics governed by the matrix $\A = M + L J R$, according to  
\req{ode}, and derive the general formulae presented in Sec.~\ref {sec:genformdynamics}.
We will first  consider  the system's response to an impulse input, $\vi(t) = \vx_0\delta(t)$, at $t=0$,  before which we assume the system was at rest in its fixed point $\vx = 0$.
We assume $\vx = 0$ is a stable fixed point, \ie\ all eigenvalues of $-\gamma\one + \A$ have negative real parts, or equivalently, all eigenvalues of $A$ have real parts less than $\gamma$ (more precisely, we assume that as $N\to \infty$, this will be the case almost surely, \ie\ for any typical realization of $J$; in particular, the vertical line of $z$'s with real part $\gamma$ must be to the right of the support of $\rho(z)$, the average eigenvalue density for $A$, as found by solving \req{support-res}).
 This means that $\vx(t)$ decays exponentially as $t\to \infty$, and therefore its Fourier transform, $\tilde{\vx}(\omega) \defin  \int_{-\infty}^\infty e^{-i \omega t} \vx(t) dt = \int_0^\infty e^{-i \omega t} \vx(t) dt$, is well-defined. 
Fourier transformation of \req{ode} with $\vi(t) = \vx_0\delta(t)$ 
yields $i\omega\tilde{\vx}(\omega) = (-\gamma + \A) \tilde{\vx}(\omega) + \vx_0$. Solving algebraically for $\tilde{\vx}(\omega)$, we obtain $\tilde{\vx}(\omega) = (\gamma  + i \omega -A)^{-1} \vx_0$,  or using \reqs{Gj}--\myref{GFresolvant},  $\tilde{\vx}(\omega) = R^{-1} G(\gamma+i \omega;J) L^{-1} \vx_0$. 
The inverse Fourier transform, $\vx(t) =  \int_{-\infty}^{\infty}   e^{i t \omega} \tilde{\vx}(\omega)\frac{d\omega}{2\pi}$, then yields
\bea\label{invlaplace}
\vx(t) \areq  \int_{-\infty}^{\infty} \frac{d\omega}{2\pi}  e^{i t \omega} R^{-1} G(\gamma + i\omega;J) L^{-1} \vx_0.
\eea

Our goal is to study the statistics of $\vx(t)$ (\eg, its moments) under the distribution \req{jmeas}. Equation \myref{invlaplace} allows us to reduce this task to the calculation of various moments of $G(z;J)$ and its adjoint, and these can be found using the diagrammatic technique. Note that, in general, these moments involve not only the statistics of the eigenvalues, but also that of the eigenvectors of $\A= M+LJR$; this can be seen from the spectral representation $ R^{-1} G(z;J) L^{-1}=(z-A)^{-1} = V  (z-\Lambda)^{-1} V^{-1}$ where $\Lambda$ is a diagonal matrix of the eigenvalues of $\A$, and $V$ is the matrix whose columns are the eigenvectors of $\A$.
Here we will look at the simplest interesting statistic involving the eigenvectors: the average square norm of the state vector, namely, $\Javg{\|\vx(t)\|^2}$. As we discussed in  Sec.~\ref{sec:genformdynamics}, its study is also motivated by the fact that transient amplification due to nonnormality of $A$ manifests itself in the transient growth of $\|\vx(t)\|^2= \vx(t)\trans \vx(t)$. 
With a slight generalization, we derive a formula for the average of a general quadratic function $\vx(t)\trans B \vx(t)$ where $B$ is any {symmetric} matrix; the norm squared corresponds to $B = \one$.
Using, $\vx(t)\trans = \vx(t)^\dagger$ ($\vx(t)$ is {real}), the identity $\vx^\dagger B \vx = \Tr (B \vx\vx^\dagger)$, and  \req{invlaplace}, 
we obtain
\bes\label{genformul}
\hspace{-.2cm} 
& \vx(t)\trans B \vx(t)  =  \int\!\!\!  \int\! \frac{d\omega_1}{2\pi}\frac{d\omega_2}{2\pi} e^{it(\omega_1-\omega_2)}\,
\\
& \quad\qquad \Tr\!\!\lpr  \tB\,  G(\gamma + i\omega_1;J)\, \tC \, G^\dagger(\gamma + i\omega_2;J)\rpr,
 \end{split}
\ee
where we defined $\tC \defin   L^{-1} \vx_0\vx_0\trans L^{-\dagger}$ and $\tB \defin R^{-\dagger} B R^{-1}$.  
Using \req{Geta-z} and $G^\dagger(z;J) = -\lim_{\eta\to  i0^+} \Gb^{12}(\eta,z;J)$, and the $2\times 2$ matrices $\pi^{r}$ defined in \req{pidef}, we can rewrite the trace in \req{genformul} as
$ {\Tr\!\!\lpr   \pi^2\! \otimes\! \tB\, \Gb(0,z_1;J)\, \pi^1\!\otimes\! \tC\, \Gb(0,z_2;J) \rpr}$, with $z_i = \gamma +i \omega_i$, where now the trace is performed over $2N\times 2N$ matrices.
Averaging over $J$ we then obtain
\bea \label{avgformul}
&& 
\Javg{\vx(t)\trans B \vx(t)} = \!\!
\\
&& \quad
\int\!\!\!  \int\! \frac{d\omega_1}{2\pi}\frac{d\omega_2}{2\pi} e^{it(\omega_1-\omega_2)}
 \F(\gamma + i\omega_1,\gamma + i\omega_2;B,\vx_0\vx_0\trans),
 \qquad
 \nonumber
 \eea
 where, for general matrix arguments $B$ and $C$, we define
 \bea
  \label{FBdef}
&&\F(z_1,z_2;B,C) \defin 
\\
&& \qquad\qquad \qquad
\Javg{\Tr\!\!\lpr   \Bb\, \Gb(0,z_1;J) \rhob\, \Gb(0,z_2;J) \rpr}.
\nonumber
\eea
with
\bea\label{Bb}\label{Btild}
&&\Bb\defin \pi^2 \otimes \tB,
\qquad\qquad \tB \defin R^{-\dagger} B R^{-1},
 \\
&& 
 \rhob \defin \pi^1\otimes \tC,
 \qquad\qquad\,
\tC \defin L^{-1} C L^{-\dagger}.
\label{Ctild}
\eea

Before proceeding to the calculation of $\F(z_1,z_2;B,C)$ using the diagrammatic technique, we will also express the other quantities presented in Sec.~\ref {sec:genformdynamics} in terms of $\F( \gamma + i\omega,  \gamma + i\omega;B,C)$, with  appropriate $B$'s and $C$'s.
First, we obtain the desired expression for the matrix power spectrum, \req{Covxomega}, of the steady-state response to a temporally white noisy input, $\vi(t)$, with covariance \req{CovI}. Using the Fourier transform of \req{ode}, and following similar steps to those leading to \req{invlaplace}, we can write the steady-state solution for $\vx(t)$ as in \req{invlaplace} with $\vx_0$ replaced by the Fourier transform of the input, $\tilde\vi(\omega)$. Using this and exploiting 
${\vx}_j(t_2) = {\vx}_j^*(t_2)$
 we can write (after averaging over the input noise)
\be\label{CovxA}
\overline{{\vx}_i(t_1) {\vx}_j(t_2)} = \int\!\!\!  \int\! \frac{d\omega_1}{2\pi}\frac{d\omega_2}{2\pi} e^{it_1 \omega_1 - i t_2 \omega_2} K_{ij}(\omega_1,\omega_2)
\ee
where the Fourier-domain covariance matrix, $K(\omega_1,\omega_2) \equiv \overline{\tilde\vx(\omega_1)\tilde\vx(\omega_2)^\dagger}$, is given by 
\bea\label{Adef1}
&& K(\omega_1,\omega_2) \defin  
\\
&& R^{-1} G(\gamma + i\omega_1;J) L^{-1} C^\vi(\omega_1,\omega_2) L^{-\dagger} G^\dagger(\gamma + i\omega_2;J) R^{-\dagger}.
\nonumber
\eea
Here,    the bars indicate averaging over the input noise distribution, and we defined $C^\vi(\omega_1,\omega_2) \equiv \overline{\tilde {\vi}(\omega_1) \tilde{\vi}(\omega_2)^\dagger}$. 
On the other hand, the Fourier transform of \req{CovI} yields 
\be\label{CovIFourier}
C^\vi(\omega_1,\omega_2) \defin   \overline{\tilde {\vi}(\omega_1) \tilde{\vi}(\omega_2)^\dagger} = 2\pi\delta(\omega_1-\omega_2) C^{\vi},
\ee
where we also exploited $\tilde{\rm I}^*_j(\omega) = \tilde{\rm I}_j(-\omega)$ for a real $\vi(t)$.
Substituting \req {CovIFourier} into \reqs{CovxA}--\myref{Adef1} we obtain 
\be\label{InvFourCovx}
\overline{{\vx}_i(t_1) {\vx}_j(t_2)} =   \int\! \frac{d\omega}{2\pi} e^{i \omega(t_1-t_2)} C^{\vx}_{ij}(\omega),
\ee
where
\bea\label{Cxdef1}
&& C^{\vx}(\omega) = 
\\
&& \qquad R^{-1} G(\gamma + i\omega;J) L^{-1} C^\vi L^{-\dagger} G^\dagger(\gamma + i\omega;J) R^{-\dagger}.
\nonumber
\eea
Noting that \req {InvFourCovx}  expresses the covariance of the response as an inverse Fourier transform, we see that  $C^{\vx}(\omega)$ is indeed the power spectrum of the response, as defined in \req{Covxomega}. Finally note that the element,  $C_{ij}$, of any matrix can be expressed as $\Tr(\ve_j \ve_i\trans C)$, where $\ve_i$ are the unit basis vectors (\ie\ vectors whose $a$-th component is $\delta_{ia}$).
Using this trick with \req{Cxdef1}, and following the steps leading from \req{genformul} to \req{avgformul}, we see that after ensemble averaging, $\Javg{C^{\vx}_{ij}(\omega)}$ can be written in the form 
\be\label{CxFB}
\Javg{C^{\vx}_{ij}(\omega)} = \F(\gamma+i\omega, \gamma + i\omega; \ve_j \ve_i\trans,C^\vi)
\ee
 where $\F$ was defined by \reqs{FBdef}--\myref{Ctild}.

Next, consider the system \req{ode} being driven by a sinusoidal input $\vi(t) =\vi_0 \sqrt{2} \cos{\omega t}$ (the factor of $\sqrt{2}$ serves to normalize the time average of $(\sqrt{2}\cos \omega t)^2$ to one), and consider the steady state response, which will also oscillate at frequency $\omega$. Decomposing the input, $\vi(t)$, and the steady-state response, $\vx_\omega(t)$, into their positive and negative frequency components (proportional to $e^{i\omega t}$ and $e^{-i\omega t}$, respectively), from \req{ode} we obtain
\be
\vx_\omega(t) = \sqrt{2} R^{-1} \Re[e^{i\omega t} G(\gamma + i\omega;J) ] L^{-1} \vi_0.
\ee 
Thus the norm squared of the steady state response, $\|\vx(t)\|^2 = \vx(t)^\dagger \vx(t)$, will have a zero frequency component, plus components oscillating at $\pm 2\omega$. Averaging over time kills the latter, leaving the zero frequency component intact, yielding
\bea
\overline{\vx_\omega(t)\trans \vx_\omega(t)}  \areq  \vi_0\trans L^{-1} G^\dagger(z;J)R^{-\dagger}R^{-1}G(z;J) L^{-1} \vi_0 
\nl
\areq  \Tr\!\!\lpr R^{-\dagger}R^{-1} G(z;J) \rho_{\vi} G^\dagger(z;J) \rpr
\eea
where $z=\gamma + i\omega$, the bar indicates temporal averaging,  and we defined 
$\rho_{\vi} \defin   L^{-1} \vi_0\vi_0\trans L^{-\dagger}$.
 Generalizing to $\overline{\vx_\omega(t)\trans B \vx_\omega(t)}$, averaging over the ensemble, and following the steps leading from \req{genformul} to \req{avgformul}, we obtain
\be\label{powerspec}
\Javg{\overline{\vx_\omega(t)\trans B \vx_\omega(t)}}  =  \F(\gamma + i\omega,\gamma + i\omega; B,\vi_0\vi_0\trans), 
\ee
where $\F$ is given by \reqs{FBdef}--\myref{Ctild}.
Comparing \req {powerspec}  with \req {CxFB}, we also Êobtain
\be\label{powerspec2}
\Javg{\overline{\vx_\omega(t)\trans B \vx_\omega(t)}}  =  \Tr\!\!\lpr B \Javg{C^{\vx}(\omega)} \rpr
\ee
which is \req{res-powerspec} of Sec.~\ref{sec-results}, it being understood that $C^{\vi}$ in \req{CxFB} is replaced by $\vi_0\vi_0\trans$ as in \req{powerspec}. 
%

\begin{figure}[!t]
\includegraphics[height=2.15in,angle=0]{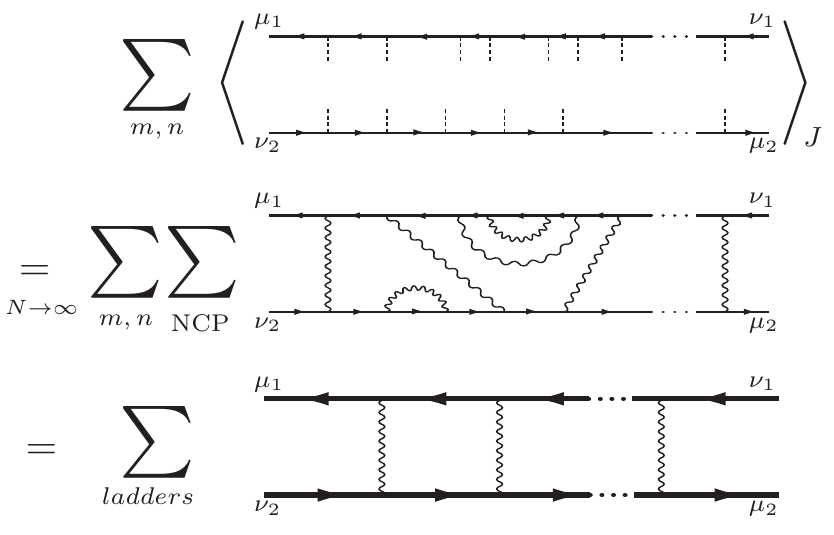}
\vspace{-.2cm}
\caption{Contribtutions to \req{Fdef} in the non-crossing approximation. The first line shows \req{Fdef} written using the expansion \req{PT}. The diagram shows the contribution of the $m$-th and $n$-th terms in the expansion for two Green's functions, respectively. Thus the top (bottom) solid line contains $m$ ($n$) factors of $\J$, shown by dashed lines. In the large $N$ limit, averaging each summand over $J$ boils down to summing all non-crossing pairings (NCP) of the dashed lines. The second row shows a specific non-crossing pairing for the diagram shown in the first line. Finally,  summing over all $m$, and $n$ and all NCP's, is equivalent to replacing all solid lines (representing $\Gb(\eta_i,z_i;J=0)$) with thick solid lines representing the non-crossing average Green's function, $\G(\eta_i,z_i)$ (calculated according to \reqs{scba}--\myref{selfen}), and summing over all NCP's with \emph{every} pairing connecting the straight lines on top and bottom (and not each to itself). This procedure yields the ladder diagrams, the sum over which is shown in the third line.
}
\label{fig-ladder}
\end{figure}

Now that we have expressed all our quantities of interest in terms of the kernel $\F$ as defined in \req{FBdef}, our task boils down to performing the average over $J$ in \req{FBdef} to obtain a closed formula for $\F$ with general arguments $B$ and $C$. 
To this end,  we now  proceed to calculate the more general object 
\be\label{Fdef}
F_{\mu_1\nu_2;\mu_2\nu_1}(1;2)
\defin \Javg{ \Gb_{\mu_1\nu_1}(1;J)\Gb_{\mu_2\nu_2}(2;J)}, 
\ee
using the diagrammatic technique. Here, we adopted the abbreviated notation $(1) \defin  (\eta_1,z_1)$ and $(2)\defin (\eta_2,z_2)$
 for the function arguments, and  
 $\mu_i = (\alpha_i,a_i)$ (similarly for $\nu_i$) for indices in the $2N$ dimensional space (as in Sec.~\ref{sec-deriv1}, $\alpha,\beta,\ldots$, and $a,b,\ldots$ denote indices in the 2 and $N$ dimensional spaces, respectively). Once we have calculated  $F_{\mu_1\nu_2;\mu_2\nu_1}(1;2)$, we can obtain $\F(z_1,z_2;B,C)$, with the appropriate $B$ and $C$, via 
 \be\label{FBF}
\F(z_1,z_2;B,C) = \Bb_{\nu_2\mu_1}F_{\mu_1\nu_2;\mu_2\nu_1}(0,z_1;0,z_2)\rhob_{\nu_1\mu_2} ,
\ee
where all indices are summed over, and $\Bb$ and $\rhob$ were defined in \reqs{Btild}--\myref{Ctild}. 

As before, we start by using the expansion \req{PT}  for the two Green's functions in \req{Fdef}. This is shown diagrammatically in the first line of Fig.~\ref{fig-ladder}, for the contribution of $m$-th and $n$-th terms in the expansion of the first and the second Green's function, respectively. 
As before, for large $N$, averaging over $J$ entails summing the contribution of all non-crossing pairings.
 This is indicated in the second line of Fig.~\ref{fig-ladder}. Finally, the third line of Fig.~\ref{fig-ladder} shows that summing over all $m$'s, and $n$'s and all non-crossing pairings, is equivalent to replacing all solid lines with thick solid lines representing the average Green's function in the non-crossing approximation, $\G(\eta_i,z_i)$ (defined diagrammatically in the third line of Fig.~\req{fig-dyson}, and given by \req{Gscba} as we found in the previous section), and summing over all  non-crossing pairings with \emph{every} pairing connecting the thick arrow lines on top and bottom (and not each to itself). This procedure yields a sum over all ladder diagrams with different number of rungs, as shown in the third line  of Fig.~\ref{fig-ladder}.

\begin{figure}[!t]\hspace{-.2cm}
\includegraphics[height=2in,angle=0]{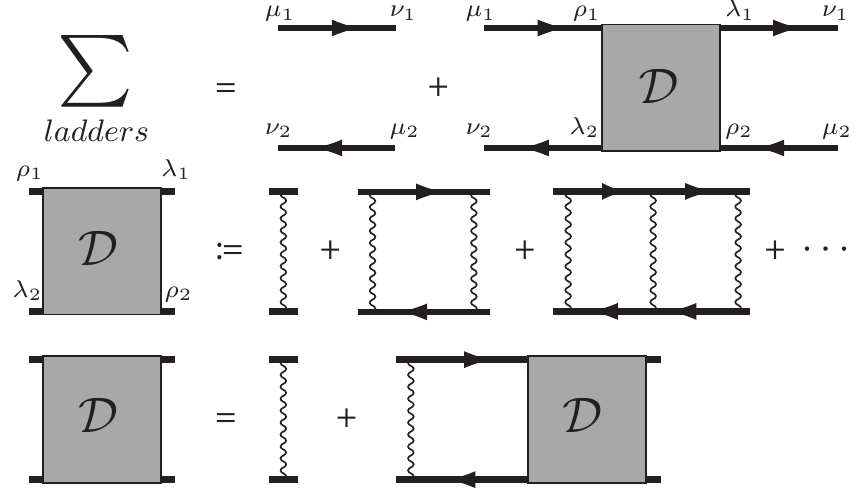}
\caption{
The first row is the diagrammatic representation of \reqs{F0Fc}--\myref{Fcdef}. In the last term,  $\rho$'s and $\lambda$'s are summed over. It shows the sum of all ladder diagram contributing to \req{Fdef} (\ie\, the last line of Fig.~\ref{fig-ladder}) in terms of $\D$, which is defined in the second row. The first term on the right side of the first row equation (the ladder with zero rungs) is the disconnected average \req{discF}; it corresponds to taking the average of each Green's function in \req{Fdef} separately and then multiplying.  The last row shows an iterative form of the  equation in the second row, which can be solved to give the expression \reqs{diffuson} and \myref{minidiffus} for $\D$.}
\label{fig-diffus}
\end{figure}

As shown in the first row of Fig.~\ref{fig-diffus}, the sum of all ladder diagrams can be written as a sum
\be\label{F0Fc}
F = F^0 + F^{\scriptscriptstyle D},
\ee
where 
\be\label{discF}
F^0_{\mu_1\nu_2;\mu_2\nu_1}(1;2)\defin  
 \G_{\mu_1\nu_1}(1) \G_{\mu_2\nu_2}(2), 
\ee
is the disconnected average of the two Green's functions, and $F^D_{\mu_1\nu_2;\mu_2\nu_1}(1;2)$ is the sum of ladder diagrams in which the two Green's function are connected by at least one wavy line. 
The latter can be written in the form
\bea\label{Fcdef}
&& F^ {\scriptscriptstyle D}_{\mu_1\nu_2;\mu_2\nu_1}(1;2) \defin  
\\
&& \quad \G_{\lambda_2\nu_2}(2)\G_{\mu_1\rho_1}(1) 
\D_{\rho_1\lambda_2;\rho_2\lambda_1}(1;2)   
 \G_{\lambda_1\nu_1}(1) \G_{\mu_2\rho_2}(2), \nonumber
\eea
where all repeated indices are summed over, and the ``diffuson", $\D$, is given by the sum of all diagrams in the second row of Fig.~\ref{fig-diffus}.

\begin{figure}[!t]\hspace{0cm}
\includegraphics[height=1.3in,angle=0]{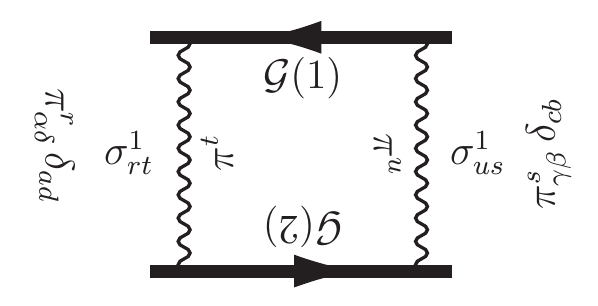}
\vspace{-.1cm}
\caption{
The contribution to $\D^{\alpha\delta;\gamma\beta}_{ad;cb}(1;2)$ from the second term in the series shown in the second row of Fig.~\ref{fig-diffus}, in more detail. The covariance of $\J$ in the form \req{jpair3} is used to write this expression in a more manageable form. The repeated indices, $r,t,u,s$, are summed over 1 and 2. The matrices inside the loop multiply each other in cyclic order, giving rise to the trace $\Tr(\G(2)\pi^t\G(1)\pi^u)$. The whole diagram gives $\frac{1}{N} \sum_{rs} (\pi^r\otimes\one)^{\alpha\delta}_{ad} \,\lbr  \sigma^1\Pi^ {\scriptscriptstyle D} \sigma^1 \rbr_{rs}\, (\pi^s\otimes\one)^{\gamma\beta}_{cb}$ where  the ``polarization matrix" $\Pi^{\scriptscriptstyle D}_{tu}$ was defined in \req{polarmat}.}
\label{fig-polar}
\end{figure}

To calculate $\D$, 
it helps to first rewrite \req{jpair2} as
\be\label{jpair3}
\Javg{ \J^{\alpha\beta}_{ab} \J^{\gamma\delta}_{cd} } = \frac{1}{N} \sum_{r,s=1}^2 (\pi^r_
{\alpha\delta}\delta_{ad})\, \sigma^1_{rs} \, (\pi^{s}_{\gamma\beta}\delta_{cb}),
\ee
where $\sigma^1  = \begin{pmatrix}
     0 &  1  \\
     1 &  0 
\end{pmatrix} 
$ is the first Pauli matrix. 
This helps us because in the expansion of Fig.~\ref{fig-diffus}, the two factors in \req{jpair2} involving $\pi^{r}$  and $\pi^{s}$ decouple and get absorbed in adjacent loops, or contribute to form factors in the left or right ends of the ladder diagrams. This is demonstrated in Fig.~\ref{fig-polar} for the second term in the series expansion of $\D$ shown in the second line of Fig.~\ref{fig-diffus}.
 Extending this similarly to all the terms in that expansion,
 we obtain 
\bes\label{diffuson}
& \D_{\mu\rho;\lambda\nu}(1;2)  =  \D^{\alpha\delta;\gamma\beta}_{ad;cb}(1;2) 
\\
&= \frac{1}{N} \sum_{r,s =1}^2 (\pi^r_{\alpha\delta}\delta_{ad})\,\, D_{rs}(1;2) \,\,(\pi^{s}_{\gamma\beta}\delta_{cb}),
\end{split}
\ee
where $\mu=(\alpha,a)$, $\nu=(\beta,b)$, $\lambda = (\gamma,c)$, $\rho = (\delta,d)$ and we defined the $2\times 2$ matrices
\be\label{minidiffusdef}
D(1;2) \defin   \sigma^1 + \sigma^1\Pi^{\scriptscriptstyle D}\sigma^1  + \cdots =  \sigma^1 \sum_{n=0}^\infty \lpr \Pi^{\scriptscriptstyle D}\sigma^1\rpr^n, 
\ee
and the ``polarization matrix"  for the diffuson
\be\label{polarmat}
\Pi^{\scriptscriptstyle D}_{rs}(1;2) \defin  \tr\!(\pi^r\G(1)\pi^s\G(2)) =  \tr\!(\G^{rs}(1)\G^{sr}(2)). 
\ee
Here, as before, with the trace performed over the $2N$-dimensional space,  and we used \req{pidef} to write the last form of $\Pi^{\scriptscriptstyle D}$. 
Summing the geometric series in \req{minidiffusdef} we obtain
\be\label{minidiffus}
D(1;2) =  \sigma^1 \lpr \one_{{\scriptscriptstyle 2\times 2}} - \Pi^{\scriptscriptstyle D}(1;2)\sigma^1\rpr^{-1}.
\ee
The $2\times 2$ matrix inversion yields
\bea\label{diffusmat}
&& D(1;2) =   
\\
&& \quad \frac{1}{(1-\Pi^{\scriptscriptstyle D}_{12})(1-\Pi^{\scriptscriptstyle D}_{21}) -\Pi^{\scriptscriptstyle D}_{11}\Pi^{\scriptscriptstyle D}_{22} }
\begin{pmatrix}
\Pi^{\scriptscriptstyle D}_{22}      &   1-\Pi^{\scriptscriptstyle D}_{12} \\
1-\Pi^{\scriptscriptstyle D}_{21}      &  \Pi^{\scriptscriptstyle D}_{11}
\end{pmatrix},
\nonumber
\eea
where all $\Pi^{\scriptscriptstyle D}$'s have arguments $(1;2) = (\eta_1,z_1;\eta_2,z_2)$ which were suppressed for clarity.

Going back to \req{FBF}, we can also break up $\F(z_1,z_2;B,C) $  into a disconnected part and a connected part mirroring the decomposition \reqs{F0Fc}--\myref{Fcdef}:
\be\label{FBcdc}
\F(z_1,z_2;B,C) = \F^0(z_1,z_2;B,C) + \Delta \F(z_1,z_2;B,C), 
\ee 
where $\F^0(z_1,z_2;B,C)$ and $\Delta \F(z_1,z_2;B,C)$ are defined as in \req{FBF}, but with $F_{\mu_1\nu_2;\mu_2\nu_1}$ on the right side replaced by $F_{\mu_1\nu_2;\mu_2\nu_1}^0$ and $F_{\mu_1\nu_2;\mu_2\nu_1}^{\scriptscriptstyle D}$, respectively. Using \reqs{discF}--\myref{Fcdef} and \myref{diffuson},  we then obtain
\bea\label{FBdiscon}
\F^0(z_1,z_2;B,C) \areq \Tr\!\lpr \Bb \G(0,z_1) \rhob \G(0,z_2)  \rpr \\
\areq \Tr\!\!\lpr \tB \G^{21}(0,z_1) \tC \G^{12}(0,z_2)  \rpr\!, 
\nonumber
\eea
and 
\bea\label{FBcon}
&& \Delta \F(z_1,z_2;B,C) = \frac{1}{N}\sum_{r,s} \Tr\!\!\lpr \tB \G^{2r}(0,z_1) \G^{r2}(0,z_2)  \rpr
\times
\nl
&&  \qquad D_{rs}(0,z_1;0,z_2)\,\, \Tr\!\!\lpr \G^{s1}(0,z_1)\tC \G^{1s}(0,z_2)  \rpr\!, 
\qquad\quad
\eea
where  $r$ and $s$ are summed over $\lcr 1,2\rcr $.

According to \req{avgformul} we are interested in $z_i = \gamma +i \omega_i$ ($i$ = 1,2) for  arbitrary real $\omega_i$. As we mentioned before \req{invlaplace}, these trace a vertical line in the complex plane that is entirely to the right of the support of the average eigenvalue density, $\rho(z)$, of $A$, \ie\, they are in the region where the  the valid solution of \req{geq} is the trivial $g(0,z)=0$. In this case, we have \req{sol1G}, and for $\eta\to i0^+$,  from \req{G0z} (replacing $\A$ with $M$, corresponding to $J=0$) we have 
\be\label{G0zi}
\G(0,z_i)  =  - \begin{pmatrix}
 0     &   M_{z_i}^{-\dagger}\\
M_{z_i}^{-1}     &  0
\end{pmatrix}.
\ee
Using this in \reqs{FBdiscon}--\myref{FBcon} we obtain 
\bea\label{FBdiscon11}
\F^0(z_1,z_2;B,C)  \areq \Tr\!\!\lpr  \tB M_{z_1}^{-1} \tC M_{z_2}^{-\dagger}  \rpr, 
\eea
and
\bea\label{FBcon11}
&& \Delta \F(z_1,z_2;B,C) =   \tr\!\!\lpr \tB \G^{21}(0,z_1) \G^{12}(0,z_2)  \rpr  \times
\nl 
&& \qquad D_{12}(0,z_1;0,z_2)\,\, \Tr\!\!\lpr \G^{21}(0,z_1)\tC \G^{12}(0,z_2)  \rpr\!. 
\qquad\quad
\eea
Using the definitions \req{Mzdef} and \reqs{Btild}--\myref{Ctild} we can simplify \req {FBdiscon11} to
\be
\F^0(z_1,z_2;B,C) = \Tr\!\!\lpr  B \frac{1}{z_1-M} C \frac{1}{\zbar_2 - M^\dagger}  \rpr.
\label{FBdiscon1}
\ee
From \reqs{polarmat} and \myref{G0zi}  we  see that (for $z_i$ of interest and for $\eta_i$ going to zero) $\Pi^{rr} = 0$ and $\Pi^{12} = \Pi^{21}$, and from \req{diffusmat} we obtain
\bea
 D_{12}(0,z_1;0,z_2) \areq \frac{1}{1-\Pi^{\scriptscriptstyle D}_{21}(0,z_1;0,z_2)} 
 \\
 \areq \frac{1}{1- \tr\!(\G^{21}(0,z_1)\G^{12}(0,z_2))}.
\nonumber
\eea
Substituting this in \req{FBcon11} and using \req{G0zi} once again, we finally obtain 
\be
 \Delta \F(z_1,z_2;B,C) =
 \frac{\tr\!\!\lpr \tB M_{z_1}^{-1} M_{z_2}^{-\dagger}    \rpr    \Tr\!\!\lpr M_{z_1}^{-1}\tC M_{z_2}^{-\dagger}   \rpr}{1- \tr\!\!\lpr M_{z_1}^{-1}  M_{z_2}^{-\dagger}    \rpr},\qquad \,\,\,\nonumber
\ee
and after simplification using \reqs{Mzdef} and \myref{Btild}--\myref{Ctild}, 
\bea\label{FBcon1}
&& \Delta \F(z_1,z_2;B,C) = \qquad \qquad
\\
 && \qquad = \frac{\tr\!\!\lpr B \frac{1}{z_1-M} L L^\dagger \frac{1}{\zbar_2 - M^\dagger}    \rpr    \Tr\!\!\lpr R^\dagger R \frac{1}{z_1-M} C \frac{1}{\zbar_2 - M^\dagger}   \rpr}{1- \tr\!\!\lpr R^\dagger R \frac{1}{z_1-M} L L^\dagger \frac{1}{\zbar_2 - M^\dagger}     \rpr}.
\nonumber
\eea

The general formulae of Sec.~\ref {sec:genformdynamics} readily follow.  
Equations \myref{res-avgnormsq}--\myref{res-FBcon1}, with $C^{\vi}$ replaced by $\vx_0\vx_0\trans$,  for the case of response to an impulse input follow from \reqs{avgformul}, \myref{FBcdc}, \myref{FBdiscon1} and \myref{FBcon1}, respectively. 
Equations \myref{Cnoise}--\myref {delCnoise} (with $\C_0^{\vx}$ and $\Delta \C^{\vx}$ defined in \reqs{res-FBdiscon1}--\myref{res-FBcon1}) for the power spectrum of the response to a temporally white noisy input, are similarly obtained 
from \req{CxFB} by using \reqs{FBcdc}, \myref{FBdiscon1} and \req{FBcon1}, after  setting $B = \ve_j\ve_i\trans$, $C=C^{\vi}$ and $z_1 = z_2 = \gamma + i\omega$ (with   the traces involving $B= \ve_j\ve_i\trans$ turned into matrices  in \reqs{Cnoise}--\myref {delCnoise}, using $\Tr(\ve_j\ve_i\trans X) = X_{ij}$). The result \req{res-powerspec} for the  steady state response to a sinusoidal input was already derived in \req{powerspec2}. 

We see that according to \reqs{avgformul} and \myref{FBcdc}
\be\label{Bavg}
\Javg{\vx(t)\trans B \vx(t)} = \Big [\vx(t)\trans B \vx(t)\Big ]_{J=0}  + \Delta f_B(t),
\ee
where the two terms on the right hand side are obtained by replacing $\F(\cdot,\cdot;B)$ in \req{avgformul} with \req{FBdiscon1} and \req{FBcon1}, respectively. The integrals over $\omega_1$ and $\omega_2$ decouple for the first term yielding the expected result for $J=0$,
\bea
\Big [\vx(t)\trans B \vx(t)\Big ]_{J=0} \areq e^{-2\gamma t}\,\Tr\!\!\lpr B e^{tM} \vx_0\vx_0\trans \,e^{tM^\dagger}  \rpr,
\nonumber
\\
\areq \vx_0\trans \,e^{t (-\gamma+ M) ^\dagger}  B e^{t (-\gamma+ M)} \vx_0.\qquad
\label{Jzero}
\eea
Unlike the $J=0$ contribution,  it is not  possible to perform the double Fourier transform,   \req{avgformul}, needed for obtaining $\Delta f_B(t)$  for arbitrary $M$, $L$ and $R$. {In the next section, we will analytically calculate this for some special examples of $M$, with $L$ and $R$ proportional to the identity matrix (\ie\ for iid quenched randomness).}
%
%

\section{Calculations for specific examples of $M$}
\label{sec-example}

In this section we give the detailed calculations of the explicit expressions for the spectral density \req{rho-res}, the power spectrum \req{res-powerspec}, and the average squared norm \reqs{res-avgnormsq} and \myref{res-F1}, for the specific examples of $M$, $L$ and $R$ presented in Sec.~\ref{sec-examples}. 

In the examples worked out in the subsections \ref{subsec-bidiagcalcs} and \ref{sec-rajanabbott}, both $R$ and $L$ are proportional to the identity matrix;  we take $L= \one$ and $R = \Jstd \one$. 
Furthermore, for such examples we will do the calculations by choosing the unit of time such that $\Jstd = 1$ (notice that given \req{ode}, the elements of $A$ and $M$ have dimensions of frequency); then at the end of our calculations using the replacements $t\to t \Jstd$, $z \to z/\Jstd$, $\gamma\to \gamma/\Jstd$,  $M\to M/\Jstd$, and $\rho \to \Jstd^2 \rho$ (with the latter applying to both the eigenvalue density and the power spectral density), we obtain the result for general $\Jstd$. 
The eigenvalue density and the norm squared $\| \vx\|^2$ are invariant with respect to unitary transforms, and, for $L$ and $R$ proportional to the identity, so is the distribution of the random part of $A$, \req{jmeas}. Thus by effecting a unitary transform $M\to U^\dagger M U$, we can assume $M$ is already in its Schur form \req{Tex2} without loss of generality.

%

\subsection{Single feedforward chain of length $N$: $M_{ij}= w\, \delta_{i+1,j} - \gamma \delta_{ij}$}
\label{subsec-bidiagcalcs}

We start with the example in Sec.~\ref{sec-res-longffwd}, where $M$ is
\be\label{Mbidiag}
M = T =  \begin{pmatrix}
     0 & w & 0 & \cdots   \\
     0 & 0 & w & \cdots   \\
     \vdots & \vdots & \vdots & \ddots   \\     
\end{pmatrix}
\ee
or  $M_{ij}= w\, \delta_{i+1,j}$. First we calculate the eigenvalue density. 
According to \reqs{rho-res}--\myref{geqn}, in order to calculate the spectral density, we need to calculate first the inverse of $M_z M_z^\dagger  + g^2 = (z-M)(z-M)^\dagger  + g^2$ (remember that we have set $\sigma = 1$,  as we explained in the beginning of the section). To this end, notice that $K_{ij}\equiv \lbr (z-M)  (z-M)^\dagger  \rbr_{ij} = Q_{ij} - |w|^2 \delta_{iN}\delta_{jN}$ where 
\be\label{Q}
Q_{ij} \defin  (|z|^2 + |w|^2)\delta_{ij}  - w \zbar \delta_{i+1,j} - \bar{w}z\delta_{i,j+1}.
\ee
As the difference  $ (z-M) (z-M)^\dagger  - Q = -|w|^2 \ve_{_N} \ve_{_N}\trans$ is single rank, we  can use the Woodbury formula for matrix inversion to write 
\bea\label{woodbury}
&&\frac{1}{K  + g^2} =   \frac{1}{Q + g^2} \, + 
\\
&& \qquad  \frac{1}{Q + g^2} \ve_{_N} \ve_{_N}\trans \frac{1}{Q + g^2}
\lpr  \frac{1}{|w|^{-2}- \ve_{_N}\trans (Q + g^2)^{-1} \ve_{_N}}  \rpr,
\nonumber
\eea
where $\ve_{_N}\trans = (0,\ldots,0,1)$.
~(The only conditions for the validity of \req{woodbury} is that
the factor in parenthesis is not singular, \ie\, $  \ve_{_N}\trans (Q + g^2)^{-1} \ve_{_N} \neq |w|^{-2}$; we will consider the validity of this condition below.)
Since $Q$ is Toeplitz and Hermitian, it can be diagonalized easily. 
Using standard methods, 
we find that the eigenvalues and eigenvectors of $Q$, satisfying $Q\vv_n = \lambda_n \vv_n$, are given by 
\bea\label{eigsQ}
\lambda_n \areq \Big | |z| - |w| e^{i \phi_n} \Big |^2, 
\qquad\qquad \phi_n \defin  \frac{\pi n}{N+1}
\\
v^j_n \areq \sqrt{\frac{2}{N+1}} \lpr \frac{\bar{w} z}{w \zbar}\rpr^{{j}/{2}} \sin \phi_n j
\label{eigvecvj}
\eea
for $n=1,\ldots,N$.
The eigenvectors are orthonormal $\vv_n^\dagger \vv_m = \delta_{nm}$, 
and we have the spectral representation 
\be\label{Qg2}
\frac{1}{Q + g^2}= \sum_{n=1}^N \vv_n \frac{1}{\lambda_n + g^2} \vv_n^\dagger.
\ee
~Using  \reqs{eigsQ}--\myref{Qg2} in  \req{woodbury} we obtain 
\be\label{trlhs}
\tr \frac{1}{K  + g^2} = I_1(g,z)  +  \frac{1}{N}  \frac{- \frac{\partial I_2(g,z)}{\partial g^2}}{|w|^{-2} - I_2(g,z)},  
\ee
where we defined 
\bea\label{I1def}
&& I_1(g,z) \defin  \tr  \frac{1}{Q + g^2} = \frac{1}{N} \sum_{n=1}^N  \frac{1}{\lambda_n + g^2} 
\\
&& I_2(g,z) \defin  \ve_{_N}\trans \frac{1}{Q + g^2} \ve_{_N} =  \frac{1}{N+1} \sum_{n=1}^N  \frac{2 \sin^2 \phi_n}{\lambda_n + g^2}. \qquad
\label{I2def}
\eea
In writing the numerator of the last term in \req{trlhs}, we used \reqs {eigvecvj}--\myref {Qg2} to write $\Tr\!\! \lpr \frac{1}{Q + g^2} \ve_{_N} \ve_{_N}\trans \frac{1}{Q + g^2}\rpr = \| \frac{1}{Q + g^2} \ve_{_N}\|^2 = \frac{1}{N+1} \sum_{n=1}^N  \frac{2 \sin^2 \phi_n}{(\lambda_n + g^2)^2} = - \frac{\partial I_2(g,z)}{\partial g^2}$.
In the $N\to \infty$ limit,  the sums in \reqs{I1def}--\myref{I2def} can be approximated  
by the integrals
\bea\label{I1int}
&& I_1(g,z) = \int_0^{2\pi}  \frac{1}{\big | |z| - |w| e^{i \phi} \big |^2 + g^2} \frac{d\phi}{2\pi},
\\
&& I_2(g,z) = \int_0^{2\pi}  \frac{2 \sin^2 \phi}{\big | |z| - |w| e^{i \phi} \big |^2 + g^2} \frac{d\phi}{2\pi}. \qquad
\label{I2int}
\eea
Some elementary contour integration then yields
\bea\label{I1fin}
&& I_1(g,z) = \lbr { (|z|^2 + |w|^2 + g^2)^2 - 4 |w|^2|z|^2}\rbr^{-1/2}, \qquad
\\
&& I_2(g,z) = \frac{{|z|^2 + |w|^2 + g^2} - I_1(g,z)^{-1}}{2|w|^2|z|^2}.  \qquad
\label{I2fin}
\eea
In particular, we see that $I_2(0,z) = \min(|w|^{-2},|z|^{-2})$, so that the  condition for the validity of \req{woodbury}  would be violated  for $|z|<|w|$, if $g$ turns out to be zero.
However, note that $I_2(g,z)$ is a decreasing function of $g^2$, so for finite $g^2>0$, the denominator in \req{trlhs} is always positive (as is its numerator, for the same reason). Thus if we follow the correct procedure of \req{Kdef1}--\myref{support-res-K}, taking the $N\to\infty$ limit before sending $g^2$ to zero, we are justified in using \reqs{woodbury} and \myref{trlhs}.
 Furthermore,  for $g^2>0$  the second term in \req{trlhs} is $O(N^{-1})$, and should be neglected. 
Solving \req{geqn} (with left hand side correctly interpreted as \req{Kdef1}), which now takes the form $I_1(g,z) = 1$, yields 
\be\label{gbidiag}
g(z)^2 = -|z|^2 - |w|^2 + \sqrt{4 |w|^2|z|^2 - 1 }.
\ee
This is positive
 if and only if 
\be\label{suppbidiag}
\sqrt{|w|^2 - 1} \leq |z| \leq \sqrt{|w|^2 + 1},
\ee
which after the proper rescaling yields \req{suppbidiag-res} for general $\sigma$. 
Note that  \req{suppbidiag}  is precisely the region given by \req{support-res-K},   which in the present case reads $I_1(0,z)  \geq 1$. It is instructive to compare this result with what we would obtain by naively using \req{support-res}, \ie\ $\tr(K^{-1}) \geq 1$, wherein $g$ is set to zero before taking the $N\to\infty$ limit; as we now show, that  only yields the right inequality in \req{suppbidiag}. 
To see this, first note that for $|w|>|z|$, we can use \req{trlhs} even for $g^2 = 0$ (since the denominator of the last term does not vanish), which yields $\tr\!\!\lbr {(M_z M_z^\dagger)^{-1}} \rbr = \tr(K^{-1})   = I_1(0,z) + o(1)$, and by \req{support-res}, the right inequality in \req{suppbidiag}. For $|z|<|w|$, however, we cannot set $g=0$ in \req{trlhs}. In fact,  when $|z|<|w|$, the matrix $z-M$ has an exponentially small singular value; 
to see this, note that the vector $\vu$ with components $u_i = (\frac{z}{w})^{i-1}$ satisfies $(z-M)\vu  = w (\frac{z}{w})^N \ve_N$,  so that $\|(z-M)\vu\| = |w| |\frac{z}{w}|^N$, and since  
$s_{\min}(z-M) \leq  \frac{\| (z-M)\vu\|}{ \|\vu\|}$ and $\|\vu\|\geq1$,  it follows that $s_{\min}(z-M) \leq |w| |\frac{z}{w} |^N$, which is $O(e^{-c N})$  for $|z|<|w|$. 
For large enough $N$, this singular value \emph{alone} suffices to make \req{support-res-3} (equivalent to \req{support-res}) hold for \emph{any} $|z|<|w|$, as $\frac{1}{N}s_{\min}(z)^{-2}$ diverges  despite its $\frac{1}{N}$ prefactor.

Let us now calculate the eigenvalue density in the annulus \req{suppbidiag}.
In order to use \req{rho-res}, we will first calculate 
\be\label{trGbidiag}
\tr \frac{\zbar - M^\dagger}{K + g(z)^2} = \zbar - \tr \frac{M^\dagger}{K + g(z)^2}
\ee
where we used \req{geqn} to write the last expression.
To obtain $ \tr \frac{M^\dagger}{K + g(z)^2}$, we will again use \req{woodbury}. In the region \req{suppbidiag}, the contribution of the second term in \req{woodbury} is again suppressed by $1/N$, and from \req{Mbidiag} and \myref{Qg2} we have $\tr M^\dagger \lbr{Q + g^2}\rbr^{-1}  = \bar{w} \frac{1}{N} \sum_{n=1}^N \lpr \sum_{j=1}^{N-1} v_n^j \bar{v}_n^{j+1} \rpr  \frac{1}{\lambda_n+g^2}$. A straightforward calculation using \req{eigvecvj} (and the orthonormality of $\vv_n$) yields $\sum_{j=1}^{N-1} v_n^j \bar{v}_n^{j+1}  = \lpr\frac{w\zbar}{\bar{w}z} \rpr^{1/2} \cos\phi_n$. Using this and approximating the sum over $n$ with an integral, we obtain 
\bea
&&  \tr\!\!\lbr \frac{M^\dagger}{Q+g(z)^2} \rbr   \approx
 \frac{ |w| |z|}{z} \int_0^{2\pi}  \frac{\cos\phi}{\big | |z| - |w| e^{i \phi} \big |^2 + g(z)^2} \frac{d\phi}{2\pi},
 \nl
 &&= \frac{1}{2z} \lbr {(|z|^2 + |w|^2 + g(z)^2) I_1(g(z),z) - 1} \rbr.
\label{tereysak}
\eea
 Using \req{tereysak}
  with $I_1(g(z),z) = 1$ (true in the region \req{suppbidiag}),  differentiating \req {trGbidiag}  with respect to $\zbar$, and substituting in \req{rho-res}, we finally obtain 
\be\label{rho-bidiag}
\rho(z) =  \frac{1}{\pi}\lbr 1 - \frac{|w|^2}{\sqrt{4 |w|^2|z|^2 + 1  } } \rbr,
\ee
for $z$ in the region \req{suppbidiag}. After the proper rescaling this yields \req{rho-bidiag-res}.

We now turn to the  calculation of $\Javg{\|\vx(t)\|^2}$, using \myref{res-F1}. 
To calculate the trace in the denominator of \req{res-F1},  first note that for \req{Mbidiag} the expansion 
$(z-M)^{-1} =  \sum_{n=0}^{N-1}  \frac{M^n}{z^{n+1}}$ terminates and is exact, yielding
\be\label{invbidiag}
\lbr \frac{1}{z-M} \rbr_{i,j} = \frac{1}{z}\lpr \frac{w}{z}\rpr^{j-i}, 
\ee
for $j\geq i$, and zero otherwise. 
Turning the sums in the trace into a sum over the nonzero diagonals of \req{invbidiag} we obtain
\be\label{kosde}
\tr\!\!\lpr  \frac{1}{\zbar_2 - M^\dagger} \frac{1}{z_1-M}    \rpr = \frac{1}{\zbar_2 z_1} \sum_{n=0}^{N-1} (1-\frac{n}{N}) q^n
\ee
 where $q \defin  |w|^2/(\zbar_2 z_1)$ and $z_i = \gamma  + i\omega_i$. The condition of stability of \req{ode} requires  the entire spectrum of $-\gamma\one + A = -\gamma \one + M+ J$ to be to the left of the imaginary axis. By \req{suppbidiag}, this requires  $\gamma > \sqrt{|w|^2 + 1} > |w|$. It follows that $|q|<1$, and therefore the geometric series \req{kosde} converges as $N\to\infty$. Summing the series and retaining terms of leading order as $N\to\infty$, we obtain
\bea\label{koskaj}
\tr\!\!\lpr  \frac{1}{\zbar_2 - M^\dagger} \frac{1}{z_1-M}    \rpr \areq \frac{1}{\zbar_2 z_1 - |w|^2}.
\eea
If we set the initial condition $\vx_0$ in \reqs{res-F1} (or the input amplitude $\vi_0$ in \req{res-powerspec}) to  $\ve_{_N}  = (0,\cdots,0,1)\trans$, and use \req{invbidiag}, we find that the numerator in \req{res-F1} is also given by the right hand side of \req{koskaj}. Using this and \reqs{koskaj}, we obtain
\be\label{doori}
F(z_1,z_2)  = \frac{1}{\zbar_2 z_1 - |w|^2 - 1},
\ee
for the integrand of \req{res-F1} which we denoted by $F(z_1,z_2)$, with $z_i = \gamma + i\omega_i$, ($i=1,2$).
By comparing the integrand of \req{res-F1} with  \req {res-powerspecBLR1}, we see that to obtain the total power spectrum for the input amplitude $\vi_0 = {\rm I}_0 (0,\cdots,0,1)\trans$, we  need to multiply  \req{doori} by ${\rm I}_0^2 = \|\vi_0\|^2$, and substitute $z_{1} = z_2 = \gamma + i\omega$.
With the proper rescaling, this yields \req{res-bidiag-powerspec} for general $\Jstd$. 
To obtain the formula for $\Javg{\|\vx(t) \|^2}$, we substitute \req{doori} with $z_i = \gamma + i\omega_i$ for the integrand of \req{res-F1}. Changing the integration variables by $\omega_1 = \Omega +\omega/2$ and $\omega_2 = \Omega -\omega/2$, we obtain
\bea
\Javg{\| \vx(t)\|^2}
\areq 
\!\int\!   \frac{d\omega}{2\pi}e^{it \omega}\!\! \int\! \frac{d\Omega}{2\pi} \frac{1}{\Omega^2 + (\gamma + i\omega/2)^2  - |w|^2 - 1},
\nl
\areq 
   \frac{1}{2} \int\! \frac{d\omega}{2\pi}\! \frac{e^{it \omega}}{\sqrt{ (\gamma + i\omega/2)^2  - |w|^2 - 1}}.
\eea
Finally consulting a table of Laplace transforms \cite{AbramowitzStegun}, we obtain
\be
\Javg{\| \vx(t)\|^2}  = e^{-2\gamma t} I_0(2 t \sqrt{|w|^2 + 1} ), \qquad (t\geq 0)
\ee
where $I_0(x)$ is the $0$-th modified Bessel function. Implementing the rescalings $t\to t \Jstd$, $\gamma \to \gamma/\Jstd$ and $w\to w/\Jstd$, we obtain \req {avgnormsq-bidiag-res}.

\subsection{$N/2$ feedforward chains of length $2$}
\label{sec-rajanabbott}

Here we carry out the explicit calculations for the example of Sec.~\ref{sec-res-rajanabbott} where $M$ is given by \req{Tex2} (without loss of generality, we assume $M$ has its Schur form), using  formulae \myref{rho-res}--\myref{geqn} for the spectral density and \req{res-F1} for $\Javg{\| \vx(t)\|^2}$. 
First we will calculate the eigenvalue density.
From \req{Tex2}, $K\defin  M_z M_z^\dagger = (z-M)(z-M)^\dagger$ (we are setting $L = R =\one$ in \req{Mzdef}; see the comments at the beginning of this section) is a block-diagonal matrix with $2\times 2$ diagonal blocks, with the $b$-th block ($b=1,\ldots,N/2$) given by 
\be\label{Kblocks}
\begin{pmatrix}
     z & \,\,\, -w_b    \\
     0 & z
\end{pmatrix}
\begin{pmatrix}
     \zbar & \,\,\, 0   \\
     -\bar{w}_b & \zbar
\end{pmatrix}
=
\begin{pmatrix}
     |z|^2 + |w_b|^2 & \,\,\, -w_b \zbar   \\
     -\bar{w}_b z & |z|^2
\end{pmatrix},
\ee
 where $w_b$ is the corresponding  Schur weight in \req{Tex2}. Likewise, $(K + g^2)^{-1}$ whose trace appears in \reqs{geqn}   is given by a block-diagonal matrix with diagonal blocks $\frac{1}{(|z|^2 + g^2)^2 + |w_b|^2 g^2} \begin{pmatrix}
     |z|^2 + g^2 & w_b \zbar   \\
     \bar{w}_b z & |z|^2 + g^2+ |w_b|^2 
\end{pmatrix}$. Taking the normalized trace we thus obtain
\be\label{Kg2}
\tr (K + g^2)^{-1} = \left\langle \frac{|z|^2 + g^2 + \frac{1}{2} |w_b|^2} {(|z|^2 + g^2)^2 + |w_b|^2 g^2} \right\rangle_b,
\ee
where $\langle \cdot \rangle_b$ means averaging over the $N/2$ blocks, \ie\ $\langle f(w_b)\rangle_b \defin  \frac{1}{N/2} \sum_{b=1}^{N/2} f(w_b)$.

Let us first calculate the support boundary of $\rho(z)$. As discussed in Sec.~\ref {subsec-specdensity}, when  (for $|z|\neq 0$) all singular values of $M_z = z-M$ are bounded from below as $N\to\infty$, the support is correctly given by \req{support-res} (we will discuss cases in which some $s_i(z)$ are $o(1)$ further below). Setting $g=0$ in  \req{Kg2}, and substituting in \req{support-res}, this yields
\be\label{Kinv}
 1 \leq \tr K^{-1} =  |z|^{-2} + \mu^2 |z|^{-4},
 \ee
where we defined $\mu^2 = \frac{1}{2} \langle |w_b|^2 \rangle_b = \tr(M^\dagger M)$. 
It follows that the support is the disk  $|z|\leq r_0$, where
\be\label{r0}
r_0^2 = \frac{1}{2}  + \sqrt{ \frac{1}{4} + {\mu^2}}.
\ee
 The replacements $\mu \to \mu/\Jstd$ and $r_0 \to r_0/\Jstd$ then yield \req{spectralrad-rajan-res}. 

From \reqs{geqn} and \myref{Kg2}, within the support, $g^2(z)$ is found by solving the equation
\be\label{Kg22}
\tr\! \frac{1}{K + g^2} = \left\langle \frac{|z|^2 + g^2 + \frac{1}{2} |w_b|^2} {(|z|^2 + g^2)^2 + |w_b|^2 g^2} \right\rangle_b =  1,
\ee
while for $|z|>r_0$ we have $g(z)=0$.  
 It is clear from \req{Kg22} that $g^2(z)$ depends on $z$ and $\zbar$ only through $|z|\defin  r$. 
From \req{rho-res}, within its support the eigenvalue density is given by
  \bea
  \pi\rho(z) \areq \frac{\partial}{\partial \zbar} \tr\!\! \lbr M_z^\dagger(K + g^2)^{-1}\rbr  
\nl
  \areq 1 -\frac{\partial}{\partial \zbar}  \tr\!\! \lbr M^\dagger(K + g^2)^{-1} \rbr,
  \label{rhoKg}
  \eea
   where we are now using the short-hand $g^2 = g^2(|z|)$ (the solution of \req{Kg22}), and in writing the second line we used $M_z^\dagger = \zbar - M^\dagger$ and \req{Kg22}. 
   From \reqs{Tex2} and \myref{Kg2} we see that $M^\dagger(K + g^2)^{-1}$ has the same block-diagonal structure as \req{Tex2}, and a  short calculation shows that $\tr\!\! \lbr M^\dagger(K + g^2)^{-1} \rbr  = \zbar I_3(|z|)$, where we defined
\be\label{I3}
I_3 (r) \defin   \left\langle \frac{\frac{1}{2} |w_b|^2} {(r^2 + g(r)^2)^2 + |w_b|^2 g(r)^2}  \right\rangle_b. 
\ee
$I_3(r)$ is manifestly positive (assuming some $w_b$ are nonzero), while when $g^2>0$, from \req{Kg22} we have $I_3(r) \leq 1$, and thus
\be\label{I3bound}
0< I_3(r) \leq 1.
\ee
Replacing this in \req {rhoKg}, and using $\zbar\frac{\partial f(|z|)}{\partial \zbar} = 2r \frac{\partial f(r)}{\partial r}\!\big\vert_{\scriptscriptstyle r=|z|} $, 
 we obtain
\be\label {1rhoN/2}
\pi\rho(z) = \frac{1}{2r} \frac{\partial \, \,}{\partial r} \lbr r^2 -  r^2 I_3 (r) \rbr,
 \ee
 for $r = |z|\leq r_0$, and zero otherwise; the spectral density is rotationally symmetric and depends only on $r= |z|$. The advantage of writing the density as a complete derivative is that it can be immediately integrated to yield $n_{_<}(r)$, the proportion of eigenvalues with modulus smaller than some radius $r$. We have $n_{_<}(r) = 2\pi \int_0^r \rho(r') r'dr' $, which upon substitution of \req{1rhoN/2}, yields
 \be\label{nofr}
 n_{_<}(r) = r^2 \lpr 1 - I_3 (r) \rpr \qquad (r\leq r_0).
 \ee
Likewise, we define  $n_{_>}(r)\equiv 1 - n_{_<}(r)$ to be the proportion of eigenvalues with modulus larger than $r$. From these definitions we have 
\be\label {rhoN/2}
\rho(r) = \frac{1}{2 \pi r} \frac{\partial n_{_<}(r)}{\partial r} = -\frac{1}{2 \pi r} \frac{\partial n_{_>}(r)}{\partial r},
\ee
and from \reqs{nofr} and $n_{_>}(r)= 1 - n_{_<}(r)$, after some manipulation exploiting \req{Kg22}, we obtain
\be\label{1n>}
n_{_>}(r)= g(r)^2 (1 + I_3(r)).
\ee
We see that beyond the radius $r$ at which $g^2$ vanishes (which when all $w_b$'s are bounded is $r=r_0$), $n_{_>} (r) $ and $ \rho(r)  = -\frac{1}{2 \pi r} \frac{\partial n_{_>}}{\partial r}$  vanish identically, while for smaller $r$  they are positive.

In cases in which some $w_b$ grow without bound as $N\to\infty$, some singular values, $s_i(z)$, of $M_z = z- M$ are $o(1)$, and more care is needed.
First, to see this, note that by definition $s_i(z)^2$ are the eigenvalues of the block-diagonal $K = M_z M_z^\dagger$; thus they come in pairs composed of the eigenvalues of $K$'s  $2\times 2$ blocks, given by \req {Kblocks}. We denote the pair of eigenvalues corresponding to block $b$ by $s_{b,\pm}(z)^2$, with the plus and minus subscripts denoting the larger and smaller singular value, respectively. The sum $s_{b+} (z) ^2 + s_{b-} (z) ^2$ and the product $s_{b+} (z) ^2  s_{b-} (z) ^2$ are given by the trace and determinant of \req{Kblocks}, \ie\ by $|w_b|^2 + 2|z|^2$ and $|z|^4$, respectively. It follows that for blocks where the feedforward weight $w_b$ is $O(1)$, both $s_{b,\pm}(z)$ will be $\Theta(1)$ for $|z|\neq 0$, while for blocks in which $w_b \to\infty$ as $N\to \infty$, we have 
\bea\label{svrajan}
s^2_{b+} (z) &=& |w_b|^2 + O(1) \to \infty \\
s^2_{b-} (z) &\approx& \frac{|z|^4}{|w_b|^2} = o(1).
\eea 
(Note that as stated after \req {frobnorm} we assume $\|M\|_{\Fr}^2 = \mu^2 = {\langle |w_b|^2\rangle_b/2}$ is $O(1)$, so that at most $o(N)$ number of weights can be unbounded, and each such $w_b$  can at most be $O(\sqrt{N})$.)
If all the $w_b$ are $O(1)$,  and hence all singular values are $\Theta(1)$ (for $|z|\neq 0$), \req{r0} yields the correct support radius  as noted above, and for  $r \leq r_0$, \req{Kg22} yields a $\Theta(1)$ solution for $g(r)^2$, which leads to a $\Theta(1)$ solution for $n_{_>}(r)$ and $\rho(r)$ via \reqs {1n>}--\myref {rhoN/2}. 
In cases in which some $w_b$ are unbounded, however,  \req{r0} (derived from \req{support-res}) may not yield the correct support boundary. Such cases are  examples of the highly nonnormal cases mentioned in the general discussion after \req{rho-svd-res},   for which the support of $\lim_{N\to\infty}\rho(z)$ must be found by using \reqs{Kdef1}--\myref{support-res-K}. This is equivalent to solving \req{Kg22} after the limit $N\to\infty$ is taken (assuming $g^2>0$), and then finding where the solution for $g^2(|z|)$ vanishes, which yields the correct support radius. From \req{1n>} this is indeed the radius at which $\lim_{N\to\infty} n_{>}(r)$ and hence $\lim_{N\to\infty} \rho(r)$ vanish as well.  This radius is in general smaller  than $r_0$  as given by \req{r0}.

We now calculate $\rho(z)$ for two specific examples of $M$ from each group. 
The first example is that of equal and $O(1)$ feedforward weights in all blocks, which we denote by $w$ (in terms of \req{Mex2}, this case corresponds to $\BB = w \one$). Here we can drop the block averages in \reqs{Kg22} and \myref{I3}, replacing $w_b$ with $w$.
 Solving \reqs{Kg22}  for $g^2(|z| = r)$ 
 we find
 \bea
g^2( r) = \frac{1}{2} - \frac{w^2}{2} - r^2 + \frac{1}{2}\sqrt{1 + w^4 + 4 w^2 r^2}.  
  \eea
 Substituting this into \req{I3} and \req{nofr} yields 
 \be
  n_{_<}(r) = {r^2}  - \frac{w^2 {r^2}}{ 1 + \sqrt {1 + w^4 + 4 w^2 r^2}}.
 \ee
The replacements $w \to w/\Jstd$ and $r \to r/\Jstd$  then yield \req{eigdens-rajan-res} for general $\Jstd$, and $\rho(r)$ can be caclulated using \req{rhoN/2}.

The second case is that of \req{rajanM}. In this case only one of the blocks has a nonzero Schur weight given by $|w_1|^2 = \Tr(M^\dagger M) = N \mu^2 = O(N)$, where $\mu = O(1)$ is given by \req {muRajan}. Equation \myref{Kg22} 
now yields
\be\label{geqnrajan1}
1 = \frac{1}{r^2 + g^2} + \frac{\mu^2} {(r^2 + g^2)^2 + N \mu^2 g^2} \cdot \frac{r^2-g^2}{r^2+g^2},
\ee
or
\bea\label{geqnrajan}
   \frac{r^2+g^2 - 1} {r^2-g^2} = \frac{\mu^2}{(r^2 + g^2)^2 + N \mu^2 g^2}.
\eea
The right hand side of this last equation is $I_3(r)$, as follows from \req{I3}; thus using \req {geqnrajan} we can rewrite \req{1n>} as
\be\label{n>}
n_{_>}(r) =
{g^2(r)} \frac{2r^{2}- 1}{r^{2} - g^2(r)}.
\ee
Let us now solve \req{geqnrajan} to find $g(r)^2$. 
As noted above, and in accordance with the general prescription given after \req{rho-svd-res}, for the purpose of obtaining  $\lim_{N\to\infty} \rho(z)$ we have to first take the $N\to \infty$ limit in \req{geqnrajan1}, keeping $g^2>0$ fixed, and only then solve for $g^2$. Doing so makes the last term in \req{geqnrajan1} vanish, and we obtain $g^2(r) = 1- r^2$. This is positive for $r\leq 1$ and vanishes at $r=1$, the correct support radius of $\lim_{N\to\infty} \rho(z)$, which is strictly smaller than $r_0$ given by \req{r0}. 
From  \req {n>} we obtain $n_{_>}(r) = g^2(r) = 1-r^2$. It then follows from \req{rhoN/2}  that the $N\to\infty$ limit of the eigenvalue density is identical with the circular law (the result for the $M=0$), \ie\
$	
\lim_{N\to\infty}\rho(r) =  \frac{1}{\pi} 
$ 
for $r\leq 1$ and zero otherwise. With the correct scaling, this yields \req{res-rajanrho}.

Contrary to the general prescription given after \req{rho-svd-res}, we will now solve equations \reqs{Kg22}, \myref{I3} and \req{1n>} for $r>1$, without taking the limit $N\to\infty$ first. As we will see, the obtained solution for $g(r)^2$, and by \reqs{I3bound} and \myref{1n>} therefore the solutions for $n_{>}(r)$ and $\rho(r)$,  will be nonzero but $o(1)$ for $1< r \leq r_0$. As discussed in Sec.~\ref{sec-res-rajanabbott}, these finite-size corrections, which in general are not trustworthy, in the present case are in surprisingly good agreement with simulations for some range of $r$'s beyond $r_{\Theta(1)}$, but deviate from the true $n_{>}(r)$ for larger $r$ (see Fig.~\ref {fig-rajanoutliers}).
At finite $N$, it can indeed be checked that \req{Kg22} has a positive solution for $g^2$ if and only if $r<r_0$, with $r_0$ given by \req{r0}. Simplifying  \req{geqnrajan} yields a cubic equation in $g^2$. However, it turns out that ignoring the cubic term in $g^2$ is harmless for large $N$; the quadratic approximation has the positive solution
\be\label{gsolrajan}
g^2(r) =  
\sqrt{\left[ \frac{1-r^2}{2} \right]^2  + \frac{ r^2  (r^2+\mu ^2) -r^6}{\mu ^2 N}} + \frac{1-r^2}{2},
\ee
and for all $r<r_0$, corrections to \req{gsolrajan} when the cubic term is reinstated decay faster than the leading contribution from \req{gsolrajan} (nevertheless we numerically solved the full cubic equation \myref{Kg22} to obtain the black curve in Fig.~\ref {fig-rajanoutliers}, and the blue trace in Fig.~\ref {fig-rajanN_>(1)}). First, analyzing \req{gsolrajan} we see that $g^2(r)$ is indeed $\Theta(1)$ only for 
$r<1$, 
where as we already found
$
g^2(r) = 1 - r^2 + o(1).
$
Furthermore, for a fixed $r>1$ (such that $r-1$ does not vanish as $N\to\infty$), the solution for $g(r)$ is $O(N^{-1})$. 
Thus from \req{n>} we thus see that $N n_{_>}(r)$, \ie\ the total number of eigenvalues with modulus larger than $r$, for $1<r<r_0$ (and $r-1 = \Theta(1)$) is only $O(1)$; the solution for $N n_{_>}(r)$ is shown in Fig.~\ref {fig-rajanoutliers}. 
Correspondingly, from \reqs{n>} and \myref{rhoN/2} we see that $\rho(r)$ is $o(1)$ in this region and vanishes in the limit $N\to\infty$, as already found.
Now let us calculate the total number of eigenvalues lying outside the circle $|z| = 1$. This is given by $N n_{_>}(1)$. From \req{gsolrajan} we find $g^2(1) = \frac{1}{\sqrt{N}}$, and substituting in \req{n>} we obtain
\be
N_{_>}(1) \defin  N n_{_>}(1)  = \sqrt{N} + O(1).
\ee
With the proper rescaling this yields \req{rajanoutliers} for general $\Jstd$.
Note that, according to \req{gsolrajan}, $g(r)$ (and hence $n_{_>}(r)$) remains $\Theta(N^{-1/2})$ (as opposed to $O(N^{-1})$) in a thin boundary layer outside of width $\Theta(N^{-1/2})$ just outside of the circle $|z| = 1$. 

We will now work out the formula for $\Javg{\|\vx(t)\|^2}$, \reqs{res-F1}--\myref {res-avgnormsq}, when the initial condition $\vx_0$ is  the second Schur-vector in block $b=a$, which we denote by $\ve_{a2}$; in the Schur representation, \req{Tex2},  we have $\ve_{a2} = (0,1)\trans$ (we only write the components of $\ve_{a2}$ in block $a$). To calculate the numerator in \req{res-F1} we first calculate $(z-T_a)^{-1}\ve_{a2}$ where $T_a = \begin{pmatrix}
 0     &   w_a \\
   0   &  0
\end{pmatrix}$ denotes the $a$-th diagonal $2\times 2$ block of \req{Tex2}. Since $T_a^2 = 0$, we have $(z-T_a)^{-1} = z^{-1} + z^{-2} T_a$, which yields $\vv_a(z)\defin  (z-T_a)^{-1}\ve_{a2} = (w_a z^{-2}, z^{-1})\trans$. We thus obtain 
\be\label{koonde1}
 {\vx_0\trans \frac{1}{\zbar_2 - M^\dagger}  \frac{1}{z_1-M}\vx_0}  =  \vv_a(z_2)^\dagger \vv_a(z_1) = \frac{1}{z_1 \zbar_2} + \frac{|w_a|^2}{z_1^2 \zbar_2^2}.
\ee
On the other hand, we have 
\bea\label{koonde2}
&&\tr \frac{1}{\zbar_2 - M^\dagger}  \frac{1}{z_1-M} =
\\&&
 \langle \frac{1}{2}\Tr_{\scriptscriptstyle{2\times 2}} (\zbar_2-T_b^\dagger)^{-1}(z_1-T_b)^{-1} \rangle_b
 = \frac{1}{z_1 \zbar_2} + \frac{\langle |w_b|^2\rangle_b/2}{z_1^2 \zbar_2^2}. 
 \nonumber
\eea
Substituting \reqs{koonde1}--\myref{koonde2} in \req{res-F1} we obtain
\be\label{booliboo}
F(z_1,z_2) = \frac{z_1 \zbar_2 + |w_a|^2 } {(z_1\zbar_2)^2 -  (z_1\zbar_2 + \mu^2)}. 
\ee
where we used $\mu^2 = {\langle |w_b|^2\rangle_b/2}$, and we denoted the integrand of \req{res-F1} by $F(z_1,z_2)$ with $z_i = \gamma + i\omega_i$, ($i=1,2$). 
By comparing the integrand of \req{res-F1} with  \req {res-powerspecBLR1}, we see that substituting $z_1 = z_2 = \gamma + i\omega$ into \req{booliboo} yields 
 the total power spectrum, $\Javg{\overline{\| \vx_\omega\|^2}}$. After the proper rescalings, this yields \req{res-powerspecrajan} for general $\Jstd$.
To obtain $\Javg{\|\vx(t)\|^2}$, on the other hand, 
we should substitute \req{booliboo} into  \req {res-avgnormsq} with 
$z_i = \gamma + i\omega_i$. 
Let us use the change of variables $\omega_1 = \Omega + \omega$ and $\omega_2 = \Omega - \omega$.
Then we have $z_1\zbar_2 = \Omega^2 + (\gamma + i\omega) ^2$, and from \req {res-avgnormsq} we obtain 
\be\label{kooni22}
\Javg{\| \vx(t)\|^2} = \int \frac{d\omega}{2\pi} e^{2it\omega} f_a (\gamma + i\omega)
\ee
where we defined
\be\label{kooni3}
f_a(u) \defin  {2} \int \frac{d\Omega}{2\pi} \frac{\Omega^2 + u^2 + |w_a|^2 } {(\Omega^2 + u^2)^2 - (\Omega^2 + u^2+ \mu^2)}.
\ee
Let us rewrite the integrand in \req{kooni3} as 
\bea
&& \frac{\Omega^2 + u^2 + |w_a|^2 } {(\Omega^2 + u^2 - r_0^2) (\Omega^2 + u^2 + r_1^2)}=  \frac{\Omega^2 + u^2 + |w_a|^2 }{r_0^2 + r_1^2 } \times \qquad
 \nl
&& \qquad\qquad\qquad  \lbr \frac{1}{\Omega^2 + u^2 - r_0^2} - \frac{1}{\Omega^2 + u^2 + r_1^2} \rbr,
\label{kooni4}
\eea
where $r_0^2$ was defined in \req{r0} and 
\be\label{r1}
r_1^2 \defin  r_0^2- 1 \geq 0.
\ee
One can calculate the integral over $\Omega$ in \req{kooni3} by contour integration, closing the contour, say, in the upper half of complex plane. The poles of the first and the second terms on the second line of \req{kooni4} are located at $\Omega_{0,\pm} = \pm i \sqrt{u^2- r_0^2}$  and $\Omega_{1,\pm} = \pm i \sqrt {u^2 + r_1^2} $, respectively. For $u = \gamma + i\omega $ ($\gamma>0$) the roots falling in the upper half plane are $\Omega_{0,+}$ and $\Omega_{1,+}$, independently of $\omega$. From their residues we obtain
\be\label{kooni5}
f_a(u) = \frac{1 }{r_0^2 + r_1^2 } \lbr \frac{r_0^2 + |w_a|^2}{\sqrt{u^2 -  r_0^2}} + \frac{r_1^2 - |w_a|^2}{\sqrt{u^2 +  r_1^2}} \rbr.
\ee
The integral of \req{kooni5} in \req{kooni22} is essentially the inverse Laplace transform of \req{kooni5}. Consulting a table of Laplace transforms \cite{AbramowitzStegun} yields 
\bea
&& \Javg{\| \vx(t)\|^2}= 
\\
&& \quad 
e^{-2\gamma t}  \lbr \frac{r_0^2 + |w_a|^2}{r_0^2 + r_1^2 } I_0(2 r_0 t) + \frac{r_1^2- |w_a|^2}{r_0^2 + r_1^2 } J_0(2r_1t)\rbr.
\nonumber
\eea
where $J_0(x)$ ($I_0(x)$) is  the $0$-th Bessel function (modified Bessel function).
From \reqs{r0} and \myref{r1} it follows that $r_0^2 + r_1^2 = \sqrt{1 + 4\mu^2}$,
 and using $\mu^2 = \langle |w_b|^2\rangle_b/2$ once again, we obtain 
\be\label{rajantransamp}
\Javg{\| \vx(t)\|^2}=  
\lbr 
 \frac{1 + C_a}{2} I_0(2 r_0 t) +  \frac{1 - C_a}{2}  J_0(2 r_1 t)
\rbr e^{-2 \gamma t}
\ee
where we defined
\be
C_a \defin  \frac {1+ {2 |w_a|^2}}{\sqrt {1+ {2 \langle |w_b|^2\rangle_b }}}.
\ee
Effecting the proper rescalings we obtain the result for general $\Jstd$,  \reqs{res-rajantransamp}--\myref{Cbdef}.

\subsection{Network with different neural types and independent, factorizable weights}
\label{sec-typenet}

Here we carry out the explicit calculations for the network with $C$ neural types presented Sec.~\ref{sec-res-rajanabbott}, with $M$, $L$ and $R$ are given by \reqs{typesMLR}--\myref{typesM}. 
From \reqs{Mzdef} and \myref{typesMLR}--\myref{typesM} we obtain
$
M_z = z(RL)^{-1}  - s \vu \vu\trans,
$
and 
\be\label{mzmz1}
M_z M_z^\dagger = |z|^2 (RL)^{-2} - z s \vcv \vu\trans - \zbar s   \vu \vcv\trans + s^2 \vu \vu\trans,
\ee
where we defined $\vcv \equiv (RL)^{-1}  \vu$. Using the Woodbury matrix identity we can write 
\be\label{typedwood}
\frac{1}{g^2 + M_z M_z^\dagger }  = Q - Q U \frac{1}{D^{-1} + U^\dagger Q U} U^\dagger Q
\ee
where we defined the $N\times 2$ matrix $U = \begin{pmatrix}\vu\, ,
      &\!    \vcv        
\end{pmatrix}$, and
\bea\label{Qdef-typesec}
Q &\equiv& \frac{1}{g^2 + |z|^2 (RL)^{-2}}
\\
D &\equiv& \begin{pmatrix}
    s^2  &   -\zbar s \\
-z s      &  0
\end{pmatrix}.
\eea
We will argue that for $g>0$, $\tr({g^2 + M_z M_z^\dagger })^{-1} = \tr Q$, up to $o(1)$ corrections. From \req{typedwood}, for the remainder $\Delta(g,z)\equiv \tr({g^2 + M_z M_z^\dagger })^{-1} - \tr Q$, we obtain
\be
\Delta(g,z) = -\frac{1}{N} \Tr\!\! \lbr \frac{ U^\dagger Q^2 U}{D^{-1} + U^\dagger Q U} \rbr 
\ee
where the trace is now over $2\times 2$ matrices. We have 
\be
D^{-1} = -\begin{pmatrix}
    0  &  (z s)^{-1}   \\
(\zbar s)^{-1}      &  |z|^{-2}
\end{pmatrix}
\ee
and for $n=1,2$ we obtain
\bea
U^\dagger Q^n U  = \begin{pmatrix}
    \vu^\dagger Q^n \vu\,\,  &   \vu^\dagger Q^n \vcv \\
\vu^\dagger Q^n \vcv\,\,      &  \vcv^\dagger Q^n \vcv
\end{pmatrix} = 
\begin{pmatrix}
    I_{n,0}  & I_{n,1}    \\
I_{n,1}         &  I_{n,2}
\end{pmatrix},
\eea
where
\bea
I_{n,k}(g,z) &\equiv & \frac{1}{N} \sum_{i = 1}^N \frac{(l_{c(i)}r_{c(i)})^{-k} }{\lbr g^2 + |z|^2 (l_{c(i)}r_{c(i)})^{-2}\rbr^n}
\\
\areq \left\langle  \frac{\sigma_c^{ -k} }{\lpr g^2   + \sigma_c^{-2} |z|^2\rpr^n}\right\rangle_c
\eea
and we are using the notation \req{typesavgdef} 
(we will drop the explicit $g$ and $z$ dependence of $I_{n,k}$ when convenient).
Note that all $I_{n,k}(g,z)$ are $O(1)$ and for even $k$   are  positive. 
Inverting $D^{-1} + U^\dagger Q U$ we obtain 
\be\label{Deltagz}
\Delta(g,z) = \frac{1}{N} \frac{T(g,z)}{ - \det(D^{-1} + U^\dagger Q U)} 
\ee
where 
\bea
&&\!\! T(g,z) \equiv 
\Tr\!\! \lbr \begin{pmatrix}
    I_{2,0}  & I_{2,1}    \\
I_{2,1}         &  I_{2,2}
\end{pmatrix}
\begin{pmatrix}
    I_{1,2}-|z|^{-2}\,\,  & (zs)^{-1}-I_{1,1}    \\
(\zbar s)^{-1}-I_{1,1}         &  I_{1,0}
\end{pmatrix}\rbr
\nl
&& \quad = 
I_{2,2}I_{1,0} + I_{2,0} \!\lpr  I_{1,2}-\frac{1}{|z|^{2}}\rpr  - 2 I_{2,1}\! \lpr I_{1,1}   -\frac{1}{s\Re\, z}\rpr 
\nonumber
\eea
and 
\bea
- {\det(D^{-1} + U^\dagger Q U)} =  I_{1,0} \!\lpr \frac{1}{|z|^{2}} -  I_{1,2}\rpr  + \left | I_{1,1}   -\frac{1}{s z} \right |^2
&&
\nl
 =  
\frac{g^2}{|z|^2}I_{1,0}^2 + \left | I_{1,1}   -\frac{1}{s z} \right |^2. \qquad\qquad && 
\eea
We see that both $T(g,z)$ and $- {\det(D^{-1} + U^\dagger Q U)}$ are $O(1)$ (to obtain their limits as $N\to\infty$ we can set $s^{-1} = O(N^{-1/2})$ equal to zero) and since $- {\det(D^{-1} + U^\dagger Q U)} \geq \frac{g^2}{|z|^2}I_{1,0}^2$ and $I_{1,0}^2>0$, we see that for $g>0$,  the denominator in \req{Deltagz} is bounded away from zero, and hence $\Delta(g,z) = O(N^{-1})$ and can be safely ignored for  $g>0$.

We will thus use $\tr({g^2 + M_z M_z^\dagger })^{-1} = \tr Q + o(1)$. From \req{Qdef-typesec} we obtain $\tr Q = I_{1,0}(g,z)$ and hence from \reqs{Kdef1}, 
\be\label{chiruchiru}
\K(g,z) = \lim_{N\to\infty} \tr Q =  \left\langle  \frac{1}{ g^2   + \sigma_c^{-2} r^2}\right\rangle_c
\ee
where $r \equiv |z|$.
Note that the approximation $\tr({g^2 + M_z M_z^\dagger })^{-1} = \tr Q$ is equivalent to using $M_zM_z^\dagger = |z|^2 (RL)^{-2}$ instead of the full expression \req{mzmz1} and hence to setting $M=0$.
 Accordingly, the support of the eigenvalue distribution is given by \req{boundaryM0}, or equivalently by \req{boundary-types}, and within this support, $g^2$ is  depends only on $|z| = r$ and is found by solving \req{geqnM0}, or equivalently  \req{geqn-types}.  Similar considerations show that in using \req{rho-res} to obtain $\lim_{N\to\infty}\rho(z)$ we can set $M = 0$,  yielding an isotropic eigenvalue density. From \reqs{rhoM0}--\myref{n>M0}, the proportion, $n_>$, of eigenvalues lying a distance larger than $r$ is equal to $g^2(r)$, which is found by solving \req{geqn-types}. 
 The results \reqs{rhoM00r}--\myref{rhoM0r0} also hold, wherein the normalized sums over $i$ can be replaced with appropriate averages $\langle \cdot\rangle_c$.

Let us now go back to the expression for $\Delta(g,z)$, and consider the case $g=0$. In this case
\be
I_{n,k}(0,z) = |z|^{-2n} \langle \sigma_c^{2n-k}\rangle_c,
\ee
and we obtain 
\be
T(g,z) = |z|^{-6}\lpr \langle\sigma_c^2\rangle_c^2 -2 \langle\sigma_c^3\rangle_c\langle\sigma_c\rangle_c\rpr  + 2 \langle\sigma_c^3\rangle_c \frac{s^{-1}}{|z|^{4} \Re\, z}
\ee
and 
\be
- {\det(D^{-1} + U^\dagger Q U)} = \left | |z|^{-2} \langle \sigma_c\rangle_c - \frac{s^{-1}}{z}\right |^2.
\ee
In the special case in which $\langle \sigma_c\rangle_c = 0$ (this corresponds to the special case of the example \req{rajanM} with  $f \mu_E - (1-f)\mu_I \propto \vu\cdot \vcv =0$, which we considered above), the determinant will have a vanishing limit as $N\to\infty$ (or $s^{-1}\to 0$). This leads to a finite limit for $\Delta(0,z)$ and we obtain
\be\label{Deltag0}
\Delta(0,z) =  \frac{s^2 \langle \sigma_c^2\rangle^2}{N |z|^4}  = \frac{\xi^2 \langle \sigma_c^2\rangle^2}{ |z|^4}, \qquad (\langle \sigma_c\rangle_c = 0).
\ee
Adding this to $\tr Q$ in the right side of \req{chiruchiru}, and using the naive formula  \req{support-res} or $\K(0,z) =1$ for the spectral boundary, we would have obtained the equation 
\be
1 = \frac{\left\langle {\sigma_c^{2} }\right\rangle_c}{r^2} + \frac{\xi^2 \langle \sigma_c^2\rangle^2}{ r^4}.
\ee
 This in turn yields the radius \req{wrongradius-types}
which is larger than the true boundary of the support of $\lim_{N\to\infty} \rho(z)$ given by  \req{boundary-types}.

\section{Conclusions}
\label{sec-concl}

We have provided a general formula for the eigenvalue density of partly random matrices, \ie\, matrices with general mean and non-trivial covariance structure. General formulae have also been derived for the magnitude of impulse response and frequency power spectrum in an 
$N$-dimensional linear dynamical system with a coupling given by such partly random matrices. Our theory makes no requirement on the normality of matrices; its applications include therefore the stability and linear response analysis of neural circuits, whose linearized dynamics is always nonnormal. We have demonstrated our theory by tackling analytically two specific neural circuits: a feedforward chain of length $N$, and a set of randomly coupled feedforward subchains of length 2. A connection has also been revealed between  the eigenvalue spectra of dense random matrix perturbations, and the theory of pseudospectra. 

The non-crossing diagrammatic method can be used to calculate other quantities of interest for matrix ensembles of the form $A = M+LJR$, considered here as well; possible examples are direct statistics of eigenvectors \cite{Mehlig:1999}, or the correlations of the random fluctuations of the eigenvalue density $\Javg{\delta \rho_{_J}(z) \delta \rho_{_J}(z+w)}$ for macroscopic $w$ (\ie\ for $|w| = \Theta(1)$). 
On the other hand, quantities  such as the microscopic structure of $\Javg{\delta \rho_{_J}(z) \delta \rho_{_J}(z+w)}$, \eg\ for $|w| = \Theta(N^{-1/2})$ with $z$ inside the support, which could be of interest in the study of eigenvalue repulsion are not accessible to the non-crossing approximation. 
This is  also the case,  in general, for the statistics of the ``outlier" eigenvalues that we discussed after \req{rho-svd-res} and in the examples of Sec.~\ref {sec-res-longffwd} and Sec.~\ref {sec-res-rajanabbott}, which may be of  importance in practical applications.  
The calculation of such quantities is possible, for example, by using the replica technique (see \eg\ 
Ref.~\cite{NishigakiKamenev:2002}).

Finally, there are important forms of disorder which are not covered by the general ensemble $A = M + LJR$ with iid, and hence dense, $J$. 
Examples of relevance to neuroscientific applications include sparse $A$ \cite{RogersCastillo:2009, SlaninaSlanina:2011,Neri:2012aa} (note that, \eg, binary matrices with probability of a nonzero weight, $p$, which is small but $\Theta(1)$ as $N\to\infty$ are covered by our formulae; by ``sparse" disorder we refer, \eg\, to the case $p = o(1)$), or more general structure of correlations between the elements of $A$ (in the ensemble considered in this article, and for real $J$, the covariance $\Javg{ \delta A_{ij}\delta A_{i'j'} } = (LL\trans)_{ii' }(R\trans R)_{jj'}$ is single rank); the latter is of importance in considering networks with local topologies where, \eg, the matrix $A$ has a banded structure.
Generalization to other forms of random disorder is thus an important direction for future research.

\acknowledgements

We thank Larry Abbott  and Merav Stern for helpful discussions. Y.A. was supported by the Kavli Institute for Brain Science Postdoctoral Fellowship and by the Swartz Program in Computational Neuroscience at Columbia University, supported by a gift from the Swartz Foundation. 
K.D.M. was supported by grant R01-EY11001 from the NIH and by the Gatsby Charitable Foundation through the Gatsby Initiative in Brain Circuitry at Columbia University.

\appendix

\section{Validity of the non-crossing approximation}
\label{app-nonxing}
 
\begin{figure}[!t]\hspace{-0.2cm}
\includegraphics[width=3.2in,angle=0]{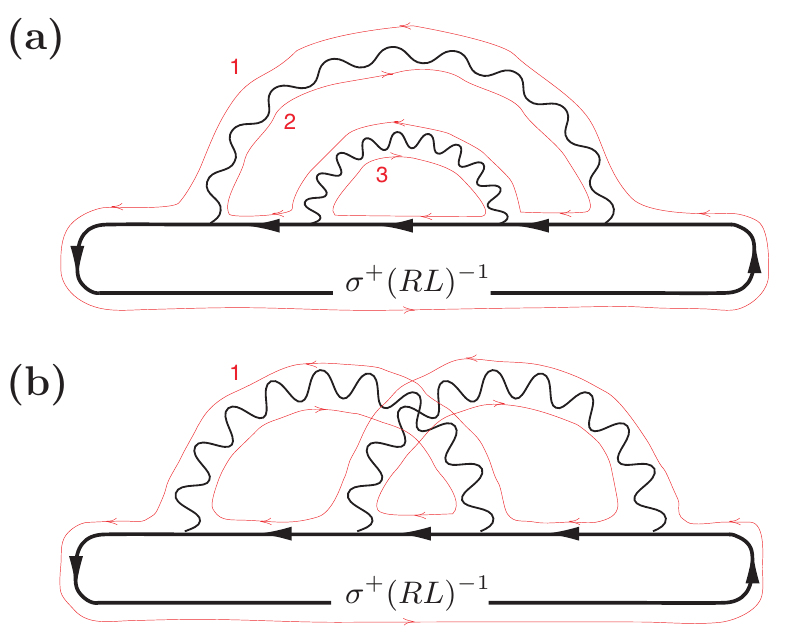}
\vspace{-0cm}
\caption{(Color online) The orbits (shown by thin red paths) for two diagrams for the spectral density in a complex $J$ ensemble. The non-crossing diagram on top has three orbits: orbit (1) is the external orbit connecting the two ends of the Green's function, while orbits (2) and (3) are the internal orbits. As in \reqs{INmk0} and \myref{ENmk0}, they contribute $\tr\!\lpr\sigma^+  \Gb(\eta,z;0)\pi^{r_1}  \Gb(\eta,z;0)\rpr$, $\Tr\!\lpr\pi^{3-r_1}\Gb(\eta,z;0)\pi^{r_2}  \Gb(\eta,z;0)\rpr$ and $\Tr\!\lpr\pi^{3-r_2}  \Gb(\eta,z;0)\rpr$, respectively, with $r_1$ and $r_2$ summed over 1 and 2 (cf. \req{jpair2}). The trace contributed by each of the three orbits Êis $O(N)$,  which when combined with the three factors of $1/N$ accounting for the two wavy lines and the normalization of  the external orbit's trace,  yield an $O(1)$ expression for this diagram. By contrast, the crossing diagram on the right has no internal orbits. Its only external orbit contributes $\Tr\!\!\lpr\sigma^+  \Gb(\eta,z;0)\pi^{r_2} \Gb(\eta,z;0)\pi^{3-r_1} \Gb(\eta,z;0)\pi^{3-r_2}\right.$ $\left.\Gb(\eta,z;0)\pi^{r_1}  \Gb(\eta,z;0)\rpr$ which after normalization is $O(1)$. Accounting for two factors of $1/N$ coming from the wavy lines, we then see that this crossing diagram is $O(N^{-2})$ and hence is suppressed as $N\to\infty$.}
\label{fig-orbs}
\end{figure}

In this appendix we will give the justification for the non-crossing approximation used in Sec. \myref {sec-deriv1} and \myref {sec-deriv2}. That is, we will show that the only diagrams not suppressed by inverse powers of $N$ are the non-crossing diagrams.   We will limit our discussion to the case of the eigenvalue density considered in Sec. \myref {sec-deriv1}, but the {generalization to the quantities calculated in \myref {sec-deriv2} is straightforward}. 
As explained after \req{PT}, averaging of $\Gb(\eta,z;J)$ over the disorder $J$ involves summing over all complete pairings of the factors of $J$ in every term of the expansion \req{PT}, with each pairing of each term represented by a diagram as shown in Fig.~\ref{fig-dyson}. 
Each such diagram is composed of a solid directed line (each segment of which represents a factor of $\Gb_{ab}^{\alpha\beta}(\eta,z;0)$), with a number of wavy lines (each representing the expression \req{jpair2}, with different indices) connecting different points on the solid arrow line, and all the internal matrix indices summed over. 
For the purpose of calculating the eigenvalue density, according to \reqs{rho1}--\myref{gbar}, what we need to calculate is actually $\tr\!\!\lpr \sigp (RL)^{-1} \Javg{\Gb(\eta,z;J)} \rpr$; thus we can imagine the solid arrow making a loop by closing-in on itself sandwiching $\sigp \otimes (RL)^{-1}$ (see Fig. \ref{fig-orbs}).
%

Given the structure of the Kronecker deltas in \req{jpair2},  it is more convenient for our purpose here, however, to think of each diagram  as a number of ``orbits," each formed by starting somewhere on the solid line and moving on it always along its arrow until the next wavy line is encountered, whereby we leave the solid line, continuing on the wavy line \emph{without} crossing it 
 {(because \req{jpair2} is composed of two Kronecker deltas, one for each side of the wavy line, enforcing index identification at the corresponding ends on each side 
\footnote{This structure is a consequence of using a complex ensemble for $J$, for which the covariances
 $\langle J_{ab}J_{cd}\rangle$  vanishes. 
For the real Gaussian ensemble, by contrast, the latter do not vanish; in this case \req{jpair2} becomes $\langle \J^{\alpha\beta}_{ab} \J^{\gamma\delta}_{cd} \rangle = \frac{1}{N}\,\lbr \delta_{ad}\delta_{bc} \lpr \sigp_{\alpha\beta}\sigm_{\gamma\delta} +  \sigm_{\alpha\beta}\sigp_{\gamma\delta}  \rpr
+\delta_{ac}\delta_{bd} \lpr \sigp_{\alpha\beta}\sigp_{\gamma\delta} +  \sigm_{\alpha\beta}\sigm_{\gamma\delta}  \rpr
\rbr 
= \frac{1}{N} \sum_{r=1}^2 \lbr (\pi^r_{\alpha\delta}\delta_{ad}) ( \pi^{3-r}_{\beta\gamma}\delta_{bc}) + (\pi^r_{\alpha\gamma}\delta_{ac}) ( \pi^{3-r}_{\beta \delta}  \delta_{bd}) \rbr
$.
}})  
and return somewhere else on the solid line, continuing as before until we reach the initial point (see Fig. \ref{fig-orbs}). 
As we go around this orbit, for each solid line traversed we write down, from right to left, a $\Gb(\eta,z;0)$ and for each wavy line a $\pi^{r_i}$ (see \reqs{pidef}) where $i$ is the index of the wavy line.   Because all matrix indices are summed over, such adjacent factors multiply like matrices, and since the orbit forms a loop, in the end we obtain the \emph{trace} of the matrix product thus obtained. (This recipe for assigning the contribution of each orbit accounts for the Kronecker deltas and $\pi^r$'s in \req{jpair2}, but not for the factor $\frac{1}{N}$ and the sum over $r$'s; we will account for these, at the end, after \req{ENmk0}.)
A generic orbit, which we refer to as internal, closes on itself after traversing, say, $m$ wavy lines sanwiching $m$ Green's functions (\eg\ the orbits labeled 2 and 3 in panel (a) of Fig. \ref{fig-orbs}), 
 and thus contributes a trace of the form
\be\label{INmk0}
I_{m,\vct{r}} \equiv \Tr\!\!\lpr \Gb(\eta,z;0) \pi^{r_{i_1}} \cdots \Gb(\eta,z;0) \pi^{r_{i_m}} \rpr, 
\ee
where $\vct{r}$ is short-hand for $\lcr r_ {i_1},\ldots, r_{i_m}\rcr$, and $i_k$ are the indices of the wavy lines traversed in the orbit.
In every diagram, there is also exactly one orbit (\eg\ the orbits labeled 1 in both panels of Fig. \ref{fig-orbs})  which in addition includes the factor $ \sigp (RL)^{-1}$ sandwhiched between the two external Green's functions. This orbit, which we call the external orbit, contributes a trace of the form
\be\label{ENmk0}
 E_{n,\tilde{\vct{r}}} \equiv \Tr\!\!\lpr \sigma^+ (RL)^{-1} \Gb(\eta,z;0) \pi^{r_ {j_1}}  \cdots \pi^{r_ {j_n}} \Gb(\eta,z;0) \rpr
\ee
where $n$ is the number wavy lines the orbit traverses and $\tilde{\vct{r}}$ is short for $\lcr r_ {j_1},\ldots,  r_ {j_n}\rcr$, and $j_k$ are the indices of the wavy lines traversed in this orbit (in  writing \req{ENmk0} we dropped the  $\frac{1}{N}$ that normalizes the trace in \reqs{rho1},  but we will account for it below). For succinctness, in \reqs{INmk0}--\myref{ENmk0} we suppressed the arguments $(\eta,z)$ for $I_{m,r}$ and $E_{n,\tilde r}$ on which they depend.
The full expression for the diagram is obtained by multiplying all such trace factors contributed by every orbit in the diagram, as well as a factor of $N^{-w -1}$ where $w$ is the number of wavy lines in the diagrams, to account for the $N^{-1}$ in \req{jpair2} for each wavy line, as well as the extra $N^{-1}$ which normalizes the trace in the external orbit \req{ENmk0} as dictated by \req{rho1}. The obtained expression is finally summed over all the $r$-indices corresponding to each wavy line, as required by \req{jpair2}.

The justification for the non-crossing approximation is based on the claim that each trace contributed by a orbit (external or internal) as in \reqs{INmk0}--\myref{ENmk0} is $O(N)$, irrespective of $\eta$, $z$, $m$ or $\vct{r}$. We will provide justification for this claim below. However, accepting it as true, 
 we see that any diagram's scaling with $N$  solely depends on the number of orbits and wavy lines it contains. A well-known topological argument then shows that the contributions of crossing diagrams are suppressed by inverse powers of $N$ \cite{tHooft:1974,BrezinZuber:1978}; for completeness we will summarize this argument here.   First note that, assuming the claim, any diagram will yield an expression that is $O(N^\alpha)$ with
\be\label{alphaf}
\alpha = f - w - 1,
\ee
where $f$ is the number of orbits in the diagram (the sum over at most $2^w$ possible configurations of $r_i$ does not contribute to the scaling with $N$). Let  $V$ denote the total number of vertices in the diagram (\ie\, the number of intersections of wavy lines and the solid line, plus an extra one representing the insertion of $\sigp (RL)^{-1}$ in the solid line loop) and let $E$ denote its total number of edges, \ie\ $E\defin  w + s$, where $s$ is the number of solid line segments ($s=5$ in both panels of Fig. \ref{fig-orbs}). It is easy to see that $V = s$. Thus we have $E - V = w$. Formally defining the number of ``faces" in the diagram by $F\defin  f +1$, and its  ``Euler characteristic"  by
\be \label{euler}
\chi \defin  F - E + V,
\ee 
we then find that $\chi  = F - (E - V) = f + 1 - w$. From \req{alphaf} we then obtain
\be\label {chialpha}
\alpha = \chi - 2.
\ee	
Thus  the contribution of a diagram is $O(N^\alpha)$, with $\alpha$ determined solely by the diagram's formal ``Euler characteristic" via \req{chialpha}.
	{It can be shown that}  a diagram with  $F$ formal ``faces" and a formal ``Euler characteristic" $\chi$ as defined above, can be drawn on (embedded in) a two-dimensional oriented surface with Euler characteristic $\chi$, such that no edges (solid or wavy) cross to create new vertices, and each face   created on the surface by its partitioning by the drawn diagram, 
 {a) is topologically a disk, and b)  }
has a one-to-one correspondence with and is encircled by an orbit in the diagram, where we now count among the orbits, also the loop formed by the solid arrow line. 
Thus the number of faces on the surface is indeed $F = f + 1$, and the $\chi$, as defined above for the diagram, indeed agrees with the Euler characteristic of the surface, as conventionally defined. Topologically, such a surface is a generalized torus with $g$ holes, satisfying $\chi  = 2 - 2g$; the surface with zero holes is the sphere, or after decompactification, the plane (\eg\ the diagram in panel (b) of Fig. \ref{fig-orbs} can be drawn in this manner on a torus). We thus see that 
\be
\alpha= -2 g,
\ee
and therefore the only diagrams that are not suppressed by inverse powers of $N$ are those that can be drawn,  as described above, on the plane. Since we took the area enclosed by the solid arrow line loop as a face by itself, this means that the diagram should be drawable with the wavy lines remaining outside this area (in order not to partition it into several faces) without crossing each other; this is the precise definition of the diagram being non-crossing 
\footnote{Notice that this is a more restrictive property than planarity of the diagram; for example the graph in panel (b) of Fig. \ref{fig-orbs} is planar, as one of the wavy lines can be drawn inside the solid loop without crossing any other line, but it is not non-crossing as defined here.}.

Let us now go back to justifying the claim that the traces contributed by the orbits as in \reqs{INmk0}--\myref{ENmk0} are $O(N)$. For this purpose we will make use of the singular value decomposition of $M_z$ introduced in \req{svd}. 
Defining the unitary matrix 
\be\label{Uz}
\U_z \defin  
 \begin{pmatrix}
   U_z   & 0    \\
   0   &  V_z
\end{pmatrix},
\ee
and using \req{svd}, we can write $H_0(z)$, defined in \req {H0def}, as
\be\label{diago}
H_0(z) = 
\U_z \tH_0(z)
\,\U_z^{\dagger},
\qquad
\ee
where
\be\label{hamil22}
\tH_0(z) \defin  
 \begin{pmatrix}
   0   & S_z    \\
   S_z   &  0
\end{pmatrix}.
\ee
Let us also define $\tilde \Gb(\eta,z;0) \defin  \U^{\dagger}_z
 \Gb(\eta,z;0)
\U_z $, such that 
\be\label{tG-G}
\Gb(\eta,z;0) =
\U_z
\tilde \Gb(\eta,z;0)
\U^{\dagger}_z. 
\ee
Then using the definition $\Gb(\eta,z;0) = (\eta - H_0(z))^{-1}$, we see that 
\be\label{G0tild}
\tGb(\eta,z;0) = \frac{1}{\eta-\tH_0(z)} = \begin{pmatrix} \frac{\eta}{\eta^2 - S_z^2} &  \frac{S_z}{\eta^2 - S_z^2} \\
\frac{S_z}{\eta^2 - S_z^2} & \frac{\eta}{\eta^2 - S_z^2}\end{pmatrix}, 
\ee
where we used \req{hamil22} to write the last equality. Given the block-diagonal nature of \req{Uz} and the definitons \req{pidef}, we also have 
\be\label{pirpir}
\pi^r = \U_z
\pi^r
\U^{\dagger}_z. 
\ee
We now substitute $\Gb(\eta,z;0)$ and $\pi^{r_i}$ in \reqs{INmk0}--\myref{ENmk0} with the right hand sides of \reqs{tG-G} and \myref{pirpir}, respectively. After canceling the $\U_z$'s we obtain 
\bea\label{IENmk0}
I_{m,\vct{r}} \areq \Tr\!\!\lpr \tilde \Gb(\eta,z;0) \pi^{r_{i_1}} \cdots \tilde  \Gb(\eta,z;0) \pi^{r_{i_m}} \rpr, 
\\
 E_{n,\tilde{\vct{r}}} \areq \Tr\!\!\lpr \sigma^+\! A(z) \tilde  \Gb(\eta,z;0) \pi^{r_ {j_1}}  \cdots \pi^{r_ {j_n}}\tilde \Gb(\eta,z;0) \rpr\qquad\,\,
\eea
where we defined 
\be\label{Adefapp}
A(z) \defin  U_z^\dagger (RL)^{-1} V_z,
\ee
such that $\U_z^\dagger \lbr \sigma^+\otimes  (RL)^{-1} \rbr \U_z  = \sigma^+\otimes A(z) \defin  \sigma^+ A(z) $. 
For the internal orbits,  we see from \req {G0tild} that each $\tilde \Gb(\eta,z;0)$, depending on whether it is sandwiched between the same projectors $\pi^r$, or between two opposite projectors, $\pi^r$ and $\pi^{3-r}$,  contributes a \emph{diagonal} factor equal to $\eta/(\eta^2 - S^2_z)$ or $S_z/(\eta^2 - S^2_z)$, respectively. Thus, for any configuration of $r_{i}$'s, if the number of Green's functions sandwiched the second way is $k$ ($1\leq k \leq m$),  we obtain
\be\label{INmk}
I_{m,\vct{r}}(\eta,z) = \sum_{i=1}^N \frac{\eta^{m-k} s_i(z)^k}{(\eta^2 - s_i(z)^2)^m}\qquad\qquad (1\leq k \leq m)
\ee
for the internal orbits (in particular, we see that the sole dependence of $I_{m,\vct{r}}(\eta,z)$ on $\vct{r}$ is via the number $k$). We therefore have 
\bea\label{Ibound0}
| I_{m,\vct{r}}(\eta,z) | &\leq &  N  \max_{i}
 \left \vert \frac{\eta^{m-k} s_i(z)^k}{(\eta^2 - s_i(z)^2)^m} \right \vert.
\eea
When the imaginary part of $\eta$ is nonzero, the denominator in the right hand side of \req{Ibound0} cannot vanish for any value of $s_i(z)$ {(while as $\eta\to i0$, which is the limit we have to take after summing up the relevant diagrammatic series,  $s_i(z)$ that approach zero as $N$ grows can make this expression unbounded as $N\to \infty$)}. Assuming $\Im\, \eta >0$, it will be sufficient for our purposes to substitute \req{Ibound0} with the weaker bound 
 \be
| I_{m,\vct{r}}(\eta,z) | \leq  N  \max_{s}
 \left \vert \frac{\eta^{m-k} s^k}{(\eta^2 - s^2)^m} \right \vert,\qquad (\Im\, \eta >0)
 \ee
where now the maximum is taken for $s$ ranging over the whole  $[0,\infty)$. Since $\Im\, \eta>0$ the expression has no singularities at finite real $s$, and since $2m>k$, it cannot diverge as $s\to\infty$ either; thus it has a finite maximum independent of $N$. More precisely, it is easy to show that $\max_{s}
 \left \vert \frac{\eta^{m-k} s^k}{(\eta^2 - s^2)^m} \right \vert \leq \lbr \frac{\sqrt{2}}{|\Im\, \eta |} \rbr^m$, irrespective of $k$ as long as $1\leq k \leq m$, yielding
 \bea\label{Ibound}
| I_{m,\vct{r}}(\eta,z) | &\leq &  N  \lbr \frac{\sqrt{2}}{|\Im\, \eta |} \rbr^m, \qquad (\Im\, \eta >0).
\eea

Similarly, the trace for the external orbit can be written in the new basis \req{G0tild} as
\be\label{ENmk}
 E_{n,\tilde{\vct{r}}}(\eta,z) =  \sum_{i=1}^N A_{ii}(z)\frac{\eta^{n-\tilde k} s_i(z)^{\tilde k}}{(\eta^2 - s_i(z)^2)^n},\qquad (1\leq \tilde k \leq n),
 \ee
where $\tilde k$ is the number of Green's functions in \req{ENmk0} sandwiched between two $\pi^r$'s with different superscripts; this convention works correctly for the external orbit as well, if we account for the presence of $\sigma^+$ by imagining a $\pi^2$ ($\pi^1$) to the left (right) of the leftmost (rightmost) Green's function. 
 From \req{Adefapp}, we can write $A_{ii}(z) = \vu_i(z)^\dagger (RL)^{-1} \vcv_i(z)$, where we defined the vectors $\vu_i(z)$ and $\vcv_i(z)$ to be the $i$-th column of $U_z$ and $V_z$, respectively.  By the Cauchy-Schwartz inequality we then have 
 \bea
 |A_{ii}(z)| &\leq& \| \vu_i(z)\| \| (RL)^{-1} \vcv_i(z) \|
 \nl
 & \leq & \| \vu_i(z)\| \| \vcv_i(z) \| \| (RL)^{-1} \|,
 \eea
 where $\| (RL)^{-1} \|$ is the operator norm, or the maximum singular value, of $(RL)^{-1}$.
 But since $U_z$ and $V_z$ are unitary matrices, $\vu_i(z)$ and $\vcv_i(z)$ are unit vectors, and we obtain
\be\label{Aineq}
 | A_{ii}(z) | \leq \| (RL)^{-1} \|.
\ee
Going back to \req{ENmk}, this yields the bound
\be
| E_{n,\tilde{\vct{r}}}(\eta,z) | \leq  N \| (RL)^{-1} \| \max_{i}
 \left \vert \frac{\eta^{n-\tilde k} s_i(z)^{\tilde k}}{(\eta^2 - s_i(z)^2)^n} \right \vert.
\ee
The only difference with the inequality for $I_{m,\vct{r}}$ is the factor  $\| (RL)^{-1} \|$. Repeating the same argument as for the internal traces, we therefore see that
\be\label{Ebound}
| E_{n,\tilde{\vct{r}}}(\eta,z) | \leq   N  \lbr \frac{\sqrt{2}} {|\Im\, \eta |} \rbr^n  \| (RL)^{-1} \|, \qquad (\Im\, \eta >0),
\ee
and thus a sufficient condition for $E_{n,\tilde r}$ to be $O(N)$ for $\Im\, \eta>0$, is that $\| (RL)^{-1}\|$ remains bounded as $N\to \infty$, \ie\
\be
\| (RL)^{-1} \| = O(1).
\ee

Combining \reqs{Ibound} and \myref{Ebound}, and given the prescription after \req{ENmk0}, we can bound the  absolute value of the contribution of a diagram with genus $g$ (or $g$ crossings), $w$ wavy lines, and $s$ solid lines, by $2^w  \lbr \frac{\sqrt{2}} {|\Im\, \eta |} \rbr^s \| (RL)^{-1} \| N^{-2 g} $ (the power of $s$ is obtained by noting that the powers of $m$ and $n$ in the bounds \reqs{Ibound} and \myref{Ebound}, when summed over all orbits must equal $s$, since every Green's function or solid line appears in exactly one orbit).
  Hence for a fixed, nonzero $\Im\, \eta$,   the contribution of crossing diagrams (\ie\, those with $g\geq 1$) goes to zero as $N\to\infty$.
Thus if we take the limit $N\to\infty$ before the limit $\eta\to i0^+$, 
 ignoring the crossing diagrams is safe, and the expression for $\rho(z)$ obtained from \req{rho1} after analytic continuation of $\tr\!\!\lpr \sigp (RL)^{-1} \G(\eta,z) \rpr$ to $\eta = i0$, with $\G(\eta,z)$ given by the contribtiuon of non-crossing diagrams to $\Javg{\Gb(\eta,z;J)}$, gives the correct result for $\lim_{N\to\infty} \rho(z)$.
We mention that when the smallest singular value $s_i(z)$ remains bounded away from zero as $N\to\infty$,  even at $\eta = 0$ the traces \reqs{INmk} and \myref{ENmk} are $O(N)$, as is not hard to check, justifying the non-crossing approximation at $\eta =0$. Thus it is only when some $s_i(z)$ are $o(1)$ that it becomes important to  send $\eta$ to $i0^+$ only after the limit $N\to\infty$ has been taken. 
In particular, in such cases, applying the limit $\eta \to i0^+$ to the results obtained using the non-crossing approximation before taking the limit $N\to\infty$, may yield finite-size contributions to $\lim_{N\to\infty} \rho(z)$, which in general may yield incorrect subleading corrections.

\section{$\rho(z)$ vanishes in the region \req{reg1} }
\label{app-rho}
\vspace{-.1cm}

In this appendix we prove more rigorously that in the region \req{reg1}, the eigenvalue density vanishes. More precisely, we prove that $\rho(z)\defin \lim_{\epsilon\to 0^+}  \lim_{N\to \infty} \rho_{_N}(z, \epsilon) = 0$, where 
 \be\label{B1}
 \rho_{_N}(z,\epsilon) \defin \frac{1}{\pi} \frac{\partial \,}{\partial \zbar}\,\, \tr\!\!\lbr \frac{(\sR \sL)^{-1} M_z^\dagger}{M_z M_z^\dagger + \gamma^2} \rbr,
\ee
is obtained by  substituting \req{Gscba} into \reqs{rho11}. 
Here, 
$
\gamma = g(z,\epsilon) + \epsilon, 
$ 
 is the solution of \req{gKeq0}, which as we argued in Sec.~\ref{sec-deriv1}, vanishes as $\epsilon \to 0^+$ when $z$ is in the region \req{reg1} (note that since \req{gKeq0} is defined in the limit $N\to\infty$, $\gamma$ has no dependence on $N$). Recall that for $\epsilon>0$, $g(z,\epsilon)$ is positive and therefore $\gamma >\epsilon >0$.   
 Expanding the derivative in \req{B1}  we obtain 
\bea
&& \pi \rho_{_N}(z,\epsilon) = \tr\!\!\lbr \frac{(\sR \sL)^{-1} (\sR \sL) ^{-\dagger}}{M_z M_z^\dagger + \gamma^2} \rbr
\nl
&& - \tr\!\!\lbr {(\sR \sL)^{-1} M_z^{\dagger}} \frac{1}{M_z M_z^\dagger + \gamma^2} M_z (\sR \sL) ^{-\dagger}\frac{1}{M_z M_z^\dagger + \gamma^2} \rbr 
\nl
&& - \tr\!\!\lbr \frac {(\sR \sL)^{-1} M_z^{\dagger}}{M_z M_z^\dagger + \gamma^2} \frac{1}{M_z M_z^\dagger + \gamma^2} \rbr \partial_{\zbar}( \gamma^2)
\nl
&&
= \tr\!\!\lbr \frac{(\sR \sL)^{-1} Q (\sR \sL) ^{-\dagger}}{M_z M_z^\dagger + \gamma^2} \rbr
\nl
&& \qquad - \tr\!\!\lbr \frac {(\sR \sL)^{-1} M_z^{\dagger}}{M_z M_z^\dagger + \gamma^2} \frac{1}{M_z M_z^\dagger + \gamma^2} \rbr \partial_{\zbar}( \gamma^2),
\label{rho5}
\eea
where we defined
$
Q = \one - M_z^{\dagger} \frac{1}{M_z M_z^\dagger + \gamma^2} M_z
$
(we suppress the explicit dependence of $\gamma$ on $z$ for simplicity).
By the Woodbury matrix identity
$
Q = \frac{\gamma^2}{M_z^\dagger M_z + \gamma^2},
$
which upon substitution in \req{rho5} yields
\bea\label{rho6}
\pi \rho_{_N}(z,\epsilon) \areq  \, \tr\!\!\lbr \frac{(\sR \sL)^{-1}} {M_z^\dagger M_z + \gamma^2} \frac{ (\sR \sL) ^{-\dagger}}{M_z M_z^\dagger + \gamma^2} \rbr \gamma^2
\\
&& - \tr\!\!\lbr \frac {(\sR \sL)^{-1} M_z^{\dagger}}{M_z M_z^\dagger + \gamma^2} \frac{1}{M_z M_z^\dagger + \gamma^2} \rbr \partial_{\zbar}( \gamma^2). \nonumber
 \eea
Differentiating  \req{gKeq0} with respect to $\zbar$ yields
\be
- \partial_{\zbar}( \gamma^2) = \frac{-2\gamma^2  \partial_{\zbar}\K}{1-\K -2 \gamma^2\partial_{\gamma^2} \K},
\ee
with the partial derivatives of $\K(\gamma,z)$ given by 
\bea\label{Kz}
-\partial_{\gamma^2}\K \areq  \T^{^>}_{\infty}(\gamma) 
\defin \lim_{N\to\infty}  \T^{^>}_{N}(\gamma) 
\nl
-\partial_{\zbar}\K \areq \V_\infty(\gamma) \defin 
\lim_{N\to\infty} \V_{N} (\gamma),
\eea
where we defined 
\bea
\T^{^>}_{N} (\gamma) &\defin & 
\tr\!\lbr \frac{1}{(M_z M_z^\dagger + \gamma^2)^2}\rbr
\\
\V_{N} (\gamma) &\defin & 
\tr\!\lbr \frac{1}{M_z M_z^\dagger + \gamma^2} \frac{M_z (\sR \sL) ^{-\dagger}}{M_z M_z^\dagger + \gamma^2} \rbr.
\eea
We thus obtain 
\bea\label{rho7}
\pi \rho_{_N}(z,\epsilon) \areq 
  \gamma^2\, \T_N(\gamma) + \frac{2\, \gamma^2\, \V_\infty (\gamma)\, \V_N (\gamma)^*}{1 - \K(\gamma) + 2\gamma^2\, \T^{^>}_{\infty}(\gamma)}\qquad
  \eea
  where we defined 
\be
\T_N(\gamma) \equiv \tr\!\!\lbr \frac{(\sR \sL)^{-1}} {M_z^\dagger M_z + \gamma^2} \frac{ (\sR \sL) ^{-\dagger}}{M_z M_z^\dagger + \gamma^2} \rbr.
\ee
Having eliminated derivatives of $\gamma$, we now simply need to show that
$
  \lim_{\gamma\to 0^+}  \lim_{N\to\infty}
$
of the right side of \req{rho7} vanishes for $z$ is in the region \req{reg1} (where $\gamma = 0^+$ is the solution of \req{gKeq0} as $\epsilon \to 0^+$).

We will start by bounding the traces $\T_N(\gamma)$ and $\V_N(\gamma)$ in \req{rho7}. 
For $\V_N(\gamma)$ we use the singular value decomposition \req{svd}:
\bea\label{2ndtrace}
 |\V_N(\gamma)| \areq \left | \tr\!\!\lbr \frac {(\sR \sL)^{-1} M_z^{\dagger}}{M_z M_z^\dagger + \gamma^2} \frac{1}{M_z M_z^\dagger + \gamma^2} \rbr\right |
\\
\areq \tr\!\!\lbr U_z^\dagger (RL)^{-1} V_z \frac{S_z}{(S_z^2 + \gamma^2)^2} \rbr
 \nl
&\leq & \| (RL)^{-1}\|  \tr\!\! \lbr \frac{S_z}{(S_z^2 + \gamma^2)^2} \rbr
\nonumber
\eea
(where in the last line we used \req {Aineq}), \ie\
\be\label{Vineq}
| \V_N(\gamma) | \leq \| (RL)^{-1}\| \V_N^{^>}(\gamma)
\ee
 where we defined
\be
 \V_N^{^>}(\gamma) \defin  \frac{1}{N} \sum_{i=0}^N \frac{s_i(z)}{(s_i(z)^2 + \gamma^2)^2}.
\ee
Taking the limit $N\to\infty$, we obtain from \req{Vineq} 
\be\label{Vineqinf}
| \V_\infty(\gamma) | \leq C  \V_\infty^{^>}(\gamma)
\ee
where $\V_\infty^{^>}(\gamma) \defin \lim_{N\to\infty} \V_N^{^>}(\gamma)$, and $C$ is an upper bound on $\| (LR)^{-1}\|$ (which we have assumed is $O(1)$ as $N\to\infty$). 
To bound $\T_N(\gamma)$, we use the inequality
\be\label{abcdineq}
| \tr\!(ABCD)| \leq \|A\| \|C\| \tr(BB^\dagger )^{\frac{1}{2}}\tr(DD^\dagger )^{\frac{1}{2}}.
\ee
This can be derived by first using the Cauchy-Schwartz inequality, $|\tr(AB)|^2 \leq \tr(AA^\dagger)\tr(BB^\dagger)$, and then using the inequality $|\tr(AB)| \leq \| B\| \tr(A)$, valid for positive semi-definite $A$ (which in turn follows from the definition of $\| B\|$ after unitary diagonalization of $A$).
Using \myref{abcdineq} we obtain
\bea\label{1sttrace}
&& \left |\tr\!\!\lbr \frac{(\sR \sL)^{-1}} {M_z^\dagger M_z + \gamma^2} \frac{ (\sR \sL) ^{-\dagger}}{M_z M_z^\dagger + \gamma^2} \rbr \right | 
\leq
\nl
&& \qquad 
\leq \| (RL)^{-1}\|^ {2} \, \tr\!\!\lbr \lpr \frac{1} {M_z M_z^\dagger + \gamma^2}\rpr^2\rbr
\eea
or 
\be\label{Tineq}
|\T_N(\gamma)| \leq  \| (RL)^{-1}\|^ {2} \T^{^>}_N(\gamma).
\ee

Using the inequalities, \myref {Vineq}, \myref {Vineqinf} and \myref {Tineq} in \req{rho7} we obtain
\be
\pi | \rho_{_N}(z,\epsilon)| \leq C^2 \lbr \gamma^2\, \T^{^>}_N(\gamma) 
+ \frac{2\, \gamma^2\, \V_\infty^{^>} (\gamma)\, \V_N^{^>} (\gamma)}{1 - \K(\gamma) + 2\gamma^2\, \T^{^>}_{\infty}(\gamma)}
  \rbr
\ee
Taking the $N\to\infty$ limit (while keeping $\gamma$ finite), and defining $\rho(z,\epsilon)\defin \lim_{N\to\infty} \rho_{_N}(z,\epsilon)$, we obtain 
\be\label{rho71}
\pi \rho(z,\epsilon) \leq C^2 \lbr \gamma^2\, \T^{^>}_\infty(\gamma) 
+ \frac{2\, (\gamma\, \V_\infty^{^>} (\gamma))^2}{1 - \K(\gamma) + 2\gamma^2\, \T^{^>}_{\infty}(\gamma)}
  \rbr
\ee
where we defined
\bea\label{T>V>}
\T^{^>}_\infty(\gamma) &\defin & \lim_{N\to\infty} \frac{1}{N} \sum_{i=0}^N \frac{1}{(s_i(z)^2 + \gamma^2)^2}
\\
\V^{^>}_\infty(\gamma) &\defin &  \lim_{N\to\infty} \frac{1}{N} \sum_{i=0}^N \frac{s_i(z)}{(s_i(z)^2 + \gamma^2)^2}.
\eea
Thus, to show that $\lim_{\epsilon\to 0^+} \rho(z,\epsilon) = 0$, it suffices to show that 
$
 \gamma^2 \T^{^>}_\infty(\gamma) 
$
and
$
 \gamma\, \V_\infty^{^>} (\gamma) 
$ vanish as ${\gamma\to 0^+}$ (since $z$ is in the region \req{reg1}, $1-\K(\gamma)$ and hence the denominator in the last term in \req{rho71} remains positive as $\gamma\to 0^+$).
Let us rewrite \req {T>V>} as 
\bea\label{T>V>1}
\T^{^>}_\infty(\gamma) \areq  \int_0^\infty \frac{ \rho_ {_\mathrm{S}}(s;z)ds}{(s^2 + \gamma^2)^2}
\\
\V^{^>}_\infty(\gamma) \areq \int_0^\infty \frac{s \rho_ {_\mathrm{S}}(s;z)ds}{(s^2 + \gamma^2)^2}
\eea
where we defined 
\be\label{limitrhosvd}
\rho_ {_\mathrm{S}}(s;z) = \lim_{N\to\infty} \frac{1}{N} \sum_{i=0}^N \delta(s - s_i(z))
\ee
 as the limit of the density of the singular values of $M_z$ 
 \footnote{More precisely, we only need to define the limit \req{limitrhosvd} in the sense of distributions, \ie\ such that for any regular test function, $f(s^2)$, bounded at infinity  and regular everywhere, including at $s^2\to 0^+$, we have $\lim_{N\to\infty} \frac{1}{N}\sum_{i=1}^N f(s_i(z)^2) = \int_0^\infty f(s^2)\rho_ {_\mathrm{S}}(s;z)ds$. We do not assume any smooth form for $\rho_{_\mathrm{S}}(s;z)$; in particular, $\rho_{_\mathrm{S}}(s;z)$ may have delta function singularities when an $O(N)$ singular values converge to the same value as $N\to\infty$. Also note that this assumption does not forbid the possibility that some $s_i(z)$ diverge as $N\to\infty$; our requirement that $\| M\|_{_{\mathrm{F}}}$ remain bounded automatically guarantees that  these will not be numerous enough to contribute to $\rho_{_\mathrm{S}}(s;z)$ at infinity.}.
Note that contributions to $\T^{^>}_\infty(\gamma)$ and $\V^{^>}_\infty(\gamma)$ from integration on $[s_0,\infty)$ for any fixed, nonzero $s_0$ remain finite as $\gamma \to 0^+$; only singular contributions arising from the region $s= O(\gamma)\ll 1$ can contribute to $
 \gamma^2 \T^{^>}_\infty(\gamma)$ and $\gamma\, \V_\infty^{^>} (\gamma)$ as $\gamma \to 0^+$. 
 Thus we only need concern ourselves with the portion of integrals from $0$ to some arbitrary small, but fixed $s_0$, and show that $\gamma^2   \int_0^{s_0} \frac{ \rho_{_\mathrm{S}}(s;z)ds}{(s^2 + \gamma^2)^2}$ and $\gamma\int_0^{s_0}  \frac{s \rho_{_\mathrm{S}}(s;z)ds}{(s^2 + \gamma^2)^2}$ vanish as $\gamma\to 0^+$.
 Let us first consider the situation similar to that in the two examples \reqs{Mbidiag} and \myref{rajanM}. For those examples, there is a region of $z$ outside \req{reg1}, where a single (more generally $O(1)$) singular value $s_i(z)$ vanishes as $N\to\infty$, while all the other $s_i(z)$ remain bounded from below. But an $O(1)$ set of (vanishing) singular values does not contribute to the density \req{limitrhosvd} and since the other $s_i(z)$ are bounded from below, there is an $s_0$ below which $\rho_{_\mathrm{S}}(s;z)$ identically vanishes. So the claim is clearly true for such cases. 
More generally, we exploit the fact that $z$ is in the region \req{reg1}, so that
\be\label{K<1}
\lim_{\gamma \to 0^+} 
  \int_0^\infty\frac{\rho_{_\mathrm{S}}(s;z) ds}{s^2 + \gamma^2} <1.
\ee
We conclude that as $s\to 0^+$ the density, $\rho_{_\mathrm{S}}(s;z)$, must vanish at least as fast as $s^\alpha$, \ie\ it must be $O(s^\alpha)$, for some $\alpha>1$; otherwise the integral in \req{K<1} diverges in the limit. Let us therefore choose $s_0$ to be small enough such that for $s\leq s_0$, $\rho(s;z) < c s^\alpha$ for some constant $c$ and $\alpha >1$. It is then an elementary exercise to show that $\gamma^2 \int_0^{s_0} \frac{s^\alpha ds}{(s^2 + \gamma^2)^2}$ and $\gamma \int_0^{s_0} \frac{s^{\alpha+1} ds}{(s^2 + \gamma^2)^2}$ are $O(\gamma^{\mathrm{min}(2,\alpha -1)})$ and $O(\gamma^{\mathrm{min}(1,\alpha -1)})$, respectively, as $\gamma \to 0^+$, and since $\alpha >1$, they both vanish in the limit,  proving the claim.


\end{document}